\documentclass[aoas]{imsart}

\RequirePackage{amsthm,amsmath,amsfonts,amssymb}
\usepackage[title]{appendix}
\RequirePackage[authoryear]{natbib}
\usepackage{booktabs}
\usepackage{caption}
\usepackage{centernot}
\usepackage{color}
\RequirePackage[colorlinks,citecolor=blue,urlcolor=blue]{hyperref}
\usepackage{colortbl}
\usepackage{enumerate}
\usepackage{float}
\usepackage{graphicx}
\usepackage{listings}
\usepackage{longtable}
\usepackage{lscape}
\usepackage{mathtools}
\usepackage{multirow}
\usepackage{placeins}
\usepackage{rotating}
\usepackage{soul}
\usepackage{tabu}
\usepackage{tikz}
	\usetikzlibrary{positioning}
	\usetikzlibrary{shapes}
	\tikzstyle{gene}=[circle,
	                  draw,
	                  thick,
	                  inner sep = 0pt,
	                  minimum size = 8mm]
	\tikzstyle{variant}=[diamond,
	                     draw,
	                     thick,
	                     inner sep = 0pt,
	                     minimum size = 9.5mm]
	\tikzstyle{eqtl}=[ellipse,
	                  draw = none,
	                  fill = eqtl_color,
	                  inner sep = 2pt,
	                  align = center]
	\tikzstyle{express}=[rectangle,
	                     draw = none,
	                     fill = express_color,
	                     inner sep = 2pt,
	                     align = center]
\usepackage{txfonts}
\usepackage{url} 
\usepackage{xcolor}
	\definecolor{garnet}{RGB}{210,38,48}
	\definecolor{sweetpotato}{RGB}{204,85,0}
	\definecolor{royalblue}{RGB}{0,119,204}
	\definecolor{eqtl_color}{RGB}{135,206,250}
	\definecolor{express_color}{RGB}{252, 181, 163}
\usepackage[all]{xy}

\startlocaldefs

\theoremstyle{plain}
\newtheorem{theorem}{Theorem}
\newtheorem{case}{Case}

\usepackage{newfloat}
\DeclareFloatingEnvironment[name={Supplementary Figure},fileext=lsf,listname={List of Supplementary Figures}]{suppfigure}
\DeclareFloatingEnvironment[name={Supplementary Table},fileext=lsf,listname={List of Supplementary Tables}]{supptable}

\newcommand{\indep}{\perp\!\!\!\!\perp}
\newcommand{\dep}{\not\!\perp\!\!\!\!\perp}
\newcommand{\msetrue}{MSE$_2$ }

\newcommand{\mc}{MC$^3$ }

\newcommand{\boldtheta}{\boldsymbol \theta}
\newcommand{\boldS}{\textbf{S}}
\newcommand{\boldT}{\textbf{T}}

\endlocaldefs

\begin{document}

\begin{frontmatter}
\title{Approximate Bayesian inference of directed acyclic graphs in biology with flexible priors on edge states}
\runtitle{Edge-level inference of directed acyclic graphs}

\begin{aug}
\author[A]{\fnms{Evan A}~\snm{Martin}}
\and
\author[B]{\fnms{Audrey Qiuyan}~\snm{Fu}\ead[label=e2]{audreyf@uidaho.edu}}
\address[A]{The Bioinformatics and Computational Biology Program, University of Idaho}

\address[B]{The Bioinformatics and Computational Biology Program, \\
    Department of Mathematics and Statistical Science, \\
    Institute for Modeling Collaboration and Innovation, \\
    Institute for Interdisciplinary Data Sciences, University of Idaho \printead[presep={,\ }]{e2}}
\end{aug}

\begin{abstract}
Graphical models or networks describe the statistical dependence among multiple variables and are widely used in biology (e.g., gene regulatory networks). Under appropriate assumptions, directed edges may represent causal relationships. A key feature of a biological network is sparsity, defined by how likely an edge is present, of which we often have some knowledge. However, most existing Bayesian methods use priors for the entire graph, making it difficult to specify the level of sparsity. The few methods that use priors on edges estimate the two directions independently; the sum of the two probabilities can exceed 1. Here, we present baycn (BAYesian Causal Network), a novel approximate Bayesian method that represents a graph in terms of three states of edges: the two directions and edge absence, and specifies priors on these edge states. We design a pseudo Bayesian sampling algorithm for efficient inference. We apply baycn to two genomic problems: i) distinguishing direct and indirect target genes of genetic variants, using these variants as instrumental variables, and ii) inferring combinatorial binding of highly-correlated transcription factors in Drosophila. In both cases and in extensive simulations, our method demonstrates much improved accuracy over existing methods for the whole graph and for individual edges.  
\end{abstract}


\end{frontmatter}


Graphical models (or networks), which may include directed and undirected edges, can be used to represent the statistical dependence among multiple variables and have wide applications in biology, such as gene regulatory networks \citep{friedman2004, rau2012} and protein-protein interaction networks \citep{gomez2003, zhang2012}.  Under appropriate assumptions, directed edges in a network may represent causal (or regulatory) relationships \citep{spirtes2000, guyon2010, dawid2010} (also see Discussion). 

A key feature of graphs is their sparsity, which measures how many edges are present in a graph, or how likely an individual {edge} is expected to be present.  
We often have some knowledge about the sparsity of a graph, especially in biology
where it remains a challenge to infer a reliable gene regulatory network due to complexity in biological processes and high dimension of biological
variables involved \citep{mcguire2020road, davidson2010emerging}.
For example, \citet{guelzim2002topological} estimated that
there are about 6000 genes in the yeast genome, and that on average each gene is regulated by only 2.76 regulators (such as transcription factors), 
meaning that the probability of edge presence in the graph 
is around 2.76/6000 a priori.  This example shows that we describe sparsity more easily at the edge
level than for the whole graph, and that it is often of interest to infer the probability of a state of an edge: the two directions if present, and absence.

Directed acyclic graphs (DAGs), often also called Bayesian networks, are graphs with only directed edges and no directed cycles.  
Many Bayesian methods have been developed for DAG inference, which explore the graph space and draw a sample graph, either directly from a closed-form posterior distribution \citep{schwaller2019, leday2019, yeung2011, lo2012, young2014, fronczuk2015, hung2017} or more often use a Markov Chain Monte Carlo (MCMC) algorithm \citep{madigan1995, friedman2003, giudici2003, grzegorczyk2008, he2013, mohammadi2015, goudie2016, su2016, kuipers2017, castelletti2019, rezaei2019, viinikka2020, castelletti2020, kuipers2022, castelletti2022, castelletti2022bcdag} 
to sample the posterior distribution. 

Most of these methods estimate the probabilities for edge directions and absence, 
although the priors are typically for all DAGs or for node orderings/partitions \citep{madigan1995, goudie2016, friedman2003, su2016, kuipers2017, rezaei2019, viinikka2020, castelletti2020, kuipers2022}. 
Such priors do not translate easily to priors on the states of an individual edge.  {For example, a uniform prior is often assumed for all
the DAGs in these methods.  
In principle, one can calculate the overall probability of edge presence versus absence, dividing the total number of edges in all DAGs by the number
of DAGs.  The probability of each direction is then half the 
probability of edge presence.  However, this calculation is not trivial, as the total number of DAGs is generally difficult to obtain.}
An exception is \citet{castelletti2022}, which constructed an edge-level prior as a proxy for the prior on the entire graph. 

Several methods estimate the probability of the two directions, but only for the edges that are deemed present \citep{yeung2011, lo2012, young2014, hung2017}.   
In particular, \citet{young2014} used the prior above on gene regulators in their scanBMA method for gene expression data (see Section~\ref{sec:relation}).
However, the probability of edge absence is not directly calculated.  The posterior probabilities of the two possible directions are also treated as independent of each other.  As a result, both probabilities can sum to above 1.  
  
In our development, we formulate priors by considering three states for an edge: two directions and edge absence.
We introduce a new Bayesian method that substantially improves the prior specification and estimates the posterior 
probabilities of three possible states of an edge.
We represent a Bayesian network by specifying the graph in terms of the edge states (i.e., two directions and edge absence). 
This framework allows the user to specify a prior distribution for the edge states, 
which is easier to formulate with biological information and to specify the level of sparsity in the graph.
We take a pseudo Bayesian approach to designing the sampling algorithm for efficient inference.
To reduce the size of the search space and speed up inference, our method can further take as input a graph 
from another more efficient graph inference method, e.g., constraint-based methods such as \citet{badsha2021, kalisch2012} and \citet{scutari2009}.

When instrumental variables are available, our method may be used for inferring causal networks.  In genetics, for example, a genetic variant (denoted 
$T_1$) regulating the expression of a gene ($T_2$) can be used as an instrumental variable to better understand the regulatory relationships among genes, 
as the genotypes of individuals in a natural population are randomized.  
If the expression of another gene $T_3$ tracks (i.e., is associated with) the changes in $T_2$, and if no other processes can explain such ``tracking" (i.e., no confounding 
variables), then $T_2$ is the cause of $T_3$; in other words, this reasoning leads to the inference of a small graph $T_1 \rightarrow T_2 \rightarrow T_3$.  
This is the essence of the principle of Mendelian randomization \citep{davey2014mendelian, 
emdin2017mrreview, badsha2019}.  When instrumental variables are available and confounding variables are appropriately accounted for, this principle 
translates to a 
constraint in network inference, which requires the edge connecting the variant and the gene to point only to the gene, and not the other way around \citep{badsha2019, badsha2021}.  
Our method allows for this constraint, which often greatly reduces the graph space, and makes it possible to orient other edges.  

We demonstrate the performance of our method through extensive simulation and in two biological applications. The first application examines the regulatory relationships among the target genes of genetic variants, while accounting for confounding variables in the inference.  
The second application disentangles combinatorial transcription factor binding during Drosophila mesoderm development, using a dataset with known edges supported by numerous biological experiments.  
In both cases, our method demonstrates much improved accuracy over existing methods.

\section{Methods}
\label{sec:meth}
\subsection{A Bayesian graphical model for edge states}

A graph $\mathcal{G} = (\bold V, \; \bold E)$ is a set of vertices (nodes) $\bold V = \{1, 2, ..., b\}$ and edges $\bold E \subseteq \bold V \times \bold V$, where $\bold V \times \bold V$ is the set of all ordered pairs of nodes, such as $(j, k)$, which denotes an edge pointing from node $j$ to node $k$ where $j, k \in \bold V$. 
The structure (or topology) of the graph is typically defined by the (deterministic) adjacency matrix $\bold A$ of dimension $b \times b$, where $A_{\scriptstyle {j k}} = 1$ and $A_{\scriptstyle {k j}} = 0$ represent the edge $j\rightarrow k$, and $A_{\scriptstyle {j k}} = 0$ and $A_{\scriptstyle {k j}} = 1$ represent $k \rightarrow j$. If $A_{\scriptstyle {j k}} = A_{\scriptstyle {k j}} = 0$, there is no edge between nodes $j$ and $k$. 

We introduce baycn (BAYesian Causal Network), an alternate representation of the graph to directly describe the states of individual edges with the vector $\bold S = (S_1, S_2, \dots, S_m)$, where $m$ is the number of edges.  Note that the single subscript here denotes the index of the edge.  An edge between nodes $j$ and $k$ (without loss of generality, we will assume $j<k$)  is in state $S_i$, which can take on three values: $S_i=0$ if $j \rightarrow k$, $S_i=1$ if $k \rightarrow j$, and $S_i=2$ if the edge is absent.  
The edge state probability is then $p_s \equiv \Pr(S_i=s)$
and
$\sum^2_{s=0} p_s = 1$.
If $p_0=p_1=0.5$, {we consider the edge bidirected with both directions being equally likely}.

When we want to emphasize the two nodes of an edge, we use the notation $S_{jk}$ with the double subscript for nodes $j$ and $k$.  This notation allows us to connect the edge vector with the adjacency matrix, which is defined on the nodes:
$\Pr (S_{jk}=0) = \Pr (A_{jk}=1)$, 
$\Pr (S_{jk}=1) = \Pr (A_{kj}=1)$,
and $\Pr (S_{jk}=2) = 1 - \Pr (A_{jk}=1) - \Pr (A_{kj}=1)$.

{When data is available at all the $b$ nodes, we denote the data as $\bold T = (T_1, T_2, ..., T_b)^T$, where each $T_j$ represents the random variable at node $j$}.
We aim to infer the posterior edge state probability
$\Pr (S_i | \bold T)$ for all edges. 
The posterior probabilities of the three states for an edge 
also add up to 1.  We can also represent these posterior probabilities in a probabilistic adjacency matrix, where each entry is the probability 
of one of the two possible directions: $\Pr (S_{jk}=1 | \boldT)$ or $\Pr (S_{kj}=1 | \boldT)$.  Existing Bayesian methods generally produce such a posterior probabilistic adjacency matrix.

\begin{figure}[H]
\centering
\includegraphics[scale = .6]{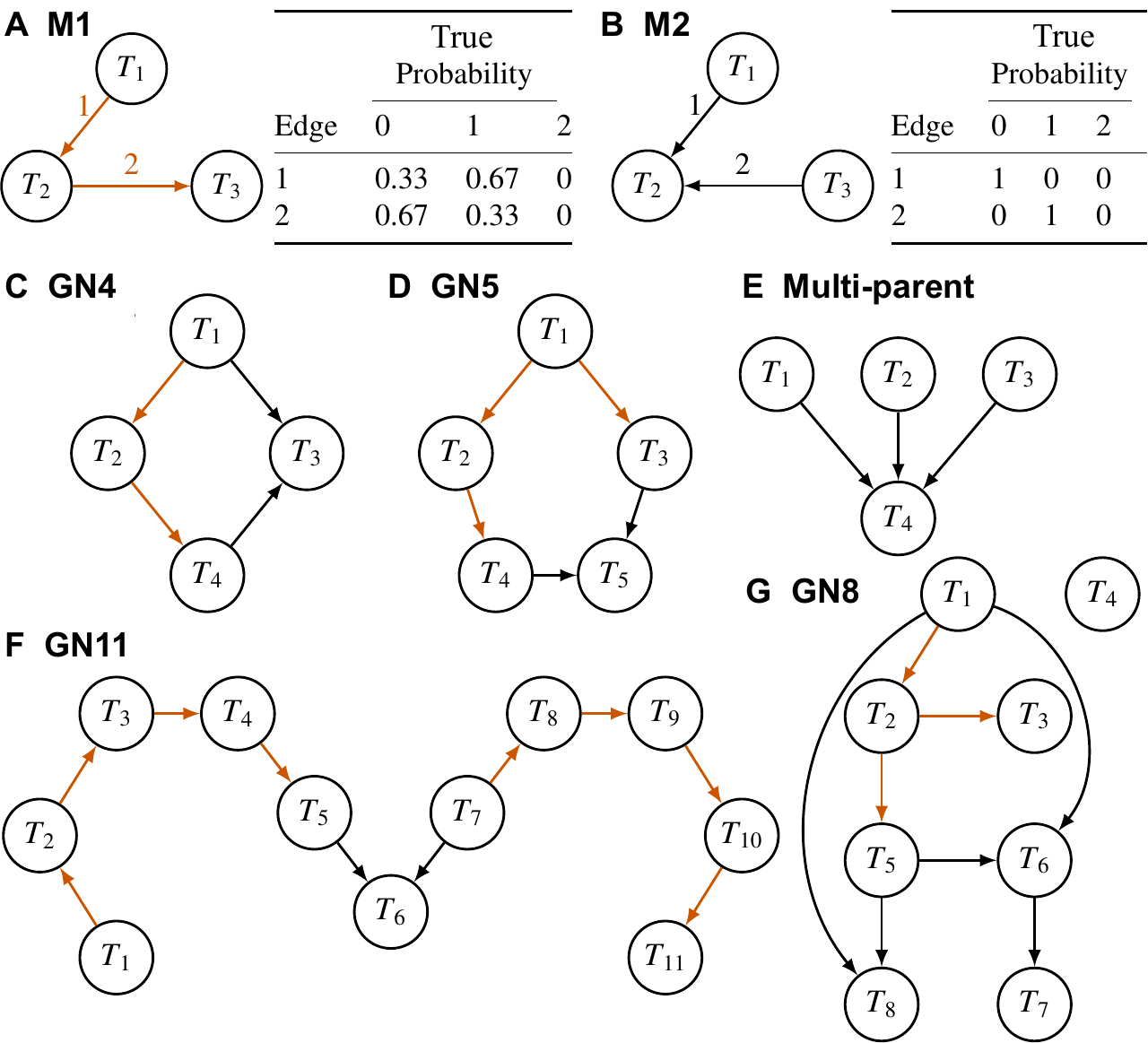}
\caption{\label{fig:allTopo} {\bf Seven topologies used in simulation studies.} Orange edges have Markov equivalent edges and cannot be deterministically inferred. 
({\bf A}) A mediation model where $T_2$ is the mediator. ({\bf B}) A v-structure.  {True probabilities on edge states account for 
Markov equivalence.  The values show the existence of Markov equivalent graphs for M1 and no Markov equivalent graphs for M2}.  
({\bf C})-({\bf G}) Larger networks that contain M1 and M2 as subgraphs. 
See Supplementary Figures~\ref{sfig:gn4}-\ref{sfig:gn8} for the true edge state probabilities of the orange edges.}
\end{figure}

Edge state probabilities need to account for Markov equivalence.  Two graphs are Markov equivalent if they have the same likelihood and 
represent the same conditional independence \citep{verma1990}.  A set of Markov equivalent graphs form a Markov equivalence class. 
For example, the canonical mediation model of $T_1 \rightarrow T_2 \rightarrow T_3$ has two Markov equivalent graphs: 
$T_1 \leftarrow T_2 \rightarrow T_3$, and $T_1 \leftarrow T_2 \leftarrow T_3$ (Figure~\ref{fig:allTopo}A).  All three graphs depict marginal 
dependence between $T_1$ and $T_3$ (i.e., $T_1 \dep T_3$) and conditional independence given $T_2$ (i.e., $T_1 \indep T_3 \; | \; T_2$).
Accounting for these Markov equivalent graphs, the true edge state probabilities are $(0.33, 0.67, 0)$ for the edge 
$T_1 \rightarrow T_2$ and $(0.67, 0.33, 0)$ for $T_2 \rightarrow T_3$.  In contrast, the graph $T_1 \rightarrow T_2 \leftarrow T_3$  
(Figure~\ref{fig:allTopo}B), also known as a v-structure, represents marginal independence between $T_1$ and $T_3$ and their conditional 
dependence ($T_1 \dep T_3 \; | \; T_2$).  This graph has no Markov equivalent graphs, and the true edge state probabilities are either 
0 or 1.  

The probability of the graph can be written as a product of conditional probabilities where each node is conditioned on its parents
\begin{align}
\Pr (\bold T \; | \; \bold S, \boldsymbol \theta) = \prod_{j = 1}^b \Pr (T_j \; | \; pa (T_j), \boldsymbol \theta_j),
\label{eqn:cond}
\end{align}
where $pa (T_j)$ is the set of parent nodes of $T_j$, $\boldsymbol{\theta}_j$ is the parameter vector for the distribution of $T_j$. 
When $pa (T_j) = \emptyset$, where $\emptyset$ is an empty set, the probability is reduced to a marginal probability $\Pr (T_j \; | \; \boldsymbol \theta_j)$.

When we assume normality for the data at each node: 
\begin{align}
T_j \sim \text{N}(\mu_j, \sigma^2_j), \;\; \text{and} \;\;
\mu_j = \beta_{0j} + \sum_{k \in pa (T_j)} \beta_{kj}  T_k,
\label{eq:lm}
\end{align}
where $\mu_j$ is the mean and $\sigma^2_j$ the variance. If there are no parents, 
$\mu_j = \beta_{0j}$.

Alternatively, if the data is binary:
\begin{align}
T_j  \sim \text{Bernoulli}(p_j),  \;\; \text{and} \;\;
\text{log} \Bigg( \frac{p_j}{1 - p_j} \Bigg)  = \beta_{0j} + \sum_{k \in pa (T_j)} \beta_{kj}  T_k,
\end{align}
where $p_j$ is the ``success'' probability.
Without parents, $\text{log}\big[ p_j / (1 - p_j) \big] = \beta_{0j}$.


\subsection{The prior distribution and a Metropolis-Hastings-like iterative algorithm using pseudo-likelihood} \label{sec:MHalg}

We aim to develop an iterative sampler that proposes changes to edge states and estimates the 
posterior probabilities, while accounting for Markov equivalence.
Since our interest here is in learning the graph structure, but not the means and variances of the effect sizes of the parent nodes on their child nodes, we view 
the parameters represented by $\boldtheta$ (as in Equation~\ref{eqn:cond}) as nuisance parameters.  
This means that we are interested in the posterior distribution only for $\boldS$:
$\Pr (\boldS | \boldT) \propto \Pr (\boldT | \boldS) \Pr (\boldS)$, 
in which the likelihood $\Pr (\boldT | \boldS)$ of the current graph involves $\boldtheta$.
Bayesian methods using pseudo-likelihoods, such as conditional, marginal, and profile likelihoods, 
have been widely used for elimination of nuisance parameters in a computationally 
efficient way (see the reviews in \citet{ventura2016pseudo} and \citet{severini2000likelihood}, although \citet{ventura2016pseudo} considers
such methods ``hybrid/quasi/pseudo Bayesian" methods and the posteriors ``pseudo-posteriors").  
{Here, we calculate the likelihood $\Pr (\boldT | \boldS)$ by plugging the maximum likelihood estimates of $\boldtheta$ for the given graph into Equation~\ref{eqn:cond}.}
As a result, we call our sampler a ``Metropolis-Hastings-like iterative algorithm".  We demonstrate
through simulations the accuracy and computational efficiency of this algorithm.

To infer the posterior of $\Pr (\boldS | \boldT)$,
we will use a prior that assumes all the edges to be independent:
$\Pr (\boldS) = \prod_m \Pr (S_i = s_i)$.
We can construct the prior based on our knowledge or belief on graph sparsity.  For example, as mentioned in the Introduction,
in the yeast genome we expect the prior probability for each direction to be 2.76/6000/2, and that for edge absence 
to be 1-2.76/6000.
Once the probabilities for each edge are specified, we can multiply the probabilities of all the edges as in the equation above
to obtain the probability of all the edges.
Although the construction of $\Pr (\boldS)$ treats the edges independently,
the posterior probabilities $\Pr (\boldS | \boldT)$ account for the dependence among the edges through the likelihood.  
Such a prior is easy to construct: for example, in biology 
one can easily find information on the mean number of connections certain genes may have.  A similar prior has been used in the scanBMA 
method \citep{young2014}; see Section~\ref{sec:relation}.  

The input of our algorithm is the binary adjacency matrix of a candidate graph and the data at the nodes.  The candidate graph may be a fully connected graph, where all nodes are connected.  A more efficient approach is to run a fast graph inference algorithm to produce a candidate graph and use it as the input, even if this graph contains false edges.


At the $t$th iteration, the key steps of the Metropolis-Hastings-like algorithm are:
\begin{enumerate}


\item \label{item:proposal} Generate a proposal graph $\bold S^\prime_{\scriptstyle (t)}$ from the current graph $\boldS_{(t-1)}$.  When $t=1$, the current graph is randomly generated from the candidate graph in the input.


\item \label{item:cycles} Check for and remove directed cycles in $\bold S^\prime_{\scriptstyle (t)}$.


\item \label{item:ratio} Calculate the acceptance probability $\alpha_{\scriptstyle (t)}$
\begin{equation} \label{eq:alpha}
\alpha_{\scriptstyle (t)}  = \min \bigg\{ 
		\frac{\Pr (\bold S^\prime_{\scriptstyle (t)})  
			\Pr (\bold T \; | \; \bold S^\prime_{\scriptstyle (t)}, \boldtheta_{\scriptstyle (t)})   
			\Pr (\bold S_{\scriptstyle (t - 1)} \; | \; \bold S^\prime_{\scriptstyle (t)})}
		{\Pr (\bold S_{\scriptstyle (t - 1)})  
			\Pr (\bold T \; | \; \bold S_{\scriptstyle (t - 1)}, \boldtheta_{\scriptstyle (t - 1)})   
			\Pr (\bold S^\prime_{\scriptstyle (t)} \; | \; \bold S_{\scriptstyle (t - 1)})}, \; 1 \bigg\},
\end{equation}
where $\Pr (\boldS)$ is the prior probability of the edge states, $\Pr (\boldT \; | \; \boldS, \boldtheta)$ the graph likelihood, and $\Pr (\boldS \; | \; \boldS')$ the transition probability.  {Here, we estimate $\boldtheta$ by maximum likelihood estimation in the corresponding graph.}


\item \label{item:ar} Generate a random probability $u$ from the uniform distribution U$(0, 1)$. 
	Accept the proposal and set $\bold S_{\scriptstyle (t)} = \bold S^\prime_{\scriptstyle (t)}$ if $u < \alpha_{\scriptstyle (t)}$; 
	or stay at the current graph and set $\bold S_{\scriptstyle (t)} = \bold S_{\scriptstyle (t-1)}$ otherwise.
\end{enumerate}

To generate the proposal graph in step \ref{item:proposal}, we first determine the number of edges to change states by sampling from a binomial distribution { Bin$(m,1/m)$, where  $m$ is the number of edges in the network and $1/m$} is then the ``success" probability.  For each of the selected edges, we then sample from a Bernoulli distribution with probability $p$ to decide which edge state to change to. Since we do not allow for an edge to remain in the same state, $p$ is determined by the prior of the two remaining edge states. For example, if an edge in state 0 is selected to change states, and if the prior probability for the edge states is $(p_0, p_1, p_2) = (0.05, 0.05, 0.9)$, then the probability of switching to state 1 is $p = 0.05/(0.05 + 0.9)$. 

To ensure that the proposal graph does not contain directed cycles, we have further designed two algorithms as part of our algorithm { (Supplementary Note S1)}: the ``cycle finder'' algorithm finds all possible cycles given the input graph; and the ``cycle remover'' algorithm removes all directed cycles from the proposal graph.

The binomial and Bernoulli probabilities above are then used to calculate the transition probability in step \ref{item:ratio}.  We show that the transition probabilities do not depend on the path taken from the current graph to the proposed graph (the process of introducing and removing directed cycles) but only on the edges that have different states between the two graphs (see the proof in Supplementary Note S2 and examples in Supplementary Note S3):  

\begin{theorem}
When calculating the acceptance probability, the transition probabilities between the current graph $\boldS$ and proposed graph $\boldS^{\boldsymbol \prime}$, $\Pr (\boldS^{\boldsymbol \prime} | \boldS)$ and $\Pr (\boldS | \boldS^{\boldsymbol \prime})$, depend only on the edges whose states are different between the two graphs.
\end{theorem}

\begin{figure}[H]
\centering
\includegraphics[width = \textwidth]{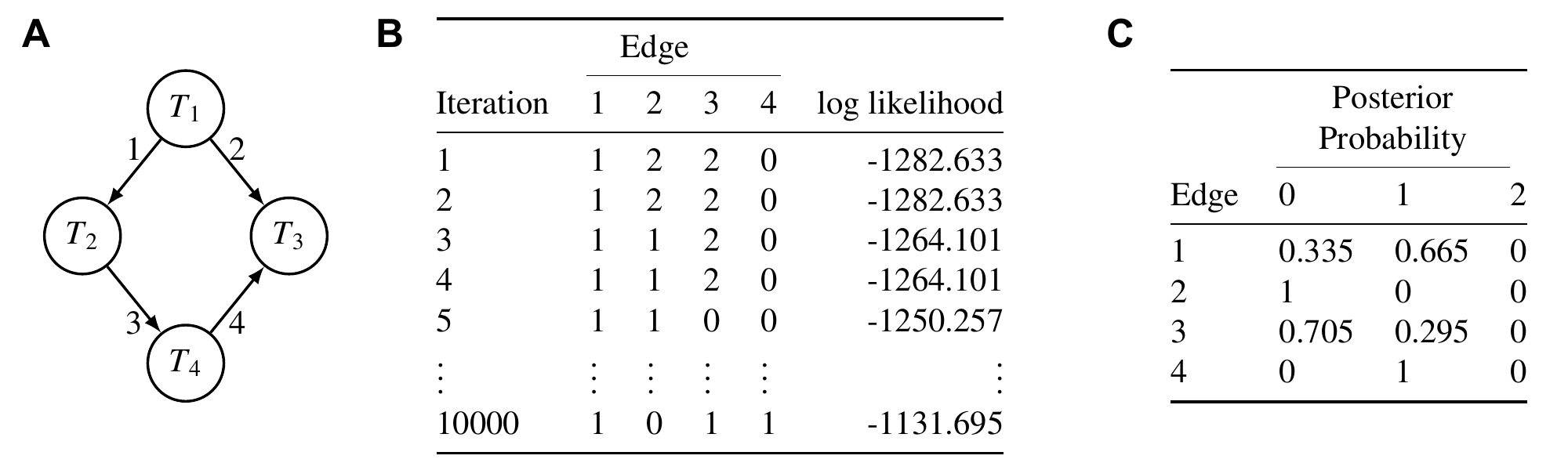}
\caption{{\bf An example of the output from baycn}. ({\bf A}) The true graph GN4. The candidate graph used for inference will also consider only these four edges.  ({\bf B}) Edge states and log likelihood for the graph accepted at each iteration of the Metropolis-Hastings-like algorithm. ({\bf C}) The proportion of each edge state in the sample provides an estimate of the posterior probability of the edge state.}
\label{fig:output}
\end{figure}

This algorithm generates a sample of graphs represented by edge states (Figure~\ref{fig:output}).  For each edge, the relative frequencies of the three states in the sample provide {an estimate of the marginal posterior probabilities of edge states $\Pr (S_i \; | \; \bold T)$, marginalized across the sampled DAGs}.

Through changes in edge states, our algorithm can sample from multiple Markov equivalent graphs and thus produce posterior probabilities that account for Markov equivalence.  With sufficient data, the posterior probabilities of edge states should be the same asymptotically as expected under Markov equivalence.

In causal inference with instrumental variables, we need to restrict the direction of the edge connecting the instrumental variable and other nodes.  Our 
algorithm achieves this by generating proposals with this restriction.  For example, when the input is the undirected network $T_1-T_2-T_3$, it may imply four possible directed networks as mentioned earlier: i) $T_1 \rightarrow T_2 \rightarrow T_3$, ii) $T_1 \leftarrow T_2 \leftarrow T_3$, iii) $T_1 \leftarrow T_2 \rightarrow T_3$, and iv) $T_1 \rightarrow T_2 \leftarrow T_3$.  
If $T_1$ is a potential instrumental variable, an edge between $T_1$ and $T_2$, when existing, can point only to $T_2$.  Then the possible directed networks are reduced to only i) and iv), which have different likelihoods and 
are therefore distinguishable.  In this simple example, our algorithm can propose only two moves for the edge $T_1-T_2$ ($T_1 \rightarrow T_2$, and no edge) but three moves for the edge $T_2-T_3$ ($T_2 \rightarrow T_3$, $T_2 \leftarrow T_3$, and no edge).


\subsection{Relationship to existing Bayesian methods}
\label{sec:relation}
We compare baycn to five methods (Table \ref{tab:methods}), four of which are MCMC methods. Two methods are based on the graph structure: Gibbs \citep{goudie2016} and MC$^3$ \citep{madigan1995}, and two methods are based on node orderings or node partitions (i.e., subsets of nodes): order MCMC \citep{friedman2003, kuipers2022} and partition MCMC \citep{kuipers2017, kuipers2022}. These methods assign a prior for the entire graph or for the node ordering/partition.


\begin{landscape}
\begin{table}
\caption{\label{tab:methods} {\bf Summary of our method and other Bayesian methods for network inference.}}
\begin{tabu}spread 0pt{@{} >{\raggedright\arraybackslash}p{1.85cm} >{\raggedright\arraybackslash}p{3cm} >{\raggedright\arraybackslash}p{3.25cm} >{\raggedright\arraybackslash}p{3.25cm} >{\raggedright\arraybackslash}p{3.75cm} >{\raggedright\arraybackslash}p{3.75cm} >{\raggedright\arraybackslash}p{2.5cm} @{}}
\toprule
 & baycn & Gibbs \citep{goudie2016} & MC$^3$ \citep{madigan1995} & order MCMC \citep{friedman2003, kuipers2022} & partition MCMC \citep{kuipers2017, kuipers2022} & scanBMA \citep{young2014}\\
\midrule

Input & Data matrix, adjacency matrix, prior on edge states & Data matrix, edge list (optional), prior on graph space & Data matrix, edge list (optional), prior on graph space & Data matrix, edge list (optional), parameters for graph score & Data matrix, edge list (optional), parameters for graph score & Data matrix, prior for edge presence \\

\\

Output & Posterior probabilities of three edge states & Posterior probabilistic adjacency matrix & Posterior probabilistic adjacency matrix & Posterior probabilistic adjacency matrix  & Posterior probabilistic adjacency matrix & Posterior probabilities of parent nodes\\

\\

Sampling space & Candidate edges and their direction & Parent nodes & Candidate edges and their direction & Node orderings & Node partitions & Parent nodes\\

\\

Sampling steps & Change states for $>= 1$ edges & Change parents for $>= 1$ nodes & Add or remove an edge & Move a node to a new position & Split or merge existing partitions & Closed form for posterior \\

\\

Type of prior & Edge states & Parameters and graph structure & Parameters and graph structure & Parameters and node orderings & Parameters and node partitions & Parameters and edges\\

\\

Type of data & Discrete, continuous, or mixed & Discrete or continuous & Discrete or continuous & Discrete or continuous & Discrete or continuous & Continuous \\

\\

{R package} & baycn & structmcmc & structmcmc & BiDAG & BiDAG & networkBMA \\

\bottomrule
\end{tabu}
\end{table}
\end{landscape}

We also compare with scanBMA \citep{young2014}, a Bayesian model averaging method that specifies a prior probability for individual edges \citep{yeung2011, lo2012, young2014}.  {As mentioned in the Introduction, one of the priors assigns the same probability to all edges, and this probability 
is $g_1/g_2$, where $g_1$ is the expected number of the regulators for any gene, and $g_2$ the number of all possible regulators (they used 2.76/6000).  
They further allow for this probability to vary for different genes in another prior.
scanBMA searches the space of parents for each node in the network independently.  As a result, the two probabilities it estimates for the two directions are independent of each other and can sum to a value above 1.  Additionally, it does not estimate the probability of edge absence.   


\subsection{Simulation studies}
\label{sec:simu}
We simulated data under seven different topologies (Figure \ref{fig:allTopo}). Each node was simulated under a linear model in Equation~\ref{eq:lm} with the variance set to 1 and the intercept $\beta_0$ set to 0.
All the other coefficients take the same value, which is referred to as the signal strength.  We considered three levels of the signal strength:
0.2 (weak), 0.5 (moderate), and 1 (strong), and three levels of the sample size: 100, 200 and 600. For each topology, we simulated 25 datasets under each of the nine combinations of signal strength and sample size. When summarizing the results, if the results are grouped by, for example, topology, we use the output from all the datasets with different sample sizes and signal strengths.  

We evaluate the performance using the following metrics:
\begin{enumerate}


\item The edgewise Mean Squared Error (eMSE): This is the MSE between the true probabilities and posterior probabilities of the three states for edge $i$:
\begin{equation} \label{eq:emse}
\text{eMSE}_i = \frac{1}{3} \sum_{s = 0}^2 \big[\Pr (S_i=s)^{\text{true}} - \Pr (S_i=s \; | \; \bold T) \big]^2,
\end{equation}
where $\Pr (S_i=s)^{\text{true}}$ is the true edge state probability under Markov equivalence. 
This metric informs us which edges are more accurately inferred and which ones are not.  
{Note that the true probabilities can be calculated only when the structure of the true graph is known: we identify all possible graphs in the Markov equivalence class of the true graph, and then calculate the frequency of an edge being in each of the three states (see Supplementary Figures \ref{sfig:gn4}--\ref{sfig:gn8})}. 


  \item MSE$_1$: This is the MSE for the whole graph based on three possible edge states. It is the eMSE averaged over all $m$ edges in the graph: 
\begin{align} \label{eq:mse1}
\text{MSE}_1 = \frac{1}{m} \sum_{i = 1}^m \text{eMSE}_i = \frac{1}{3m} \sum_{i = 1}^m \sum_{s = 0}^2 \big[\Pr (S_i=s)^{\text{true}} - \Pr (S_i=s \; | \; \bold T) \big]^2.
\end{align}


  \item MSE$_2$: This is the MSE between the true and posterior probabilistic adjacency matrices on all $m$ edges.  In other words,
  this metric uses the probability only of the two edge directions.
  \begin{align}\label{eq:mse2}
  \begin{split}
  \text{MSE}_2 = \frac{1}{2m} \sum_{\substack{(j,k) \; \text{or} \\ \; (k,j) \in \bold E}} & \bigg\{ \big[\Pr (A_{jk}=1)^{\text{true}} - \Pr (A_{jk}=1 \; | \; \bold T) \big]^2 \\
  &\quad + \big[\Pr (A_{kj}=1)^{\text{true}} - \Pr (A_{kj}=1 \; | \; \bold T) \big]^2 \bigg\}.
  \end{split}
  \end{align}  


  \item Precision and power for the whole graph. Precision, or $1 - \text{False Discovery Rate (FDR)}$, measures how many of the inferred edges are true, and power measures how many of the true edges are correctly inferred.
 For calculation, we apply a cutoff to the posterior probability of edge presence.  These metrics ignore the nuances in the probabilities, but are easy to interpret and provide a quick indication of the inference accuracy.
\end{enumerate}

\section{Results}

\subsection{Results from simulation studies}

\subsubsection{Estimating posterior probabilities of edge states using true edges as the input}

We ran baycn once per simulated dataset, used a burn-in of 20\%, and used a sparse prior, $(p_0, p_1, p_2) = (0.05, 0.05, 0.9)$, on edge states. For M1, M2, GN4, GN5, and multi-parent topologies (Figure \ref{fig:allTopo}A-E), we ran baycn for 30,000 iterations with a step size of 120.  For GN8 and GN11 (Figure \ref{fig:allTopo}F-G), we ran baycn for 50,000 iterations with a step size of 200. 

As expected, MSE$_1$ (based on three edge states) decreases as $\beta$ and $N$ increase for each topology, and is particularly affected by $\beta$ (Supplementary Table~\ref{stab:mse1}).  
When MSE$_1 <0.1$, the inference is typically accurate: the direction of each edge is correctly inferred, and the posterior probabilities of each edge state is similar to that of the true probabilities. 
Using this cutoff, we observed that baycn performs well in general when the signal strength is not weak.
For both M2 and multi-parent topologies that only contain v-structures, however, MSE$_1$ is 
much larger at $\beta=0.2$. This is consistent with previous observation that it is generally difficult for existing graph inference algorithms to correctly identify v-structures with a weak signal \citep{badsha2019, badsha2021}. 


\subsubsection{Identification of false positive edges and the choice of priors}

For this assessment we include a false edge in M1, M2 and GN4, and two false edges in GN11 (Supplementary Figure~\ref{sfig:falseEdges}).
We used the same data generated in the previous section for these topologies (without false edges) and ran baycn with the input being the true edges plus the false edges. 
{The user can specify their prior belief on sparsity through the prior edge state probabilities.}  Here, we explored the impact of three edge state priors on the inference, with an increasing probability of edge absence: prior 1: $(p_0, p_1, p_2) = (1/3, 1/3, 1/3)$; prior 2: $(p_0, p_1, p_2) = (0.25, 0.25, 0.5)$, and prior 3: $(p_0, p_1, p_2) = (0.05, 0.05, 0.9)$.  

We calculated the eMSE for each edge
and again used $0.1$ as the cutoff. 
In all four graphs, the eMSE for the false edges decreases as $p_2$ increases, but the edge probabilities of the true edges are generally estimated correctly, even when the false edges are not properly identified and have a large eMSE (Supplementary Tables~\ref{stab:m1fe} - \ref{stab:gn11fe}).  On the other hand, baycn can identify false positive edges under prior 3
 (Supplementary Tables~\ref{stab:m1fe} - \ref{stab:gn11fe}), confirming that prior 3 is the prior of choice, as it balances the need to detect false positive edges and correctly infer true edges. 


\subsubsection{Comparison with existing Bayesian methods on simulated data}

To compare baycn with other methods, we focus on GN4, GN8 and GN11, which have different levels of complexity (Figure~\ref{fig:allTopo}). Using the true edges as the input to all the methods, we ran each method for 30,000 iterations on GN4 and for 50,000 iterations on GN8 and GN11 and used a burn-in of 20\%. We set the step size in baycn to 120 for GN4 and 200 for GN8 and GN11. We used the default step size for order and partition MCMC, which results in the step size being 30 for GN4 and 50 for GN8 and GN11. Gibbs and \mc use all of the iterations after the burn-in.

Across topologies and simulation scenarios, 
baycn and Gibbs show similar and better performance than other methods (Figure~\ref{fig:boxplot_mse}; Supplementary Table~\ref{stab:mse}). 
Partition and order MCMC are slightly worse in some cases, whereas \mc is the least accurate and least stable among the MCMC methods (Figure~\ref{fig:boxplot_mse}).

\msetrue (between the true and posterior probabilistic adjacency matrices) 
used in the comparison above does not apply to scanBMA, as the posterior probabilities from scanBMA 
have a different interpretation from the other methods (see Section~\ref{sec:relation}).
We therefore compare precision and power, 
in addition to MSE$_2$, on GN4 using a fully connected graph as the input (Figure~\ref{fig:boxplot_pp_mse}).
To calculate precision and power, we applied a cutoff of 0.5 to the probability of edge presence for all methods 
and thus derived the binary inference (presence or absence) for each edge.   
Precision is nearly perfect for baycn and the MCMC methods in most simulations (Figure~\ref{fig:boxplot_pp_mse}; Supplementary Table~\ref{stab:pr}). However, scanBMA has lower precision under strong and moderate signal, mainly due to its inability to infer the v-structure.
 Power is nearly 1 for all methods at $\beta=0.5\text{ and }1$ (Figure~\ref{fig:boxplot_pp_mse}; Supplementary Table~\ref{stab:pr}), 
 but is lower at $\beta=0.2$.  Variation in power is larger for order and partition MCMC, as they tend to infer a higher posterior probability for edge absence (Supplementary Tables~\ref{stab:gn4.n100}-\ref{stab:gn4.n600}).  In terms of MSE$_2$, order MCMC has a large variability at weak signal, whereas Gibbs and \mc are more variable at strong signal.


\begin{figure}[h!]
\centering
\includegraphics[width = \textwidth]{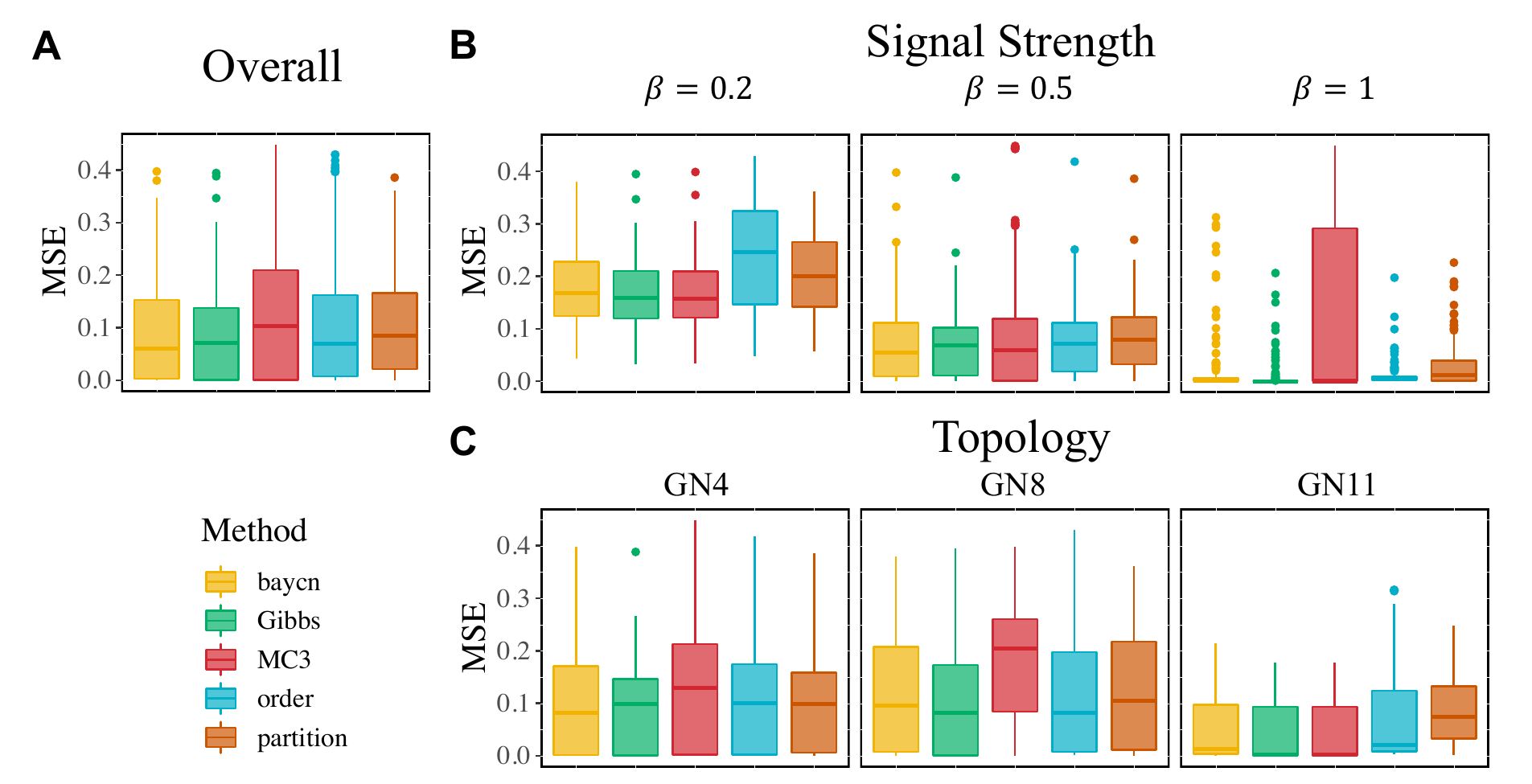}
\caption{\label{fig:boxplot_mse} {\bf Boxplots of \msetrue (on the posterior probabilistic adjacency matrix) for baycn and other Bayesian methods on simulated data from GN4, GN8, and GN11}. The true edges were used as the input to each method. 
({\bf A}) The overall \msetrue grouped by method. This plot combines the \msetrue for all topologies, signal strengths, and sample sizes. ({\bf B}) \msetrue grouped by method and signal strength $\beta$. 
({\bf C}) \msetrue grouped by method and topology. 
}
\end{figure}

To compare the runtime of these methods, we ran each algorithm on an Intel Xeon D-1540 2.00 GHz processor with 128 GB of memory on data from GN4, GN8, and GN11 with $\beta = 1$ and $N = 600$ (Table~\ref{tab:runtime}). 
For all three topologies, scanBMA is the fastest, and Gibbs and \mc the slowest.  Order MCMC, partition MCMC and baycn 
are in the middle: they are an order of magnitude slower than scanBMA, but an order of magnitude faster than Gibbs and \mc.  Among these three methods in the middle, 
order MCMC is faster, whereas {baycn and partition MCMC} are similar.  This comparison shows another advantage of baycn: apart from a more coherent framework, as well as 
improved accuracy and stability, baycn is also computationally efficient.  

\begin{table}[h!]
\centering
\caption{{\bf The mean runtime in seconds across 25 datasets}. For each topology, 25 datasets were generated with $\beta = 1$ and $N = 600$.  
Each algorithm was run once per dataset, and the runtime in seconds was recorded. 
All algorithms were run on an Intel Xeon D-1540 2.00 GHz processor with 128 GB of memory.}
\begin{tabu}{@{} l r r r r r r @{}}
\toprule
& \multicolumn{6}{c}{Runtime (seconds)} \\
\cmidrule(lr){2-7}
Topology & baycn & Gibbs & MC$^3$ & order & partition & scanBMA\\
\midrule
GN4 & 4.49 & 235.00 & 11.94 & 1.78 & 3.93 & 0.02\\
GN8 & 8.11 & 363.91 & 22.49 & 2.88 & 7.97 & 0.05\\
GN11 & 7.07 & 380.97 & 23.52 & 3.10 & 9.39 & 0.09\\
\bottomrule
\end{tabu}
\label{tab:runtime}
\end{table}


\begin{figure}[H]
\includegraphics[width = \textwidth]{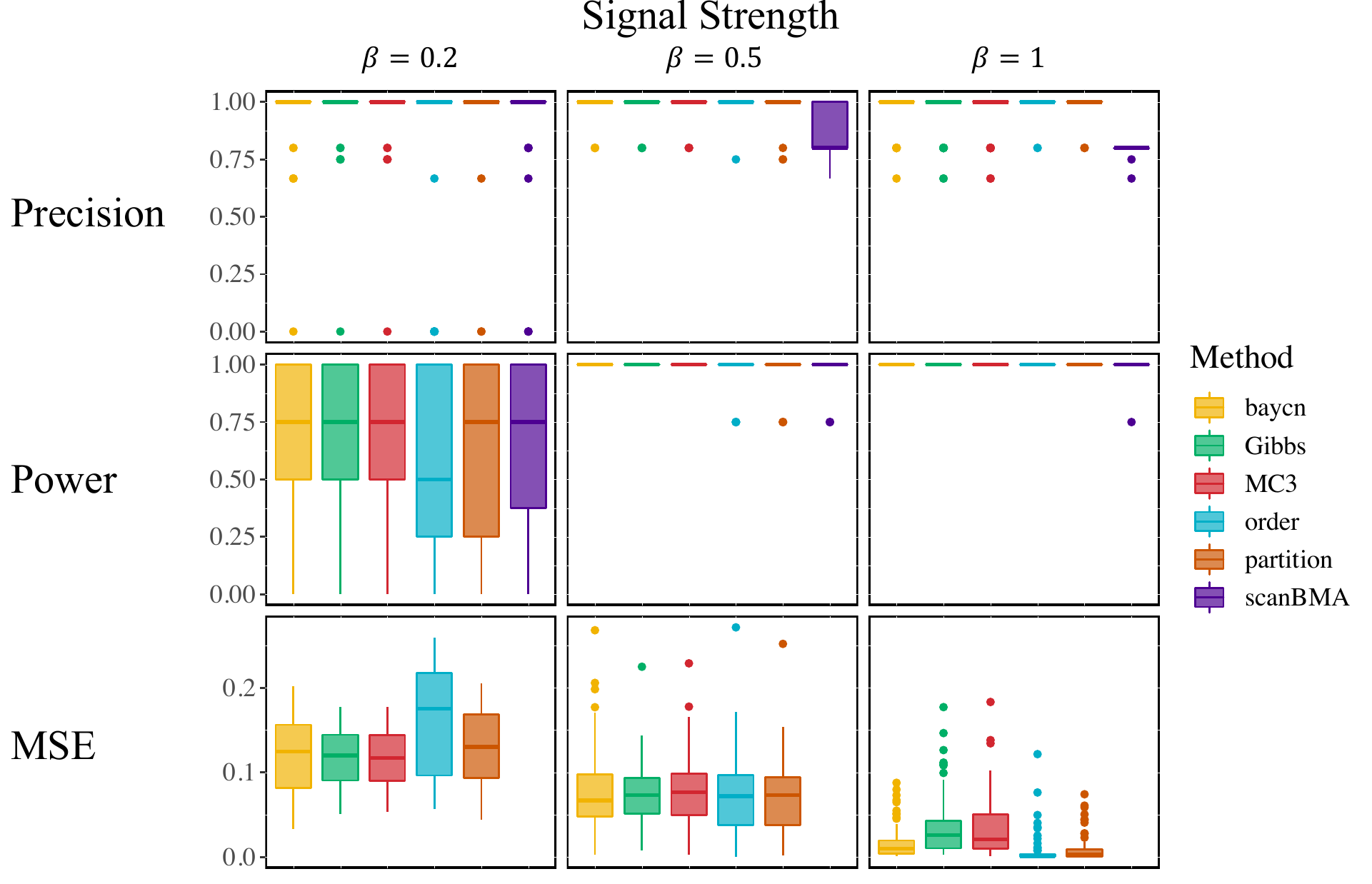}
\caption{\label{fig:boxplot_pp_mse} { Boxplots of precision, power, and \msetrue (on the posterior probabilistic adjacency matrix) for baycn and other Bayesian methods on simulated data from GN4}. A fully connected graph was used as the input to each method. 
\msetrue was not calculated for scanBMA, since the 
posterior probabilities from scanBMA have a different interpretation from the MCMC methods 
(see Section~\ref{sec:relation}).   
}
\end{figure}


\subsection{Application: causal inference of transcription regulation with genetic variants}

Genetic variants play an important role in regulating gene expression \citep{cheung2009}.  
The GEUVADIS (Genetic European Variation in Disease) project 
identified a large number of variants that are associated with the expression of one or more genes in Lymphoblastoid Cell Lines (LCLs) in Europeans and Africans \citep{lappalainen2013}.  Among these variants, which are termed expression quantitative trait loci (eQTLs),
62 are associated with more than one gene.  However, since the association analysis examined one eQTL-gene pair at a time, it is unclear which associated genes are more likely to be {direct targets (i.e., having an edge with the variant), and which ones indirect targets (i.e., not having an edge with the variant)}.  

To address this question, we infer graphs for these eQTL-gene sets, using each eQTL as an instrumental variable 
and applying the principle of Mendelian randomization \citep{sanderson2022mendelian}.  Under this principle, an edge connecting an eQTL and 
a gene points only to the gene, but not the other way around, since the DNA (eQTL here) regulates the RNA (expression). 
We focus on the larger European sample of 373 individuals.  
The gene expression data in GEUVADIS had been normalized using the PEER method \citep{stegle2012} to remove potential impact of demographic variables, batch effect, and other covariates.  
To account for confounding from genes outside the small graph, 
we performed principal component analysis on the gene expression from all genes in the GEUVADIS data. 
We used Holm's method to control the familywise type I error rate across all the p-values at 5$\%$ \citep{holm1979test},
and identified principal components (PCs) that are highly significantly associated with the eQTLs or genes in all the trios of that tissue.  
These PCs were then included in the network as confounding variables. The eQTL-gene sets Q8, Q21, Q23, Q37, and Q50 all have at least one PC associated with them while the eQTL-gene sets Q20 and Q62 do not have any PCs associated with them.

On each eQTL-gene-PC set, we used a fully connected graph as the input to baycn with a sparse prior $(p_0, p_1, p_2) = (0.05, 0.05, 0.9)$ .
We also included in baycn the constraint that a gene cannot be the parent of an eQTL. We ran baycn for 50,000 iterations with a burn-in of 20\% and a step size of 200. 

Based on the simulation results, we also ran order MCMC, partition MCMC, and scanBMA on the eQTL-gene set Q8 from GEUVADIS to compare with baycn (Figure~\ref{fig:q8_combined}). In all the methods, we used a fully connected graph as the input and applied the same constraint that a gene cannot be the parent of an eQTL.  Since scanBMA allows for a prior on edges, we set the probability of edge presence to be 0.1, which is the same as our sparse prior for baycn. We ran each MCMC method for 30,000 iterations and used a burn-in of 20\%.  We set the posterior probability cutoffs for edge presence to be 0.5, and considered an edge directed if the difference between the posterior probabilities for the two directions is greater than 0.2. }

Inferred graphs with the posterior probabilities  are shown in Figure \ref{fig:q8_combined} (see tables of posterior probabilities 
for all edges in 
Supplementary Tables~\ref{stab:q8.baycn}-\ref{stab:q8.scan}). The inference from baycn, order MCMC and partition MCMC is similar for most of the edges 
(Figure \ref{fig:q8_combined}A-C).  
However, order MCMC did not infer any edges  involving
the PCs, and partition MCMC inferred only one edge (RP11-203M5.8 - PC6) that has a relatively strong correlation of 0.25 (Figure \ref{fig:q8_combined}E).  
Since the PCs were treated the same as the
other nodes by these three methods, the results indicate that order and partition MCMC can identify only edges with strong correlations,
and are insensitive to moderate and weak correlations.
On the other hand, scanBMA is as sensitive to correlations as baycn is, but cannot infer edge absence or direction (Figure \ref{fig:q8_combined}D). 


\begin{figure}[h!]
\centering
\includegraphics[scale=.8]{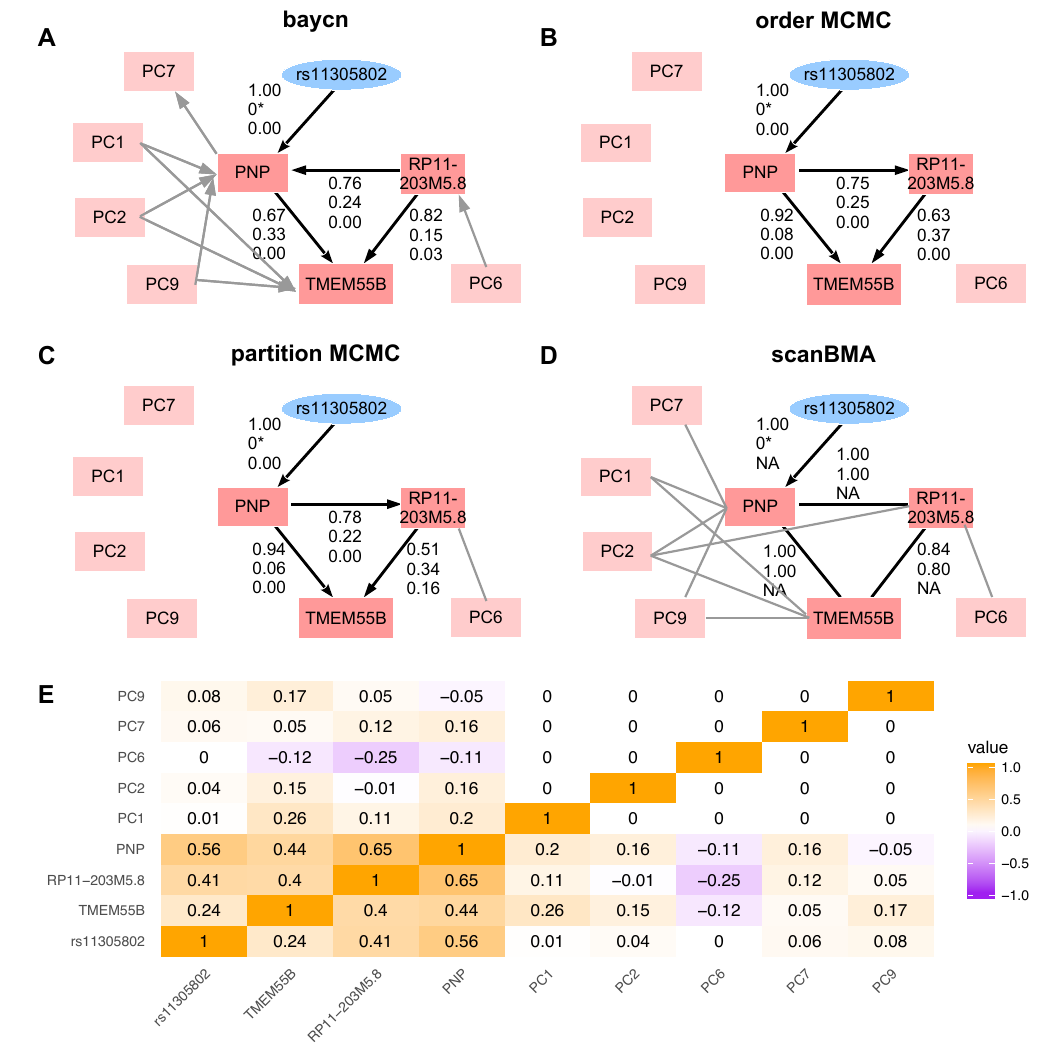}
\caption{\label{fig:q8_combined} {\bf Inference of the GEUVADIS eQTL-gene set Q8}. 
This eQTL is inferred to be associated with three genes.  ({\bf A})-({\bf D}) 
Graphs inferred by different methods with posterior probabilities shown for edges of biological interest.
The three probabilities are for the inferred direction shown in the figure, the opposite direction, 
and absence of the edge, respectively.  For an undirected edge, the two posterior probabilities for direction are close.
0* indicates that the directed edge pointing toward the eQTL was prohabited during inference.  
scanBMA is unable to infer the probability of edge absence (hence the NAs in {(\bf D)}) or distinguish 
the two directions.
({\bf E}) Heatmap of Pearson correlations in the data.
The posterior probability of all the edges are in Supplementary Tables~\ref{stab:q8.baycn}-\ref{stab:q8.scan}.
}
\end{figure}

The inference of the direction of the edge between PNP and RP11-203M5.8 is different under baycn and under order/partition MCMC (Figure \ref{fig:q8_combined}A-C).
This result is driven by PCs: when regressing RP11-203M5.8 on all the other variables, PC2 has a coefficient significantly different 
from 0 (p-value: 0.001), which means that RP11-203M5.8 and PC2 are conditionally dependent given other variables.  Since they are 
both connected to PNP, this conditional dependence means that RP11-203M5.8 and PC2 are both parents of PNP, thus forming a v-structure:
RP11-203M5.8 $\rightarrow$ PNP $\leftarrow$ PC2.  This is why baycn infers an edge from RP11-203M5.8 to PNP. 
To further verify the impact of PCs, we applied baycn to Q8, while excluding the PCs.  This analysis led to the same inference  
as order/partition MCMC, which effectively ignored the PCs (Supplementary Table~\ref{stab:q8}).  
These results demonstrate the importance of accounting for confounding variables
and that of having sufficient sensitivity to the signals in data.

Analyses of all the eQTL-gene-PC sets 
show that baycn can identify the regulatory relationship among multiple genes 
associated with the same eQTL, while accounting for confounding variables  
(Supplementary Figures~\ref{sfig:q20}-\ref{sfig:q62}; Supplementary Tables~\ref{stab:q20}-\ref{stab:q62}).
The posterior probabilities estimated by baycn are largely consistent with the sample correlations 
(Supplementary Figures~\ref{sfig:q20}-\ref{sfig:q62}).


\subsection{Application: inferring combinatorial binding of transcription factors}
\label{sec:tf}

\begin{figure}[h!]
\centering
\includegraphics[width = \textwidth]{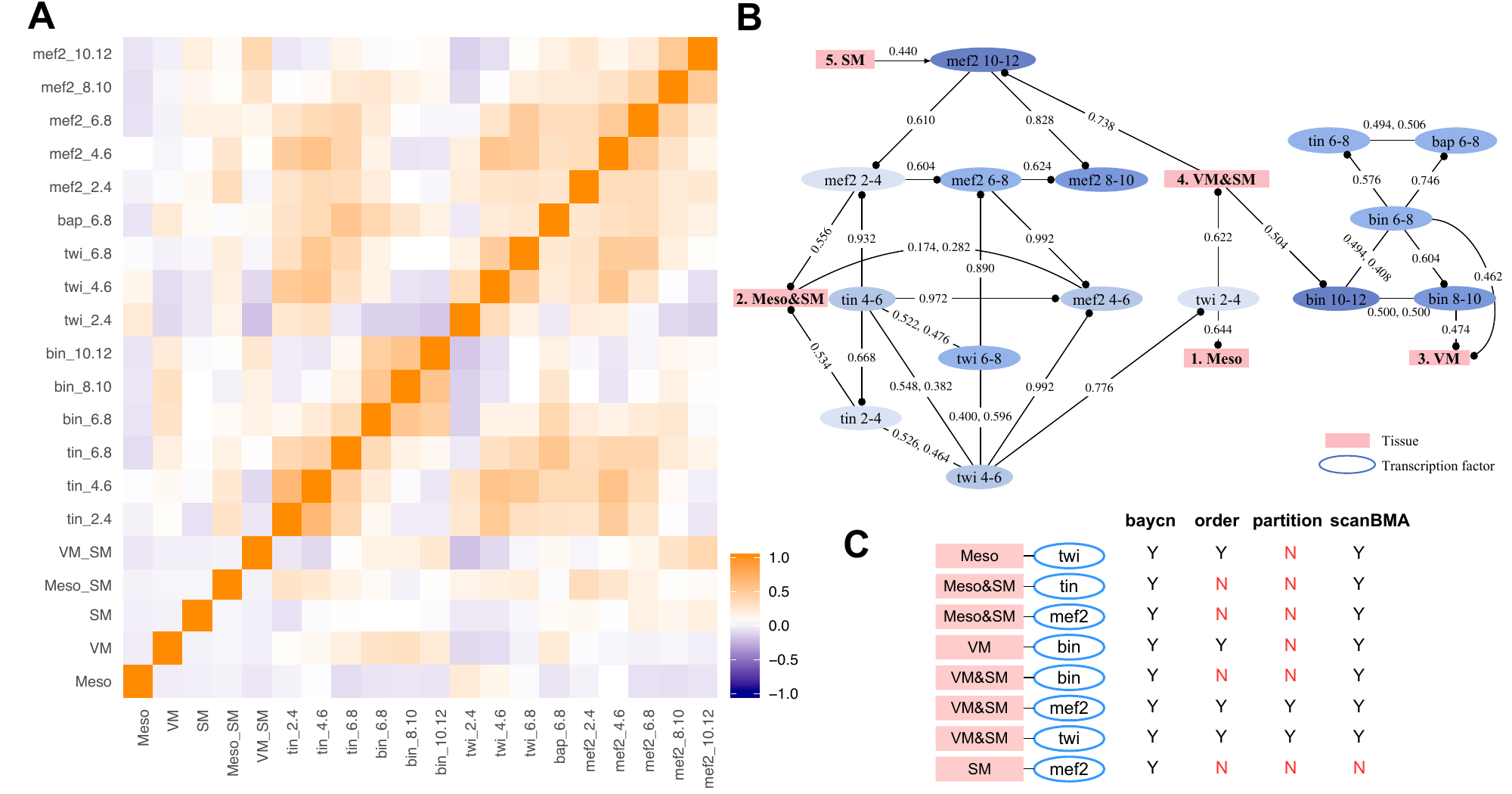}
\caption{\label{fig:drosophila} {\bf Combinatorial binding of transcription factors (TFs) in five tissue types from the benchmark dataset on 
Drosophila embryo \citep{zinzen2009}.} The TFs are: Twist (Twi), Tinman (Tin), Myocyte enhancing factor 2 (Mef2), Bagpipe (Bap), and Biniou (Bin).
 The tissue types are: mesoderm (Meso), mesoderm and somatic muscle (Meso\&SM), visceral muscle (VM), visceral muscle and somatic muscle 
 (VM\&SM), and somatic muscle (SM).  
({\bf A}) The heatmap of Pearson correlations.  ({\bf B}) The network inferred by baycn.  
To avoid potential confusion with interpreting directed edges for these time series, we used a dot in place of an arrow.
Except for bidirected edges, only the inferred direction with the corresponding posterior probability is shown.  
Shades of the TFs reflect the timing of the TF binding.
({\bf C}) Known relationships between tissues and TFs with the corresponding presence/absence inference from each method. A relationship is considered to be present if a method infers at least one edge between any of the time points for a given TF and its associated tissue.}
\end{figure}

Transcription factors (TFs) regulate the expression of target genes by combinatorial binding to regulatory sequences in the genome \citep{villar2014}. 
Here, we re-analyzed multiple highly-correlated TF binding profiles in 310 \textit{cis}-regulatory regions known as {\it cis}-regulatory modules (CRMs) during early development in Drosophila \citep{zinzen2009} (Figure~\ref{fig:drosophila}A).  These TFs have a number of edges that are well supported by experimental evidence \citep{zinzen2009, stojnic2012}, and the dataset was analyzed using a non-Bayesian graphical model approach for each of six tissues \citep{stojnic2012}.
This dataset consists of ChIP-chip binding profiles of five key TFs at two-hour intervals during mesoderm development in five tissue types in the embryos of {\it Drosophila melanogaster}: Twist (Twi), Tinman (Tin), Myocyte enhancing factor 2 (Mef2), Bagpipe (Bap), and Biniou (Bin).  The six tissue types are mesoderm (Meso), somatic muscle (SM), visceral muscle (VM), mesoderm and somatic muscle (Meso\&SM), visceral muscle and somatic muscle (VM\&SM).
Because of repeated measurements over time and combinatorial binding, there exist strong correlations among these binding profiles, and most graph inference methods could not tell them apart and tended to infer a dense network \citep{stojnic2012}.

Whereas \citet{stojnic2012} inferred a network for each tissue separately, here we applied baycn and inferred one network for the five tissues and five TFs.
We first used a machine learning network inference method MRPC \citep{badsha2019, badsha2021} to generate a candidate graph. 
Since MRPC tends to be conservative, we included in the candidate graph additional edges that could exist 
(Supplementary Figure~\ref{sfig:drosophila_skeleton}).
We ran baycn for 500,000 iterations with a burn-in of 0.2 and a step size of 800. 
An edge is considered present if the posterior probability of edge presence is greater than 0.5, and an edge is considered directed 
if the difference between the posterior probabilities for the two directions is greater than 0.2. 

Our analysis showed that baycn can identify TFs known to drive tissue differentiation, even when the inference was performed for all the tissues together (Figure~\ref{fig:drosophila}B; Supplementary Table~\ref{stab:drosophila.all}).  The inferred edges are consistent with the known relationships (Figure~\ref{fig:drosophila}C).  However, it is worth emphasizing that edges in a DAG represent conditional dependence.  The inferred edge [Twi at 4-6h $\rightarrow$ Twi at 2-4h] does not imply that later binding regulates earlier binding.  Instead, the trio [Twi at 4-6h $\rightarrow$ Twi at 2-4h $\rightarrow$ Meso\&SM] indicates that the effect of Twi binding at 4-6h on Meso\&SM is likely mediated through Twi binding at 2-4h; these inferred relationships and such interpretation are consistent with \cite{stojnic2012} (see Discussion for additional details).


We also ran order MCMC, partition MCMC, and scanBMA on this dataset to compare with baycn, using the edges confirmed and identified in \cite{stojnic2012} as the ground truth.  
Similar to the previous section, an edge is considered present if the posterior probability for edge presence is at least 0.5.
To summarize the results, we consider the inference correct if a method identifies the edge between a tissue and the binding profile of a TF at any time point (Figure~\ref{fig:drosophila}C; Supplementary Table~\ref{stab:drosophila.all}).
Both baycn and scanBMA infer almost the same known tissue-TF relationships, whereas order and partition MCMC miss many of them,
although for some edges order and partition MCMC inferred similar posterior probabilities to baycn 
(Supplementary Table~\ref{stab:drosophila.all}).  


\section{Discussion}
\label{sec:disc}
Here we present a novel approach to Bayesian inference of DAGs.  We represent a graph by a list of edge states. A prior distribution can then be assigned to an edge, which makes it easy to experiment with different beliefs about the edge states (e.g., sparsity).  
We have developed an Metropolis-Hastings-like algorithm for approximate inference, which deals with directed cycles, accounts for Markov equivalence, and is applicable to diverse data types (continuous, discrete, and mixed). 
Our method can be used for causal inference when instrumental variables are available (with confounding 
variables appropriately accounted for).
We have demonstrated through simulation studies that baycn is fast and can accurately estimate the edge probabilities in general, and that it performs as well as or better than current Bayesian methods in terms of precision, power, and mean squared error.

Our inference method takes a pseudo-likelihood approach to eliminate nuisance parameters and calculates the profile likelihood 
for a given graph.  Although not a genuine likelihood, its impact on the inference of the graph structure is minor in large samples \citep{severini2000likelihood}, 
which is the case in our simulations and applications.  Alternatively, the profile likelihood may be replaced by other pseudo-likelihoods such as 
a modified profile likelihood or a marginal likelihood (integrating out $\boldtheta$), although these pseudo-likelihoods 
each has its own computational challenges \citep{severini2000likelihood, ventura2016pseudo}.

A caveat when interpreting the inferred network is that the direction of an edge indicates statistical dependence. With additional assumptions, the direction may also indicate the actual, causal mechanism.  { In the application of the eQTLs and their target genes, for example, we imposed the constraint that an edge between an eQTL and a gene always points to the latter.  This is consistent with the biological principle that in general, DNA regulates RNA, but not the other way around.}  With this constraint, other directed edges may suggest regulatory relationships.  For example, the subgraph [rs11305802 $\rightarrow$ PNP $\rightarrow$ TMEM55B] in Figure \ref{fig:q8_combined} 
suggests that the eQTL rs11305803 likely regulates gene TMEM55B through gene PNP.  

In the application to the transcription factor binding data, however, no suitable constraint was applied. 
As a result, the direction in Figure \ref{fig:drosophila} should be interpreted carefully. 
Several directed edges, such as [Tin at 4-6h $\rightarrow$ Mef2 at 2-4h], [Tin at 4-6h $\rightarrow$ Tin at 2-4h], [Twi at 4-6h $\rightarrow$ Twi at 2-4h], and [Mef2 at 10-12h $\rightarrow$ Mef2 at 2-4h], and [Mef2 at 10-12h $\rightarrow$ Mef2 at 8-10h], are inferred by at least two of the four Bayesian methods (Supplementary Table~\ref{stab:drosophila.all}).  
However, these edges should not be interpreted as later binding regulating earlier binding.  Instead, edge directions should be interpreted as conditional (in)dependence between nodes and not biological regulation.
For example,
the subgraph [Tin at 4-6h $\rightarrow$ Mef2 at 2-4h $\rightarrow$ Meso\&SM] indicates that the association between Tin at 4-6h and the tissue can be explained away by Mef2 at 2-4h; in other words, the formation of Meso\&SM is more directly driven by Mef2 binding at 2-4h than by Tin binding at 4-6h.  
Similarly, in the subgraph [VM\&SM $\rightarrow$ Mef2 at 10-12h $\rightarrow$ Mef2 at 8-10h], our method and the two MCMC methods all inferred the latter directed edge.  This subgraph indicates that the association between Mef2 at 8-10h and VM\&SM can be explained away by the later binding of Mef2 at 10-12h, meaning that the formation of VM\&SM is more directly associated with Mef2 binding at a later time than at an earlier time.  

The graphs we have analyzed here are still fairly small, and additional work is needed for baycn to handle much larger graphs of hundreds of nodes. 
However, our Bayesian method provides the much-needed ease to experiment with different beliefs about the edge states: we can increase or 
decrease the prior probability for a certain edge state.
By comparing the posterior probabilities to the prior, the user can better assess how informative the data is and thus better interpret the inferred graph.  

\section*{Conflict of Interest Statement}

The authors declare that the research was conducted in the absence of any commercial or financial relationships that could be construed as a potential conflict of interest.

\section*{Author Contributions}

EAM and AQF conceived the project. EAM developed the R package with the help of AQF and performed the simulations and data analyses. Both EAM and AQF contributed equally to writing the manuscript.

\section*{Data Availability}

The method presented in this paper is implemented in the R package baycn available at \url{https://CRAN.R-project.org/package=baycn} or on GitHub at \url{https://github.com/evanamartin/baycn}.  

The GEUVADIS data is publicly available at \url{https://www.ebi.ac.uk/biostudies/arrayexpress/studies/E-GEUV-3}.
The Drosophila data is publicly available as Supplementary Table 8 of \citet{zinzen2009}.
All simulated and real data (GEUVADIS and Drosophila data), as well as the code used for the analysis, is available on GitHub at \url{https://github.com/evanamartin/baycn_analysis_code}. The real data is also included in the R package baycn. 

Furthermore, we used the Gibbs and MC3 samplers in the R package structmcmc (\url{https://github.com/rjbgoudie/structmcmc}), the orderMCMC and partionMCMC functions in the R package BiDAG (\url{https://cran.r-project.org/web/packages/BiDAG/index.html}), and the ScanBMA function in the R package networkBMA (\url{https://bioconductor.riken.jp/packages/3.2/bioc/html/networkBMA.html}).

\begin{acks}[Acknowledgments]
The authors thank Robert J. B. Goudie for helping with the Gibbs sampler implemented in his structmcmc R package and Md. Bahadur Badsha for preparing the GEUVADIS data for analysis. 
\end{acks}
\begin{funding}
This work was supported by the NIH grants R00HG007368 and P20GM104420.
\end{funding}

\begin{supplement}
\stitle{}
\sdescription{The supplementary materials contain details of the algorithm, additional figures, and additional tables.
}
\end{supplement}


\newpage

\begin{center}
    {\LARGE\bf Approximate Bayesian inference of directed acyclic graphs in biology with flexible priors on edge states}
\end{center}
  \begin{center}
    {\LARGE\bf Supplementary Materials}
\end{center}

\section*{Supplementary Notes}

Note: All of the references to equations and figures in Supplementary Notes refer to equations and figures in the Supplementary Materials file and not in the main text.


\section{Identifying and removing directed cycles}

In a directed cycle, one can follow the directed edges and return to the starting node (e.g., $T_1 \rightarrow T_2 \rightarrow T_3 \rightarrow T_1$).
Directed cycles can have a higher likelihood than the true graph, a graph with directed edges but no cycles, and therefore should be removed during the sampling iterations when generating a proposal graph. 

We have developed a ``cycle finder" algorithm (Section~\ref{sec:cycle.finder}) to find all directed cycles (including overlapping cycles, as well as multiple disjoint ones) in a graph, and a ``cycle remover" algorithm (Section~\ref{sec:cycle.remover}) to move out of directed cycles such that the proposed graph is free of directed cycles.
It is plausible that when proposing a new graph, our sampling algorithm may propose a graph with one or more directed cycles, try to move out of these directed cycles only to generate a graph with different directed cycles.  Therefore, our algorithm may need to repeatedly identify and remove directed cycles in one sampling iteration.  However, since we focus on relatively small graphs in this paper, this scenario is rather unlikely.


\subsection{Cycle finder algorithm}
\label{sec:cycle.finder}
\begin{enumerate}


\item Find the nodes that are connected to two or more nodes as a cycle contains at least three nodes and each node in a cycle has at least one incoming edge and one outgoing edge. To do so, we use the following steps:


\begin{enumerate}[i.]


\item Add the adjacency matrix to its transpose.


\item \label{item:drow} Sum each row and delete rows (the row indices are preserved) with a sum less than 2.


\item \label{item:dcell} Apply the following rule to the adjacency matrix
\[
\forall A_{\scriptscriptstyle j, k} = 1, 
\begin{cases}
    A_{\scriptscriptstyle j, k} = 0 \text{ if } k \notin J \\
    A_{\scriptscriptstyle j, k} = 1 \text{ if } k \in J, \\
\end{cases}
\]
where $J$ is the set of remaining row indices in the adjacency matrix.


\item Repeat steps ii and iii until $\forall A_{\scriptscriptstyle j, k} = 1$, $k \in J$ or the reduced adjacency matrix has two or fewer rows.
\end{enumerate}


\item \label{item:branches} Create a tree as deep as possible starting with the node (i.e., root node) whose index is in the first row of the reduced adjacency matrix. To do this we create a branch, which is a vector of node indices, for each of the nodes (i.e., child nodes) connected to the root. For each branch we add the index of the child node to the end of the vector, repeating the process of a child node becoming the parent node, until we add an index that has been added to the branch previously. If a parent node has two or more children a new branch is created for each child node.


\item \label{item:trim} Remove the nodes from the branches that do not belong to the cycle.
In addition to the nodes that create a cycle a branch may also contain nodes outside the cycle.
To remove these nodes, we start at the last node (i.e., the leaf) of the branches created in the previous step and work up the branch until we come to a node index that matches the leaf. Nodes above this node are then removed. For example, a branch may be $(3, 4, 1, 6, 5, 2, 1)$ and the trimmed branch would be $(1, 6, 5, 2, 1)$.


\item \label{item:coord} Convert the trimmed branches of a cycle into a vector of adjacency matrix coordinates by pairing up the adjacent nodes in each trimmed branch to produce a vector of edge coordinates. For example, if a trimmed branch is $(1, 6, 5, 2, 1)$ then the adjacency matrix coordinates for the edges between the nodes in the cycle are $\big((1, 6), (6, 5), (5, 2), (2, 1)\big)$.


\item \label{item:state} Generate a vector of edge states from the vector of edge coordinates by considering each pair of coordinates and comparing the first number to the second number. If the first number is smaller than the second number the edge state is 0, indicating that the edge points from the node with a smaller index to the node with the larger index. If the first number is larger than the second number the edge state is 1, indicating that the edge points from the node with a larger index to the node with a smaller index. The coordinates vector from the example in Step \ref{item:coord} is $\big((1, 6), (6, 5), (5, 2), (2, 1)\big)$ and the edge states that form a directed cycle are $(0, 1, 1, 1)$.


\item Use the vector of edge coordinates from step 4 to create a vector of edge indices. For example, the edge indices for the coordinates $\big((1, 6), (6, 5), (5, 2), (2, 1)\big)$ are $(2, 6, 5, 1)$.


\item Define each cycle by the edges in the cycle and the states that form a cycle Using the vector of edge states (step 5) and the vector of edge indices (step 6). The cycle vector for the $j$th cycle is:
\begin{equation} \label{eq:decimal}
(S_1, S_2, \dots, S_m)
\end{equation}
where $m$ is the number of edges in the network and each $S_k$ can take one of three values:
$$
S_k = 
    \begin{cases}
    9 & \text{if this edge is not involved in the $j$th cycle} \\
	1 & \text{if edge $S_k$ is in state 1} \\
	0 & \text{if edge $S_k$ is in state 0}.
	\end{cases}
$$

For example, Figure \ref{fig:nested} shows a graph with two cycles nested within a larger cycle. If the edges are oriented as shown in the figure the cycle vector for the cycle involving edges $(2, 3, 5, 6, 7)$ with states $\boldS = (0, 1, 0, 0, 0)$ is 
$$
(9, 0, 1, 9, 0, 0, 0),
$$
and the cycle vector for the cycle involving edges $(1, 3, 4, 5, 6, 7)$ with states $\boldS = (0, 1, 0, 0, 0, 0)$ is 
$$
(0, 9, 1, 0, 0, 0, 0).
$$

Even though the cycles in the example above have four edges in common with the same edge direction for each of these edges, we are able to distinguish between the two cycles by comparing the cycle vectors elementwise.


\begin{figure}[H]
\centering
\begin{tikzpicture}
\node[gene] (T1) {$T_1$};
\node[gene] (T2) [below left = 0.75cm and 0.5cm of T1] {$T_2$};
\node[gene] (T3) [below right = 0.75cm and 0.5cm of T1] {$T_3$};
\node[gene] (T6) [right = 3cm of T1] {$T_6$};
\node[gene] (T4) [below left = 0.75cm and 0.5cm of T6] {$T_4$};
\node[gene] (T5) [below right = 0.75cm and 0.5cm of T6] {$T_5$};

\draw[-latex, thick] (T1) to node[midway, above] {1} (T2);
\draw[-latex, thick] (T1) to node[midway, left] {2} (T3);
\draw[-latex, thick] (T6) to node[midway, above] {3} (T1);
\draw[-latex, thick] (T2) to node[midway, below] {4} (T3);
\draw[-latex, thick] (T3) to node[midway, below] {5} (T4);
\draw[-latex, thick] (T4) to node[midway, below] {6} (T5);
\draw[-latex, thick] (T5) to node[midway, right] {7} (T6);
\end{tikzpicture}
\captionsetup{width = 0.9 \linewidth}
\caption{A graph with two directed cycles one cycle is made up of edges $\{2, 5, 6, 7, 3\}$ and the other with edges $\{1, 4, 5, 6, 7, 3\}$. The two cycles have four edges in common.}
\label{fig:nested}
\end{figure}
\end{enumerate}


\subsection{Cycle remover algorithm}
\label{sec:cycle.remover}

\begin{enumerate}
\item For the proposed graph create a cycle vector, from the proposed edge states, for each set of edges that could form a directed cycle. For example, if the graph shown in Figure \ref{fig:cyclermvr}b is the proposed graph then the cycle vector would be $(0, 2, 1, 1)$.

\item For each set of edges that could form a directed cycle, compare the cycle vector of the proposed graph with the cycle vector of the directed cycles elementwise. For example, if the graph shown in Figure \ref{fig:cyclermvr}a is the true graph then the cycle vectors for the directed cycles would be $(0, 1, 0, 1)$ and $(1, 0, 1, 0)$. It is important to note that for each set of edges that could form a directed cycle there are two cycles: one with the edges oriented clockwise and another with the edges oriented counterclockwise. The cycle vector (created in the previous step) for the proposed graph is $(0, 2, 1, 1)$ which does not match either of the directed cycle vectors. Therefore, the proposed graph does not have any directed cycles.

\item For each directed cycle, randomly select an edge and change it from its current state to a different state according to the prior probability of edge states.

\item Repeat steps 1 - 3 until there are no directed cycles in the graph.
\end{enumerate}

\begin{figure}[H]
\begin{minipage}[t]{0.5 \textwidth}
a)
\centering
\begin{tikzpicture}
\node[gene] (T1) {$T_1$};
\node[gene] (T2) [below left = 0.75cm and 0.5cm of T1] {$T_2$};
\node[gene] (T3) [below right = 0.75cm and 0.5cm of T1]  {$T_3$};
\node[gene] (T4) [below right = 0.75cm and 0.5cm of T2]  {$T_4$};

\draw[-latex, thick] (T1) -- (T2) node[midway, above] {1};
\draw[-latex, thick] (T1) -- (T3) node[midway, above] {2};
\draw[-latex, thick] (T2) -- (T4) node[midway, below] {3};
\draw[-latex, thick] (T4) -- (T3) node[midway, below] {4};
\end{tikzpicture}
\end{minipage}%
\begin{minipage}[t]{0.5 \textwidth}
b)
\centering
\begin{tikzpicture}
\node[gene] (T1) {$T_1$};
\node[gene] (T2) [below left = 0.75cm and 0.5cm of T1] {$T_2$};
\node[gene] (T3) [below right = 0.75cm and 0.5cm of T1]  {$T_3$};
\node[gene] (T4) [below right = 0.75cm and 0.5cm of T2]  {$T_4$};

\draw[-latex, thick] (T1) -- (T2) node[midway, above] {1};
\draw[draw = none, thick] (T1) -- (T3) node[midway, above] {2};
\draw[-latex, thick] (T4) -- (T2) node[midway, below] {3};
\draw[-latex, thick] (T4) -- (T3) node[midway, below] {4};
\end{tikzpicture}
\end{minipage}
\caption{The graphs used for an example of the cycle remover algorithm. a) The true graph and b) the proposed graph.}
\label{fig:cyclermvr}
\end{figure}


\section{Calculating the transition probability in our sampling algorithm} \label{sec:transition}

Once a suitable proposal is generated, our sampling algorithm next calculates the acceptance probability for the proposed graph relative to the current one.  Although the (repeated) removal of directed cycles enters the calculation of the transition probability between the current and proposed graph, the probabilities involving the cycles are in the end canceled in the calculation of the acceptance ratio.  Let $\mathbf D$ be a vector of indices of the edges that {\it differ} between the current graph $\boldS$ and proposed graph $\boldS^{\boldsymbol \prime}$ and $\mathbf C$ be an integer vector where the element $c_j$ indicates the number of edges that {\it can} change state for the edge denoted by $d_j$ (see examples in Section~\ref{sec:examples}).  These two vectors have the same length, denoted by $h$. The probability of moving from $\boldS$ to $\boldS^{\boldsymbol \prime}$, i.e., $\Pr (\boldS \rightarrow \boldS^{\boldsymbol \prime})$, is the product of the probabilities of changes at individual edges in $\mathbf D$, and each of these probabilities further consists of two probabilities: the probability that an edge in the graph is chosen to change states, which is $1/c_j$, and the probability of edge $d_j$ changing from its current state $S_{d_j}$ to the state $S_{d_j}^{\prime}$, denoted by $\Pr (S_{d_j} \rightarrow S_{d_j}^{\prime})$. Therefore, 
\begin{align}
\Pr (\boldS \rightarrow \boldS^{\boldsymbol \prime}) & = \prod_{j = 1}^{h} \frac{1}{c_j}  \Pr (S_{d_j} \rightarrow S_{d_j}^{\prime}) = \prod_{j = 1}^{h} \frac{1}{c_j} \prod_{j = 1}^{h} \Pr (S_{d_j} \rightarrow S_{d_j}^{\prime}). \label{eq:transition}
\end{align}
We prove that the transition probabilities do not depend on the path taken from the current graph to the proposed graph (the process of introducing and removing directed cycles) but only on the edges that have different states between the two graphs.


\begin{theorem}
When calculating the acceptance probability, the transition probabilities between the current graph $\boldS$ and proposed graph $\boldS^{\boldsymbol \prime}$, $\Pr (\boldS \rightarrow \boldS^{\boldsymbol \prime})$ and $\Pr (\boldS^{\boldsymbol \prime} \rightarrow \boldS)$, depend only on the edges whose states are different between the two graphs.
\end{theorem}

\begin{proof}
Recall that $\boldS$ is a vector of edge states representing the current graph $\mathcal{G}$ and $\boldS^{\boldsymbol \prime}$ is a vector of edge states representing the proposal graph $\mathcal{G}^\prime$. Let $\mathbf D$ be a vector of indices of the edges that {\it differ} between the current graph $\boldS$ and proposed graph $\boldS^{\boldsymbol \prime}$ and $\mathbf C$ be an integer vector where the element $c_j$ represents the number of edges that {\it can} change state for the edge represented by $d_j$. These two vectors have the same length, denoted by $h$.

We will consider two cases: without and with potential directed cycles in the graph.


\begin{case}
Without potential directed cycles the probability of moving from the current graph to the proposed graph is
\begin{align}
\Pr (\boldS \rightarrow \boldS^{\boldsymbol \prime}) & = \prod_{j = 1}^{h} \frac{1}{c_j}  \Pr (S_{d_j} \rightarrow S_{d_j}^{\prime}) \nonumber \\
& = \prod_{j = 1}^{h} \frac{1}{c_j} \prod_{j = 1}^{h} \Pr (S_{d_j} \rightarrow S_{d_j}^{\prime}). \label{eq:cTopProof}
\end{align}
We can use the same procedure of deriving the equation for moving back to the current graph from the proposed graph. Therefore, the probability can be broken down in the same way when moving backwards:
\begin{align}
\Pr (\boldS^{\boldsymbol \prime} \rightarrow \boldS) & = \prod_{j = 1}^{h} \frac{1}{c_j} \prod_{j = 1}^{h} \Pr (S_{d_j}^{\prime} \rightarrow S_{d_j}).\label{eq:pTocProof}
\end{align}
Since there are no potential directed cycles in the network the value $c_j$ will always be $m$ which is the number of edges in the network. Therefore, $\prod_{j = 1}^{h} \frac{1}{c_j} = \prod_{j = 1}^{h} \frac{1}{m}$ whether going from $\boldS \rightarrow \boldS^{\boldsymbol \prime}$ or $\boldS^{\boldsymbol \prime} \rightarrow \boldS$ and will cancel when calculating the acceptance probability, leaving $\prod_{j = 1}^{h} \Pr (S_{d_j} \rightarrow S_{d_j}^{\prime})$ and $\prod_{j = 1}^{h} \Pr (S_{d_j}^{\prime} \rightarrow S_{d_j})$.
\end{case}


\begin{case}
With potential directed cycles there can be multiple paths when moving from the current graph to the proposed graph. Let $\mathbf C^k$ be a vector where each $c_j^k$ is the number of edges that can change state in path $k$ when moving from $\boldS$ to $\boldS^{\boldsymbol \prime}$ and $\mathbf C^{k^\prime}$ be a vector where each $c_j^{k^\prime}$ is the number of edges that can change state in path $k$ when moving from $\boldS^{\boldsymbol \prime}$ to $\boldS$. Using Equation (\ref{eq:cTopProof}) the transition probability of moving from the current graph to the proposed graph when there are multiple paths becomes
\begin{align}
\Pr (\boldS \rightarrow \boldS^{\boldsymbol \prime}) & = \sum_{k = 1}^K \prod_{j = 1}^{h} \frac{1}{c_j^k} \prod_{j = 1}^{h} \Pr (S_{d_j} \rightarrow S_{d_j}^{\prime}) = \prod_{j = 1}^{h} \Pr (S_{d_j} \rightarrow S_{d_j}^{\prime}) \sum_{k = 1}^K \prod_{j = 1}^{h} \frac{1}{c_j^k}. \label{eq:cTopSum}
\end{align}
Similarly, the transition probability when there are multiple paths of moving back to the current graph from the proposed graph is
\begin{align}
\Pr (\boldS^{\boldsymbol \prime} \rightarrow \boldS)  & = \sum_{k = 1}^K \prod_{j = 1}^{h} \frac{1}{c_j^{k^\prime}} \prod_{j = 1}^{h} \Pr (S_{d_j}^{\prime} \rightarrow S_{d_j}) = \prod_{j = 1}^{h} \Pr (S_{d_j}^{\prime} \rightarrow S_{d_j}) \sum_{k = 1}^K \prod_{j = 1}^{h} \frac{1}{c_j^{k^\prime}}. \label{eq:pTocSum}
\end{align}
In Equations (\ref{eq:cTopSum}) and (\ref{eq:pTocSum}) the summation over $K$ represents the different paths (Section \ref{sec:examples}) to get from one graph to another and the last equality holds because the edges that are different between $\boldS$ and $\boldS^{\boldsymbol \prime}$ do not depend on the path $k$.

For each path $k$
\begin{align}
\mathbf C^k & = (c_1^k, c_2^k, c_3^k, \dots, c_h^k) \label{eq:ck1} \\
& = (\underbrace{c_1^k, c_2^k, \dots, c_j^k}_{\text{create cycle(s)}}, \underbrace{c_{j + 1}^k, \dots, c_h^k}_{\text{remove cycle(s)}}) \label{eq:ck2} \\
& = (\underbrace{m, \dots, m}_{\text{$j$}}, c_{j + 1}^k, \dots, c_h^k). \label{eq:ck3}
\end{align}
The first $j$ elements can create one or more directed cycles. The remaining $h - j$ elements then remove the cycle(s) that were introduced in the network and their values are equal to the number of edges that make up the directed cycle that is being removed. The cycles that are created and removed in any path $k$ from $\boldS$ to $\boldS^{\boldsymbol \prime}$ can also be created and removed when moving from $\boldS^{\boldsymbol \prime}$ to $\boldS$. Therefore, the equations for moving from the proposed graph to the current graph will be the same as Equations (\ref{eq:ck1}) - (\ref{eq:ck3}) except for the $^\prime$ symbol indicating that we are moving backwards:
\begin{align}
\mathbf C^{k^\prime} & = (c_1^{k^\prime}, c_2^{k^\prime}, c_3^{k^\prime}, \dots, c_h^{k^\prime}) \\
& = (\underbrace{c_1^{k^\prime}, c_2^{k^\prime}, \dots, c_j^{k^\prime}}_{\text{create cycle(s)}}, \underbrace{c_{j + 1}^{k^\prime}, \dots, c_h^{k^\prime}}_{\text{remove cycle(s)}}) \\
& = (\underbrace{m, \dots, m}_{\text{$j$}}, c_{j + 1}^{k^\prime}, \dots, c_h^{k^\prime}),
\end{align}
and
\begin{align}
\sum_{k = 1}^K \prod_{j = 1}^{h} \frac{1}{c_j^k} & = \sum_{k = 1}^K \prod_{j = 1}^{h} \frac{1}{c_j^{k^\prime}}. \label{eq:coeff}
\end{align}
The terms in Equation (\ref{eq:coeff}) will cancel when calculating the acceptance probability and we will be left with $\prod_{j = 1}^{h} \Pr (S_{d_j} \rightarrow S_{d_j}^{\prime})$ and $\prod_{j = 1}^{h} \Pr (S_{d_j}^{\prime} \rightarrow S_{d_j})$. \hfill 
\end{case}
\end{proof}

See Section~\ref{sec:examples} for examples.


\section{Examples of Theorem 1} \label{sec:examples}


\subsection*{\textit{Example 1 -- one directed cycle}}

Consider the current graph in Figure \ref{fig:example1}a with states $\boldS = (0, 0, 0, 1, 1)$ and the proposed graph in Figure \ref{fig:example1}c with states $\boldS^{\boldsymbol \prime} = (0, 1, 2, 1, 1)$.
Edges \#2 and \#3 have different states between $\boldS$ and $\boldS^{\boldsymbol \prime}$ therefore $\mathbf D = 2, 3$.
There are two different paths to move from $\boldS$ to $\boldS^{\boldsymbol \prime}$. In path 1 there are two steps: i) edge \#2 changes direction which creates a directed cycle between nodes $T_1$, $T_2$, and $T_3$ (Figure \ref{fig:example1}b) and ii) the directed cycle is removed by edge \#3 changing from state 0 to 2. In path 2 edges \#2 and \#3 change states in one step.
If the prior on edge states is $p_0 = 0.05$, $p_1 = 0.05$ and $p_2 = 0.9$ then the probabilities for the two paths are
$$
\text{path 1: }
\Pr (S_{d_2} \rightarrow S_{d_2}^{\prime}) = \frac{0.05}{0.95}, \;
\Pr (S_{d_3} \rightarrow S_{d_3}^{\prime}) = \frac{0.9}{0.95}, \;
\mathbf C^1 = 5, 3
$$
and
$$
\text{path 2: }
\Pr (S_{d_2} \rightarrow S_{d_2}^{\prime}) = \frac{0.05}{0.95}, \;
\Pr (S_{d_3} \rightarrow S_{d_3}^{\prime}) = \frac{0.9}{0.95}, \;
\mathbf C^2 = 5, 5.
$$
Combining the probabilities from each path we obtain the transition probability of moving to the proposed graph:
\begin{equation} \label{eq:cToP}
\Pr (\boldS \rightarrow \boldS^{\boldsymbol \prime}) = \frac{1}{5}  \frac{1}{3}  \frac{0.05}{0.95}  \frac{0.9}{0.95} + \frac{1}{5}  \frac{1}{5}  \frac{0.05}{0.95}  \frac{0.9}{0.95} = \Big( \frac{1}{5}  \frac{1}{3} + \Big( \frac{1}{5} \Big)^2 \Big)  \frac{0.05}{0.95}  \frac{0.9}{0.95}.
\end{equation}

Any directed cycle created when moving from $\boldS$ to $\boldS^{\boldsymbol \prime}$ needs to be created when moving from $\boldS^{\boldsymbol \prime}$ to $\boldS$. Therefore, when moving from $\boldS^{\boldsymbol \prime}$ to $\boldS$ there are also two paths. Path 1 is made up of two steps: i) edge \#3 changes from state 2 to 0 creating a cycle between nodes $T_1$, $T_2$, and $T_3$ and ii) the cycle is removed by changing edge \#2 from state 1 to 0. For path 2 edges \#2 and \#3 both change states in one step. The probabilities for the paths are
$$
\text{path 1: }
\Pr (S_{d_2}^{\prime} \rightarrow S_{d_2}) = \frac{0.05}{0.95}, \;
\Pr (S_{d_3}^{\prime} \rightarrow S_{d_3}) = \frac{0.05}{0.1}, \;
\mathbf C^{1^\prime} = 5, 3
$$
and
$$
\text{path 2: }
\Pr (S_{d_2}^{\prime} \rightarrow S_{d_2}) = \frac{0.05}{0.95}, \;
\Pr (S_{d_3}^{\prime} \rightarrow S_{d_3}) = \frac{0.05}{0.1}, \;
\mathbf C^{2^\prime} = 5, 5.
$$
The transition probability of moving back to the current graph is
\begin{equation} \label{eq:pToC}
\Pr (\boldS^{\boldsymbol \prime} \rightarrow \boldS) = \frac{1}{5}  \frac{1}{3}  \frac{0.05}{0.95}  \frac{0.05}{0.1} + \frac{1}{5}  \frac{1}{5}  \frac{0.05}{0.95}  \frac{0.05}{0.1} = \Big( \frac{1}{5}  \frac{1}{3} + \Big( \frac{1}{5} \Big)^2 \Big)  \frac{0.05}{0.95}  \frac{0.05}{0.1}.
\end{equation}

The term $\frac{1}{5}  \frac{1}{3} + \Big( \frac{1}{5} \Big)^2$ in Equations (\ref{eq:cToP}) and (\ref{eq:pToC}) cancels out when calculating the acceptance probability, $\alpha$, and we are left with the probability of moving between the states that differ between the current graph (Figure \ref{fig:example1}a) and the proposed graph (Figure \ref{fig:example1}c). More generally, we can apply the same procedure to traverse the paths between any two graphs.

\begin{figure}[H]
\begin{minipage}[t]{0.33 \textwidth}
a)
\centering
\begin{tikzpicture}
\node[gene] (T1) {$T_1$};
\node[gene] (T2) [below left = 0.75cm and 0.5cm of T1] {$T_2$};
\node[gene] (T3) [below right = 0.75cm and 0.5cm of T1]  {$T_3$};
\node[gene] (T4) [below right = 0.75cm and 0.5cm of T2]  {$T_4$};

\draw[-latex, thick] (T1) -- (T2) node[midway, above] {1};
\draw[-latex, thick] (T1) -- (T3) node[midway, above] {2};
\draw[-latex, thick] (T2) -- (T3) node[midway, below] {3};
\draw[-latex, thick] (T4) -- (T2) node[midway, below] {4};
\draw[-latex, thick] (T4) -- (T3) node[midway, below] {5};
\end{tikzpicture}
\end{minipage}%
\begin{minipage}[t]{0.33 \textwidth}
b)
\centering
\begin{tikzpicture}
\node[gene] (T1) {$T_1$};
\node[gene] (T2) [below left = 0.75cm and 0.5cm of T1] {$T_2$};
\node[gene] (T3) [below right = 0.75cm and 0.5cm of T1]  {$T_3$};
\node[gene] (T4) [below right = 0.75cm and 0.5cm of T2]  {$T_4$};

\draw[-latex, thick] (T1) -- (T2) node[midway, above] {1};
\draw[-latex, thick, royalblue] (T3) -- (T1) node[midway, above] {2};
\draw[-latex, thick] (T2) -- (T3) node[midway, below] {3};
\draw[-latex, thick] (T4) -- (T2) node[midway, below] {4};
\draw[-latex, thick] (T4) -- (T3) node[midway, below] {5};
\end{tikzpicture}
\end{minipage}%
\begin{minipage}[t]{0.33 \textwidth}
c)
\centering
\begin{tikzpicture}
\node[gene] (T1) {$T_1$};
\node[gene] (T2) [below left = 0.75cm and 0.5cm of T1] {$T_2$};
\node[gene] (T3) [below right = 0.75cm and 0.5cm of T1]  {$T_3$};
\node[gene] (T4) [below right = 0.75cm and 0.5cm of T2]  {$T_4$};

\draw[-latex, thick] (T1) -- (T2) node[midway, above] {1};
\draw[-latex, thick, royalblue] (T3) -- (T1) node[midway, above] {2};
\draw[draw = none, royalblue] (T2) -- (T3) node[midway, below] {3};
\draw[-latex, thick] (T4) -- (T2) node[midway, below] {4};
\draw[-latex, thick] (T4) -- (T3) node[midway, below] {5};
\end{tikzpicture}
\end{minipage}
\caption{The graphs for example 1. a) The current graph. b) An intermediate graph between the current graph and the proposed graph where a directed cycle has been introduced into the network. c) The proposed graph.}
\label{fig:example1}
\end{figure}


\subsection*{\textit{Example 2 -- multiple directed cycles}}

We show a second more complex example below. If we start with the graph in Figure \ref{fig:example2}a with states $\boldS = (0, 0, 0, 1, 1)$ and the proposed graph in Figure \ref{fig:example2}d with states $\boldS^{\boldsymbol \prime} = (2, 1, 1, 1, 0)$ there are four edges with different states between the two graphs and $\mathbf D = 1, 2, 3, 5$. There are three different paths to move from $\boldS$ to $\boldS^{\boldsymbol \prime}$. The steps in path 1 are: i) edges \#2 and \#5 change directions creating two directed cycles, the first cycle is between nodes $T_1$, $T_2$, and $T_3$ and the second cycle is between nodes $T_2$, $T_3$, and $T_4$ (Figure \ref{fig:example2}b), ii) edge \#1 changes from state 0 to 2 removing the first cycle (Figure \ref{fig:example2}c), and iii) edge \#3 changes direction which removes the second cycle (Figure \ref{fig:example2}d). The steps in path 2 are: i) edges \#1, \#2, and \#5 all change states creating one directed cycle between nodes $T_2$, $T_3$, and $T_4$ (Figure \ref{fig:example2}c) and ii) edge \#3 changes direction removing the cycle. In path 3 edges \#1, \#2, \#3, and \#5 all change states in one step. If the prior on edge states is $p_0 = 0.05$, $p_1 = 0.05$, and $p_2 = 0.9$ then the probabilities for the three paths are: \\
\text{path 1:}
\begin{align*}
& \Pr (S_{d_1} \rightarrow S_{d_1}^{\prime}) = \frac{0.9}{0.95}, \;
\Pr (S_{d_2} \rightarrow S_{d_2}^{\prime}) = \frac{0.05}{0.95}, \\
& \Pr (S_{d_3} \rightarrow S_{d_3}^{\prime}) = \frac{0.05}{0.95}, \;
\Pr (S_{d_5} \rightarrow S_{d_5}^{\prime}) = \frac{0.05}{0.95}, \\
& \mathbf C^1 = 5, 5, 3, 3,
\end{align*}
\text{path 2:}
\begin{align*}
& \Pr (S_{d_1} \rightarrow S_{d_1}^{\prime}) = \frac{0.9}{0.95}, \;
\Pr (S_{d_2} \rightarrow S_{d_2}^{\prime}) = \frac{0.05}{0.95}, \\
& \Pr (S_{d_3} \rightarrow S_{d_3}^{\prime}) = \frac{0.05}{0.95}, \;
\Pr (S_{d_5} \rightarrow S_{d_5}^{\prime}) = \frac{0.05}{0.95}, \\
& \mathbf C^2 = 5, 5, 5, 3,
\end{align*}
and
\text{path 3:}
\begin{align*}
& \Pr (S_{d_1} \rightarrow S_{d_1}^{\prime}) = \frac{0.9}{0.95}, \;
\Pr (S_{d_2} \rightarrow S_{d_2}^{\prime}) = \frac{0.05}{0.95}, \\
& \Pr (S_{d_3} \rightarrow S_{d_3}^{\prime}) = \frac{0.05}{0.95}, \;
\Pr (S_{d_5} \rightarrow S_{d_5}^{\prime}) = \frac{0.05}{0.95}, \\
& \mathbf C^3 = 5, 5, 5, 5.
\end{align*}
Therefore, the transition probability of moving to the proposed graph is
\begin{align}
\Pr (\boldS \rightarrow \boldS^{\boldsymbol \prime}) & = \frac{1}{5}  \frac{1}{5}  \frac{1}{3}  \frac{1}{3}  \frac{0.9}{0.95}  \frac{0.05}{0.95}  \frac{0.05}{0.95}  \frac{0.05}{0.95} + \frac{1}{5}  \frac{1}{5}  \frac{1}{5}  \frac{1}{3}  \frac{0.9}{0.95}  \frac{0.05}{0.95}  \frac{0.05}{0.95}  \frac{0.05}{0.95} \nonumber \\
& + \frac{1}{5}  \frac{1}{5}  \frac{1}{5}  \frac{1}{5}  \frac{0.9}{0.95}  \frac{0.05}{0.95}  \frac{0.05}{0.95}  \frac{0.05}{0.95} \nonumber \\
& = \Bigg( \Big( \frac{1}{5} \Big)^2 \Big( \frac{1}{3} \Big)^2 + \Big( \frac{1}{5} \Big)^3 \frac{1}{3} + \Big( \frac{1}{5} \Big)^4 \Bigg)  \frac{0.9}{0.95}  \frac{0.05}{0.95}  \frac{0.05}{0.95}  \frac{0.05}{0.95}. \label{eq:cToP2}
\end{align}

As in example 1 any directed cycle that is created when moving from $\boldS$ to $\boldS^{\boldsymbol \prime}$ needs to also be created when moving from $\boldS^{\boldsymbol \prime}$ to $\boldS$. There are also three different paths to move back to $\boldS$ from $\boldS^{\boldsymbol \prime}$. In path 1 the steps are: i) edges \#1 and \#3 change states creating two directed cycles the first cycle is between nodes $T_1$, $T_2$, and $T_3$ and the second cycle is between nodes $T_2$, $T_3$, and $T_4$, ii) edge \#2 changes direction removing the first cycle, and iii) edge \#5 changes direction removing the second directed cycle. Path 2 has two steps i) edges \#1, \#2 and \#3 all change state creating a directed cycle between nodes $T_2$, $T_3$, and $T_4$ and ii) edge \#5 changes direction removing the cycle. In path 3 there is only one step where edges \#1, \#2, \#3, and \#5 all change states. The probabilities for these paths are \\
\text{path 1:}
\begin{align*}
& \Pr (S_{d_1}^{\prime} \rightarrow S_{d_1}) = \frac{0.05}{0.1}, \;
\Pr (S_{d_2}^{\prime} \rightarrow S_{d_2}) = \frac{0.05}{0.95}, \\
& \Pr (S_{d_3}^{\prime} \rightarrow S_{d_3}) = \frac{0.05}{0.95}, \;
\Pr (S_{d_5}^{\prime} \rightarrow S_{d_5}) = \frac{0.05}{0.95}, \\
& \mathbf C^{1^\prime} = 5, 5, 3, 3,
\end{align*}
\text{path 2:}
\begin{align*}
& \Pr (S_{d_1}^{\prime} \rightarrow S_{d_1}) = \frac{0.05}{0.1}, \;
\Pr (S_{d_2}^{\prime} \rightarrow S_{d_2}) = \frac{0.05}{0.95}, \\
& \Pr (S_{d_3}^{\prime} \rightarrow S_{d_3}) = \frac{0.05}{0.95}, \;
\Pr (S_{d_5}^{\prime} \rightarrow S_{d_5}) = \frac{0.05}{0.95}, \\
& \mathbf C^{2^\prime} = 5, 5, 5, 3,
\end{align*}
and 
\text{path 3:}
\begin{align*}
& \Pr (S_{d_1}^{\prime} \rightarrow S_{d_1}) = \frac{0.05}{0.1}, \;
\Pr (S_{d_2}^{\prime} \rightarrow S_{d_2}) = \frac{0.05}{0.95}, \\
& \Pr (S_{d_3}^{\prime} \rightarrow S_{d_3}) = \frac{0.05}{0.95}, \;
\Pr (S_{d_5}^{\prime} \rightarrow S_{d_5}) = \frac{0.05}{0.95}, \\
& \mathbf C^{3^\prime} = 5, 5, 5, 5.
\end{align*}
Therefore, the transition probability of moving back to the current graph is
\begin{align}
\Pr (\boldS^{\boldsymbol \prime} \rightarrow \boldS) & = \frac{1}{5}  \frac{1}{5}  \frac{1}{3}  \frac{1}{3}  \frac{0.05}{0.1}  \frac{0.05}{0.95}  \frac{0.05}{0.95}  \frac{0.05}{0.95} + \frac{1}{5}  \frac{1}{5}  \frac{1}{5}  \frac{1}{3}  \frac{0.05}{0.1}  \frac{0.05}{0.95}  \frac{0.05}{0.95}  \frac{0.05}{0.95} \nonumber \\ & + \frac{1}{5}  \frac{1}{5}  \frac{1}{5}  \frac{1}{5}  \frac{0.05}{0.1}  \frac{0.05}{0.95}  \frac{0.05}{0.95}  \frac{0.05}{0.95} \nonumber \\
& = \Bigg( \Big( \frac{1}{5} \Big)^2 \Big( \frac{1}{3} \Big)^2 + \Big( \frac{1}{5} \Big)^3 \frac{1}{3} + \Big( \frac{1}{5} \Big)^4 \Bigg) \frac{0.05}{0.1}  \frac{0.05}{0.95}  \frac{0.05}{0.95}  \frac{0.05}{0.95}. \label{eq:pToC2}
\end{align}

The term $\Big( \frac{1}{5} \Big)^2 \Big( \frac{1}{3} \Big)^2 + \Big( \frac{1}{5} \Big)^3 \frac{1}{3} + \Big( \frac{1}{5} \Big)^4$ in Equations (\ref{eq:cToP2}) and (\ref{eq:pToC2}) cancels out when calculating the acceptance probability, $\alpha$, and we are left with the probability of moving between the states that differ between the current graph (Figure \ref{fig:example2}a) and the proposed graph (Figure \ref{fig:example2}d). More generally, we can apply the same procedure to traverse the paths between any two graphs.

\begin{figure}[H]
\begin{minipage}[t]{0.25 \textwidth}
a)
\centering
\begin{tikzpicture}
\node[gene] (T1) {$T_1$};
\node[gene] (T2) [below left = 0.75cm and 0.5cm of T1] {$T_2$};
\node[gene] (T3) [below right = 0.75cm and 0.5cm of T1]  {$T_3$};
\node[gene] (T4) [below right = 0.75cm and 0.5cm of T2]  {$T_4$};

\draw[-latex, thick] (T1) -- (T2) node[midway, above] {1};
\draw[-latex, thick] (T1) -- (T3) node[midway, above] {2};
\draw[-latex, thick] (T2) -- (T3) node[midway, below] {3};
\draw[-latex, thick] (T4) -- (T2) node[midway, below] {4};
\draw[-latex, thick] (T4) -- (T3) node[midway, below] {5};
\end{tikzpicture}
\end{minipage}%
\begin{minipage}[t]{0.25 \textwidth}
b)
\centering
\begin{tikzpicture}
\node[gene] (T1) {$T_1$};
\node[gene] (T2) [below left = 0.75cm and 0.5cm of T1] {$T_2$};
\node[gene] (T3) [below right = 0.75cm and 0.5cm of T1]  {$T_3$};
\node[gene] (T4) [below right = 0.75cm and 0.5cm of T2]  {$T_4$};

\draw[-latex, thick] (T1) -- (T2) node[midway, above] {1};
\draw[-latex, thick, royalblue] (T3) -- (T1) node[midway, above] {2};
\draw[-latex, thick] (T2) -- (T3) node[midway, below] {3};
\draw[-latex, thick] (T4) -- (T2) node[midway, below] {4};
\draw[-latex, thick, royalblue] (T3) -- (T4) node[midway, below] {5};
\end{tikzpicture}
\end{minipage}%
\begin{minipage}[t]{0.25 \textwidth}
c)
\centering
\begin{tikzpicture}
\node[gene] (T1) {$T_1$};
\node[gene] (T2) [below left = 0.75cm and 0.5cm of T1] {$T_2$};
\node[gene] (T3) [below right = 0.75cm and 0.5cm of T1]  {$T_3$};
\node[gene] (T4) [below right = 0.75cm and 0.5cm of T2]  {$T_4$};

\draw[draw = none, royalblue] (T1) -- (T2) node[midway, above] {1};
\draw[-latex, thick, royalblue] (T3) -- (T1) node[midway, above] {2};
\draw[-latex, thick] (T2) -- (T3) node[midway, below] {3};
\draw[-latex, thick] (T4) -- (T2) node[midway, below] {4};
\draw[-latex, thick, royalblue] (T3) -- (T4) node[midway, below] {5};
\end{tikzpicture}
\end{minipage}%
\begin{minipage}[t]{0.25 \textwidth}
d)
\centering
\begin{tikzpicture}
\node[gene] (T1) {$T_1$};
\node[gene] (T2) [below left = 0.75cm and 0.5cm of T1] {$T_2$};
\node[gene] (T3) [below right = 0.75cm and 0.5cm of T1]  {$T_3$};
\node[gene] (T4) [below right = 0.75cm and 0.5cm of T2]  {$T_4$};

\draw[draw = none, royalblue] (T1) -- (T2) node[midway, above] {1};
\draw[-latex, thick, royalblue] (T3) -- (T1) node[midway, above] {2};
\draw[-latex, thick, royalblue] (T3) -- (T2) node[midway, below] {3};
\draw[-latex, thick] (T4) -- (T2) node[midway, below] {4};
\draw[-latex, thick, royalblue] (T3) -- (T4) node[midway, below] {5};
\end{tikzpicture}
\end{minipage}%
\caption{a) The current graph. b) An intermediate graph between the current graph and the proposed graph where two directed cycles have been introduced into the graph. c) An intermediate graph where one of the directed cycles has been removed. d) The proposed graph.}
\label{fig:example2}
\end{figure}

\newpage
\setcounter{figure}{0}
\renewcommand{\thefigure}{S\arabic{figure}}



\begin{figure}[H]
\centering
\begin{minipage}{0.25 \columnwidth}
\centering
\begin{tikzpicture}
\node[gene] (T1) {$T_1$};
\node[gene] (T2) [below left = 0.75cm and 0.5cm of T1] {$T_2$};
\node[gene] (T3) [below right = 0.75cm and 0.5cm of T1]  {$T_3$};
\node[gene] (T4) [below right = 0.75cm and 0.5cm of T2]  {$T_4$};

\draw[-latex, thick, sweetpotato] (T1) -- (T2) node[midway, above] {1};
\draw[-latex, thick] (T1) -- (T3) node[midway, above] {2};
\draw[-latex, thick, sweetpotato] (T2) -- (T4) node[midway, below] {3};
\draw[-latex, thick] (T4) -- (T3) node[midway, below] {4};
\end{tikzpicture}
\end{minipage}%
\begin{minipage}{0.25 \columnwidth}
\centering
\begin{tabu}spread 0pt{@{} l l l l @{}}
\toprule
& \multicolumn{3}{c}{True} \\
& \multicolumn{3}{c}{Probability} \\
\cmidrule(lr){2-4}
Edge & 0 & 1 & 2 \\
\midrule
1 & 0.33 & 0.66 & 0 \\
2 & 1 & 0 & 0 \\
3 & 0.66 & 0.33 & 0 \\
4 & 0 & 1 & 0 \\
\bottomrule
\end{tabu}
\end{minipage}%
\caption{\label{sfig:gn4} The true graph and probabilities for each edge in topology GN4. The edges in orange can change direction while remaining in the Markov equivalence class of the true graph -- as long as another v-structure is not created.}
\end{figure}


\begin{figure}[H]
\begin{minipage}{0.33 \textwidth}
\centering
\begin{tikzpicture}
\node[gene] (T1) {$T_1$};
\node[gene] (T2) [below left = 0.75cm and 0.5cm of T1] {$T_2$};
\node[gene] (T3) [below right = 0.75cm and 0.5cm of T1]  {$T_3$};
\node[gene] (T4) [below right = 0.75cm and 0.5cm of T2]  {$T_4$};

\draw[-latex, thick, sweetpotato] (T1) -- (T2) node[midway, above] {1};
\draw[-latex, thick] (T1) -- (T3) node[midway, above] {2};
\draw[-latex, thick, sweetpotato] (T2) -- (T4) node[midway, below] {3};
\draw[-latex, thick] (T4) -- (T3) node[midway, below] {4};
\end{tikzpicture}
\end{minipage}%
\begin{minipage}{0.33 \textwidth}
\centering
\begin{tikzpicture}
\node[gene] (T1) {$T_1$};
\node[gene] (T2) [below left = 0.75cm and 0.5cm of T1] {$T_2$};
\node[gene] (T3) [below right = 0.75cm and 0.5cm of T1]  {$T_3$};
\node[gene] (T4) [below right = 0.75cm and 0.5cm of T2]  {$T_4$};

\draw[-latex, thick, sweetpotato] (T2) -- (T1) node[midway, above] {1};
\draw[-latex, thick] (T1) -- (T3) node[midway, above] {2};
\draw[-latex, thick, sweetpotato] (T2) -- (T4) node[midway, below] {3};
\draw[-latex, thick] (T4) -- (T3) node[midway, below] {4};
\end{tikzpicture}
\end{minipage}%
\begin{minipage}{0.33 \textwidth}
\centering
\begin{tikzpicture}
\node[gene] (T1) {$T_1$};
\node[gene] (T2) [below left = 0.75cm and 0.5cm of T1] {$T_2$};
\node[gene] (T3) [below right = 0.75cm and 0.5cm of T1]  {$T_3$};
\node[gene] (T4) [below right = 0.75cm and 0.5cm of T2]  {$T_4$};

\draw[-latex, thick, sweetpotato] (T2) -- (T1) node[midway, above] {1};
\draw[-latex, thick] (T1) -- (T3) node[midway, above] {2};
\draw[-latex, thick, sweetpotato] (T4) -- (T2) node[midway, below] {3};
\draw[-latex, thick] (T4) -- (T3) node[midway, below] {4};
\end{tikzpicture}
\end{minipage}%
\captionsetup{width = 0.9 \linewidth}
\caption{The Markov equivalence class of topology GN4. The edges in orange show all possible combinations of edge directions of the Markov equivalence class. Edge 1 is oriented $T_1 \rightarrow T_2$ in one of the three graphs, giving a proportion of 0.33 for edge state 0. Edge 2 is oriented $T_2 \rightarrow T_4$ in two of the three graphs, giving a proportion of 0.66 for state 0.}
\label{sfig:gn4me}
\end{figure}


\begin{figure}[H]
\centering
\begin{minipage}{0.25 \columnwidth}
\centering
\begin{tikzpicture}
\node[gene] (T1) {$T_1$};
\node[gene] (T2) [below left = 0.75cm and 0.5cm of T1]  {$T_2$};
\node[gene] (T3) [below right = 0.75cm and 0.5cm of T1]  {$T_3$};
\node[gene] (T4) [below right = 0.75cm and -0.15cm of T2]  {$T_4$};
\node[gene] (T5) [below left = 0.75cm and -0.15cm of T3]  {$T_5$};

\draw[-latex, thick, sweetpotato] (T1) -- (T2) node[midway, left] {1};
\draw[-latex, thick, sweetpotato] (T2) -- (T4) node[midway, left] {3};
\draw[-latex, thick, sweetpotato] (T1) -- (T3) node[midway, right] {2};
\draw[-latex, thick] (T3) -- (T5) node[midway, right] {4};
\draw[-latex, thick] (T4) -- (T5) node[midway, below] {5};
\end{tikzpicture}
\end{minipage}%
\begin{minipage}{0.25 \columnwidth}
\centering
\begin{tabu}spread 0pt{@{} l l l l @{}}
\toprule
& \multicolumn{3}{c}{True} \\
& \multicolumn{3}{c}{Probability} \\
\cmidrule(lr){2-4}
Edge & 0 & 1 & 2 \\
\midrule
1 & 0.5 & 0.5 & 0 \\
2 & 0.75 & 0.25 & 0 \\
3 & 0.75 & 0.25 & 0 \\
4 & 1 & 0 & 0 \\
5 & 1 & 0 & 0 \\
\bottomrule
\end{tabu}
\end{minipage}%
\caption{The true graph and probabilities for each edge in topology GN5. The edges in orange can change direction while remaining in the Markov equivalence class of the true graph -- as long as another v-structure is not created.}
\label{sfig:gn5}
\end{figure}


\begin{figure}[H]
\begin{minipage}{0.25 \textwidth}
\centering
\begin{tikzpicture}
\node[gene] (T1) {$T_1$};
\node[gene] (T2) [below left = 0.75cm and 0.5cm of T1]  {$T_2$};
\node[gene] (T3) [below right = 0.75cm and 0.5cm of T1]  {$T_3$};
\node[gene] (T4) [below right = 0.75cm and -0.15cm of T2]  {$T_4$};
\node[gene] (T5) [below left = 0.75cm and -0.15cm of T3]  {$T_5$};

\draw[-latex, thick, sweetpotato] (T1) -- (T2) node[midway, left] {1};
\draw[-latex, thick, sweetpotato] (T2) -- (T4) node[midway, left] {3};
\draw[-latex, thick, sweetpotato] (T1) -- (T3) node[midway, right] {2};
\draw[-latex, thick] (T3) -- (T5) node[midway, right] {4};
\draw[-latex, thick] (T4) -- (T5) node[midway, below] {5};
\end{tikzpicture}
\end{minipage}%
\begin{minipage}{0.25 \textwidth}
\centering
\begin{tikzpicture}
\node[gene] (T1) {$T_1$};
\node[gene] (T2) [below left = 0.75cm and 0.5cm of T1]  {$T_2$};
\node[gene] (T3) [below right = 0.75cm and 0.5cm of T1]  {$T_3$};
\node[gene] (T4) [below right = 0.75cm and -0.15cm of T2]  {$T_4$};
\node[gene] (T5) [below left = 0.75cm and -0.15cm of T3]  {$T_5$};

\draw[-latex, thick, sweetpotato] (T1) -- (T2) node[midway, left] {1};
\draw[-latex, thick, sweetpotato] (T2) -- (T4) node[midway, left] {3};
\draw[-latex, thick, sweetpotato] (T3) -- (T1) node[midway, right] {2};
\draw[-latex, thick] (T3) -- (T5) node[midway, right] {4};
\draw[-latex, thick] (T4) -- (T5) node[midway, below] {5};
\end{tikzpicture}
\end{minipage}%
\begin{minipage}{0.25 \textwidth}
\centering
\begin{tikzpicture}
\node[gene] (T1) {$T_1$};
\node[gene] (T2) [below left = 0.75cm and 0.5cm of T1]  {$T_2$};
\node[gene] (T3) [below right = 0.75cm and 0.5cm of T1]  {$T_3$};
\node[gene] (T4) [below right = 0.75cm and -0.15cm of T2]  {$T_4$};
\node[gene] (T5) [below left = 0.75cm and -0.15cm of T3]  {$T_5$};

\draw[-latex, thick, sweetpotato] (T2) -- (T1) node[midway, left] {1};
\draw[-latex, thick, sweetpotato] (T2) -- (T4) node[midway, left] {3};
\draw[-latex, thick, sweetpotato] (T1) -- (T3) node[midway, right] {2};
\draw[-latex, thick] (T3) -- (T5) node[midway, right] {4};
\draw[-latex, thick] (T4) -- (T5) node[midway, below] {5};
\end{tikzpicture}
\end{minipage}%
\begin{minipage}{0.25 \textwidth}
\centering
\begin{tikzpicture}
\node[gene] (T1) {$T_1$};
\node[gene] (T2) [below left = 0.75cm and 0.5cm of T1]  {$T_2$};
\node[gene] (T3) [below right = 0.75cm and 0.5cm of T1]  {$T_3$};
\node[gene] (T4) [below right = 0.75cm and -0.15cm of T2]  {$T_4$};
\node[gene] (T5) [below left = 0.75cm and -0.15cm of T3]  {$T_5$};

\draw[-latex, thick, sweetpotato] (T2) -- (T1) node[midway, left] {1};
\draw[-latex, thick, sweetpotato] (T4) -- (T2) node[midway, left] {3};
\draw[-latex, thick, sweetpotato] (T1) -- (T3) node[midway, right] {2};
\draw[-latex, thick] (T3) -- (T5) node[midway, right] {4};
\draw[-latex, thick] (T4) -- (T5) node[midway, below] {5};
\end{tikzpicture}
\end{minipage}%
\captionsetup{width = 0.9 \linewidth}
\caption{The Markov equivalence class of topology GN5. The edges in orange show all possible combinations of edge directions of the Markov equivalence class. In two of the four graphs edge 1 is oriented $T_1 \rightarrow T_2$, giving a proportion of 0.5 for edge state 0. In three of the four graphs edge 2 is oriented $T_1 \rightarrow T_3$, giving a proportion of 0.75 for edge state 0. Similarly, edge 3 is oriented $T_2 \rightarrow T_4$ in three of the four graphs, giving a proportion of 0.75 for edge state 0.}
\label{sfig:gn5me}
\end{figure}


\begin{figure}[H]
\centering
\begin{minipage}[c]{0.25 \columnwidth}
\centering
\begin{tikzpicture}
\node[gene] (T1) {$T_1$};
\node[gene] (T2) [right = 0.5cm of T1] {$T_2$};
\node[gene] (T3) [right = 0.5cm of T2] {$T_3$};
\node[gene] (T4) [below = 0.75cm of T2] {$T_4$};

\draw[-latex, thick] (T1) -- (T4) node[midway, left] {1};
\draw[-latex, thick] (T2) -- (T4) node[midway, left] {2};
\draw[-latex, thick] (T3) -- (T4) node[midway, left] {3};
\end{tikzpicture}
\end{minipage}%
\begin{minipage}{0.25 \columnwidth}
\centering
\begin{tabu}spread 0pt{@{} l l l l @{}}
\toprule
& \multicolumn{3}{c}{True} \\
& \multicolumn{3}{c}{Probability} \\
\cmidrule(lr){2-4}
Edge & 0 & 1 & 2 \\
\midrule
1 & 1 & 0 & 0 \\
2 & 1 & 0 & 0 \\
3 & 1 & 0 & 0 \\
\bottomrule
\end{tabu}
\end{minipage}%
\caption{The true graph and probabilities for each edge in the multi-parent topology.}
\label{sfig:multiparent}
\end{figure}


\begin{figure}[H]
\centering
\begin{minipage}{0.65 \columnwidth}
\centering
\begin{tikzpicture}
\node[gene] (T6) {$T_6$};
\node[gene] (T5)  [above left = 0.5cm and 0.25cm of T6]  {$T_5$};
\node[gene] (T4)  [above left = 0.5cm and 0.25cm of T5]  {$T_4$};
\node[gene] (T3)  [left = 0.5cm of T4]  {$T_3$};
\node[gene] (T2)  [below left = 0.75cm and 0.25cm of T3]  {$T_2$};
\node[gene] (T1)  [below right = 0.75cm and 0.25cm of T2] {$T_1$};

\node[gene] (T7)  [above right = 0.5cm and 0.25cm of T6] {$T_7$};
\node[gene] (T8)  [above right = 0.5cm and 0.25cm of T7] {$T_8$};
\node[gene] (T9)  [right = 0.5cm of T8] {$T_9$};
\node[gene] (T10) [below right = 0.75cm and 0.25cm of T9] {$T_{10}$};
\node[gene] (T11) [below left = 0.75cm and 0.25cm of T10] {$T_{11}$};

\draw[-latex, thick, sweetpotato] (T1) -- (T2) node[near start, left] {1};
\draw[-latex, thick, sweetpotato] (T2) -- (T3) node[midway, left] {2};
\draw[-latex, thick, sweetpotato] (T3) -- (T4) node [midway, above] {3};
\draw[-latex, thick, sweetpotato] (T4) -- (T5) node [midway, left] {4};
\draw[-latex, thick] (T5) -- (T6) node [midway, left] {5};
\draw[-latex, thick] (T7) -- (T6) node [midway, right] {6};
\draw[-latex, thick, sweetpotato] (T7) -- (T8) node [near start, right] {7};
\draw[-latex, thick, sweetpotato] (T8) -- (T9) node [midway, above] {8};
\draw[-latex, thick, sweetpotato] (T9) -- (T10) node [near start, right] {9};
\draw[-latex, thick, sweetpotato] (T10) -- (T11) node [midway, right] {10};
\end{tikzpicture}
\end{minipage}%
\begin{minipage}{0.25 \columnwidth}
\centering
\begin{tabu}spread 0pt{@{} l l l l @{}}
\toprule
& \multicolumn{3}{c}{True} \\
& \multicolumn{3}{c}{Probability} \\
\cmidrule(lr){2-4}
Edge & 0 & 1 & 2 \\
\midrule
1 & 0.2 & 0.8 & 0 \\
2 & 0.4 & 0.6 & 0 \\
3 & 0.6 & 0.4 & 0 \\
4 & 0.8 & 0.2 & 0 \\
5 & 1 & 0 & 0 \\
\addlinespace
6 & 0 & 1 & 0 \\
7 & 0.2 & 0.8 & 0 \\
8 & 0.4 & 0.6 & 0 \\
9 & 0.6 & 0.4 & 0 \\
10 & 0.8 & 0.2 & 0 \\
\bottomrule
\end{tabu}
\end{minipage}%
\caption{The true graph and probabilities for each edge in topology GN11. The edges in orange can change direction while remaining in the Markov equivalence class of the true graph -- as long as another v-structure is not created.}
\label{sfig:gn11}
\end{figure}


\begin{figure}[H]
\centering
\begin{minipage}{0.25 \columnwidth}
\begin{tikzpicture}
\node[gene] (T1) {$T_1$};
\node[gene] (T2) [below left = 0.75cm and 0.25cm of T1] {$T_2$};
\node[gene] (T3) [below right = 0.75cm and 0.25cm of T1] {$T_3$};
\node[gene] (T4) [right = 0.75cm of T1] {$T_4$};
\node[gene] (T5) [below = 0.75cm of T2] {$T_5$};
\node[gene] (T6) [below = 0.75cm of T3] {$T_6$};
\node[gene] (T7) [below = 0.75cm of T6] {$T_7$};
\node[gene] (T8) [below = 0.75cm of T5] {$T_8$};

\draw[-latex, thick, sweetpotato] (T1) to node[near start, left] {1} (T2);
\draw[-latex, thick] (T1) to[bend left = 55] node[midway, right] {2} (T6);
\draw[-latex, thick] (T1) to[bend right = 50] node[midway, left] {3} (T8);
\draw[-latex, thick, sweetpotato] (T2) -- (T3) node[midway, above] {4};
\draw[-latex, thick, sweetpotato] (T2) -- (T5) node[midway, left] {5};
\draw[-latex, thick] (T5) -- (T6) node[midway, above] {6};
\draw[-latex, thick] (T5) -- (T8) node[midway, left] {7};
\draw[-latex, thick] (T6) -- (T7) node[midway, left] {8};
\end{tikzpicture}
\end{minipage}%
\begin{minipage}{0.25 \columnwidth}
\centering
\begin{tabu}spread 0pt{@{} l l l l @{}}
\toprule
& \multicolumn{3}{c}{True} \\
& \multicolumn{3}{c}{Probability} \\
\cmidrule(lr){2-4}
Edge & 0 & 1 & 2 \\
\midrule
1 & 0.25 & 0.75 & 0 \\
2 & 1 & 0 & 0 \\
3 & 1 & 0 & 0 \\
4 & 0.75 & 0.25 & 0 \\
5 & 0.75 & 0.25 & 0 \\
\addlinespace
6 & 1 & 0 & 0 \\
7 & 1 & 0 & 0 \\
8 & 1 & 0 & 0 \\
\bottomrule
\end{tabu}
\end{minipage}%
\caption{The true graph and probabilities for each edge in topology GN8. The edges in orange can change direction while remaining in the Markov equivalence class of the true graph -- as long as another v-structure is not created.}
\label{sfig:gn8}
\end{figure}


\begin{figure}[H]
\centering
\includegraphics[scale = 0.5]{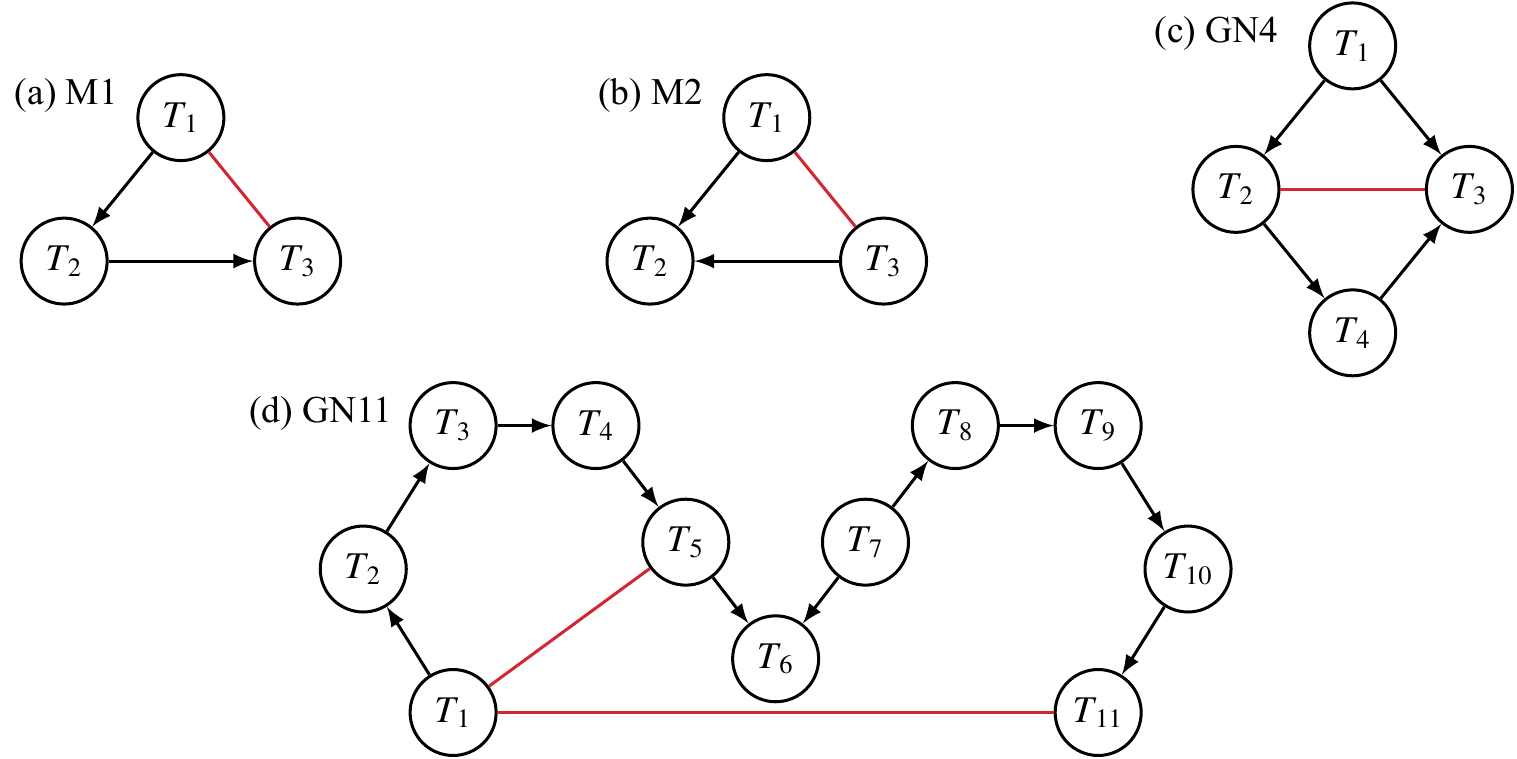}
\caption{\label{sfig:falseEdges} {The black edges (true edges) were used to simulate the data and the red edges (false edges) were added to the true adjacency matrix as input to baycn.}}
\end{figure}


\begin{figure}[H]
\centering
\includegraphics[width = \textwidth]{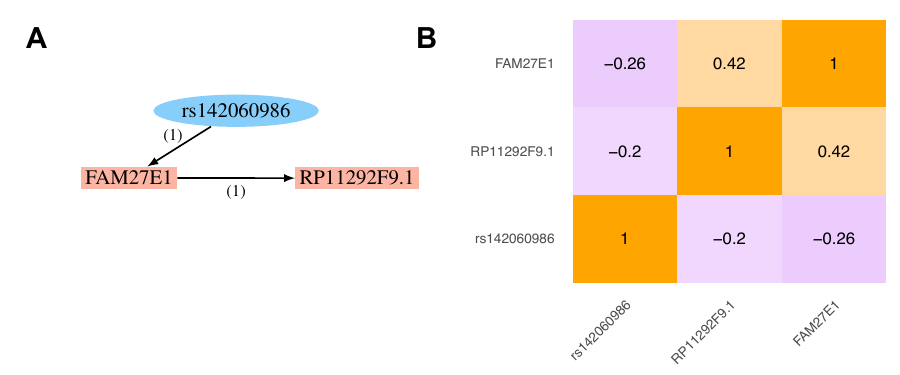}
\caption{\label{sfig:q20} Inference of the GEUVADIS eQTL-gene set Q20, which does not have associated PCs.  
(A) Inferred graph with posterior probabilities.  Numbers in parenthesis next to the edges indicate the posterior probability 
for the direction shown.  (B) Correlation heatmap.
}
\end{figure}


\begin{figure}[H]
\centering
\includegraphics[width = \textwidth]{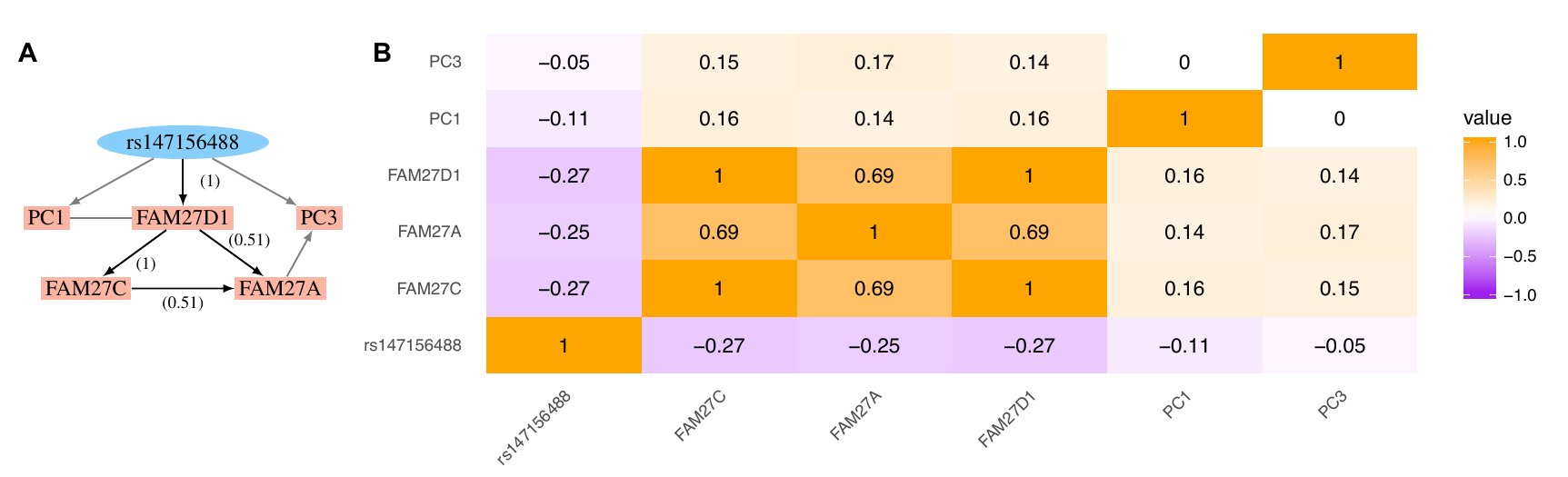}
\caption{\label{sfig:q21} Inference of the GEUVADIS eQTL-gene set Q21 with associated PCs.  
(A) Inferred graph with posterior probabilities.  Numbers in parenthesis next to the edges indicate the posterior probability 
for the direction shown.  (B) Correlation heatmap.
}
\end{figure}


\begin{figure}[H]
\centering
\includegraphics[width = \textwidth]{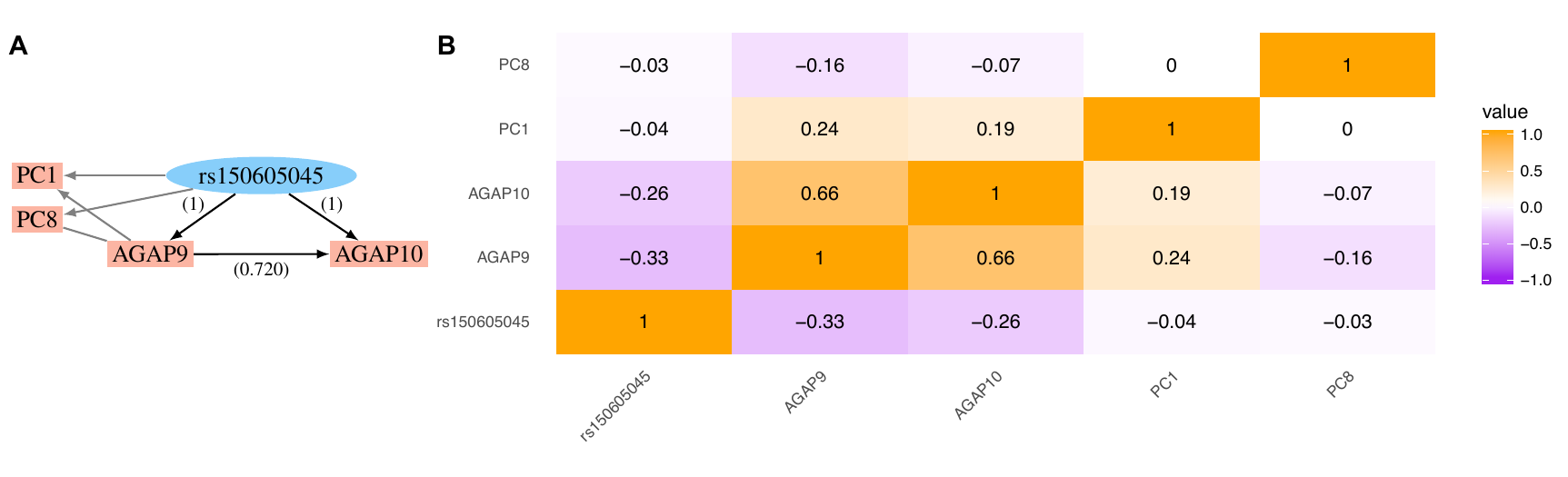}
\caption{\label{sfig:q23} Inference of the GEUVADIS eQTL-gene set Q23 with associated PCs.  
(A) Inferred graph with posterior probabilities.  Numbers in parenthesis next to the edges indicate the posterior probability 
for the direction shown.  (B) Correlation heatmap.
}
\end{figure}


\begin{figure}[H]
\centering
\includegraphics[width = \textwidth]{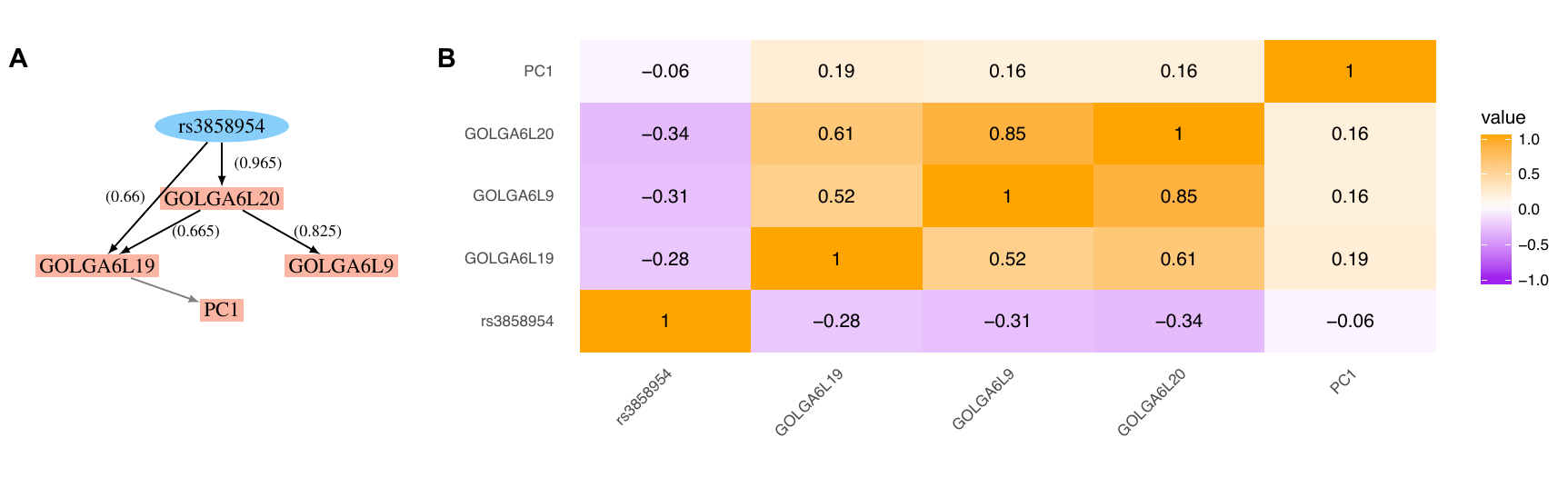}
\caption{\label{sfig:q37} Inference of the GEUVADIS eQTL-gene set Q237 with associated PCs.  
(A) Inferred graph with posterior probabilities.  Numbers in parenthesis next to the edges indicate the posterior probability 
for the direction shown.  (B) Correlation heatmap.
}
\end{figure}


\begin{figure}[H]
\centering
\includegraphics[width = \textwidth]{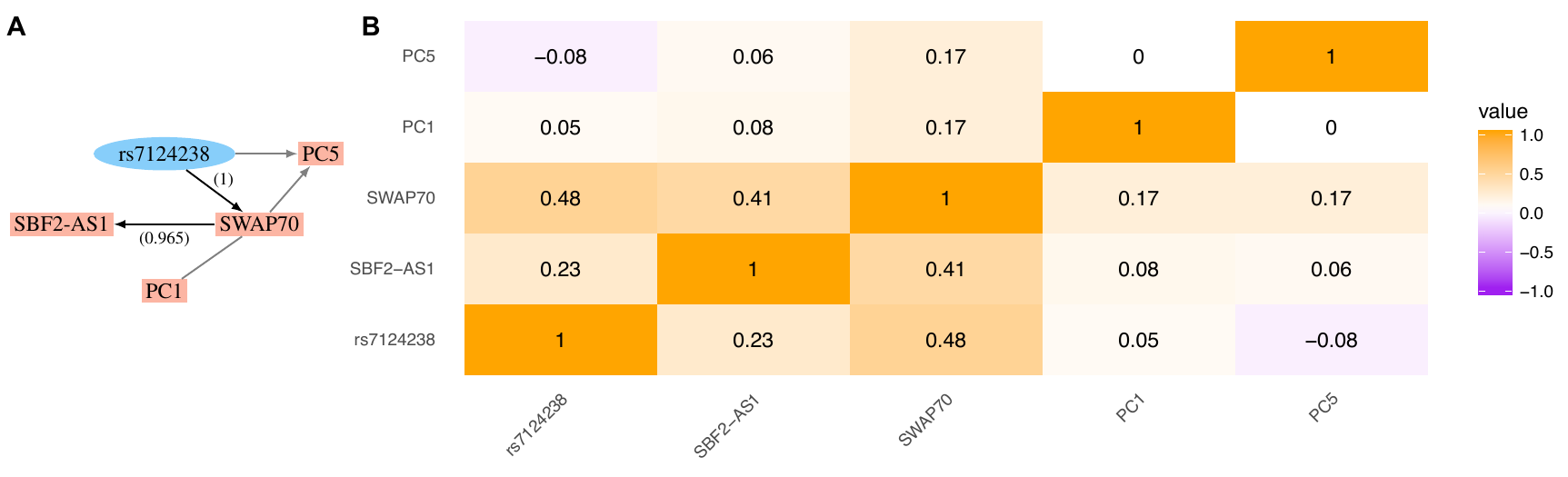}
\caption{\label{sfig:q50} Inference of the GEUVADIS eQTL-gene set Q50 with associated PCs.  
(A) Inferred graph with posterior probabilities.  Numbers in parenthesis next to the edges indicate the posterior probability 
for the direction shown.  (B) Correlation heatmap.
}
\end{figure}


\begin{figure}[H]
\centering
\includegraphics[scale=.6]{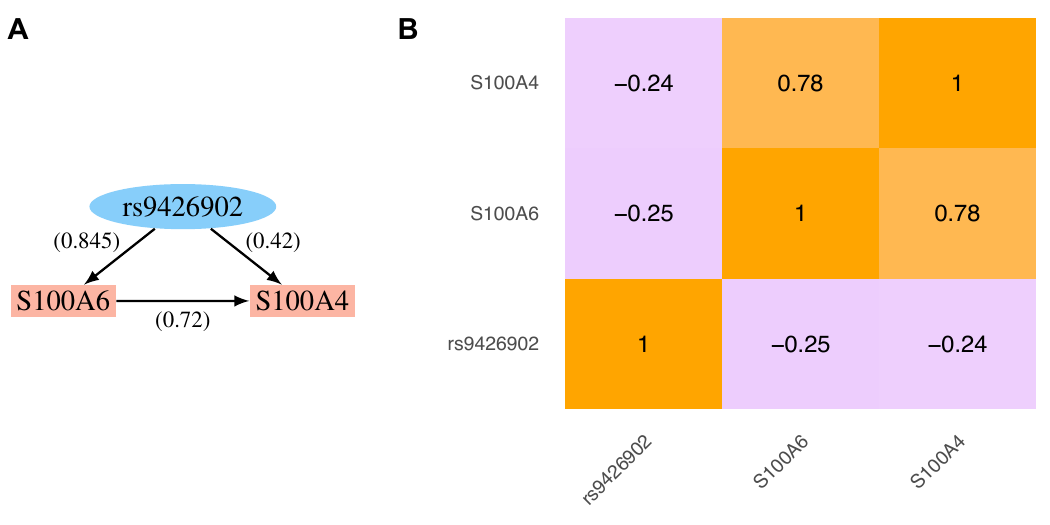}
\caption{\label{sfig:q62} Inference of the GEUVADIS eQTL-gene set Q62, which does not have associated PCs.  
(A) Inferred graph with posterior probabilities.  Numbers in parenthesis next to the edges indicate the posterior probability 
for the direction shown.  (B) Correlation heatmap.
}
\end{figure}


\begin{figure}[H]
\includegraphics[width = \textwidth]{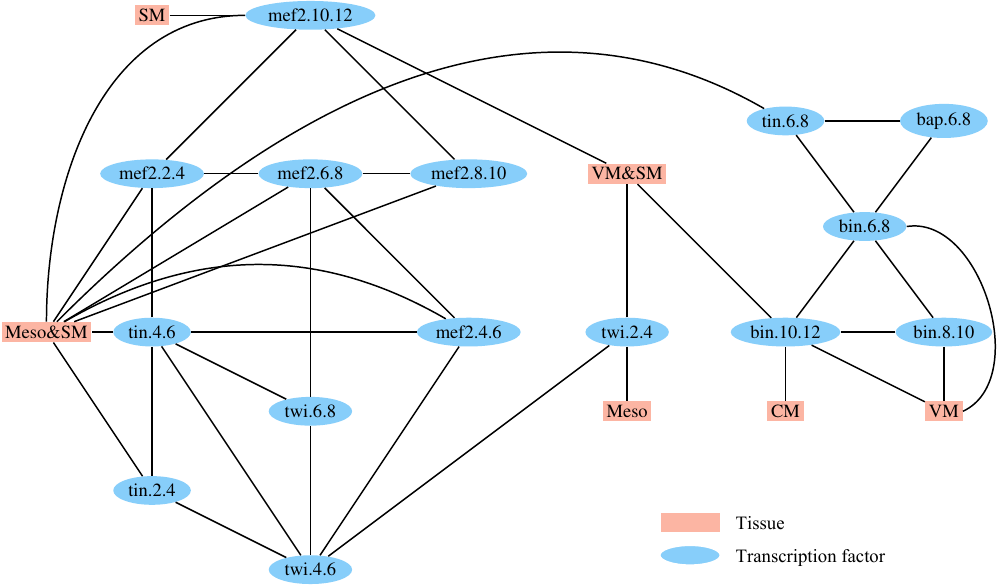}
\caption{\label{sfig:drosophila_skeleton} Graph used as the input to baycn for the Drosophila data. We first used the graph output by MRPC then added additional edges between Meso\&SM and the transcription factors Mef2 and Tin and between VM and the transcription factor Bin.}
\end{figure}


\newpage
\setcounter{table}{0}
\renewcommand{\thetable}{S\arabic{table}}


\begin{table}[H]
\centering
\caption{\label{stab:mse1} {Performance of baycn on all the graphs used in simulation studies.}  Features of the graphs in 
Figure 1 in the main text, 
such as the number of edges and v-structures, are listed.  The mean and standard deviation of MSE$_1$ (on three states of each edge), sample size $N$, and signal strength $\beta$ are also listed. For each simulation scenario we generated 25 independent data sets and ran baycn once on each data set.}
\begin{tabu}spread 0pt{@{} l l >{\raggedright\arraybackslash}p{1.75cm} l l l l l l l @{}}
\toprule
& & & & \multicolumn{6}{c}{MSE$_1$} \\
 \cmidrule(lr){5-10}
& & & & \multicolumn{2}{c}{$\beta = 0.2$} & \multicolumn{2}{c}{$\beta = 0.5$}  & \multicolumn{2}{c}{$\beta = 1$} \\
\cmidrule(lr){5-6} \cmidrule(lr){7-8} \cmidrule(lr){9-10}
Topology & \# edges & \# v-structures & $N$ & mean & sd & mean & sd & mean & sd \\
\midrule
M1 & 2 & 0 & 100 & 0.1796 & 0.0949 & 0.0127 & 0.0353 & 0.0011 & 0.0016 \\
& & & 200 & 0.0734 & 0.074 & 0.0014 & 0.0015 & 0.0012 & 0.0013 \\
& & & 600 & 0.0237 & 0.0326 & 0.001 & 0.0009 & 0.0008 & 0.0007 \\
\addlinespace
M2 & 2 & 1 &  100 & 0.3384 & 0.1081 & 0.0909 & 0.0919 & 0 & 0 \\
& & & 200 & 0.2688 & 0.0881 & 0.0597 & 0.082 & 0 & 0 \\
& & & 600 & 0.1323 & 0.0735 & 0 & 0 & 0 & 0 \\
\addlinespace
GN4 & 4 & 1 &  100 & 0.2731 & 0.0711 & 0.0674 & 0.0417 & 0.01 & 0.0276 \\
& & & 200 & 0.167 & 0.0554 & 0.0755 & 0.0639 & 0.0046 & 0.0142 \\
& & & 600 & 0.0839 & 0.0259 & 0.0503 & 0.0602 & 0.0069 & 0.0182 \\
\addlinespace
GN5 & 5 & 1 &  100 & 0.2687 & 0.0785 & 0.0396 & 0.0465 & 0.0022 & 0.0023 \\
& & & 200 & 0.1562 & 0.049 & 0.0171 & 0.0235 & 0.002 & 0.0031 \\
& & & 600 & 0.0684 & 0.0338 & 0.0114 & 0.0483 & 0.002 & 0.0024 \\
\addlinespace
Mulit-parent & 3 & 3 &  100 & 0.3361 & 0.0961 & 0.0486 & 0.0698 & 0 & 0 \\
& & & 200 & 0.237 & 0.0812 & 0.0032 & 0.0135 & 0 & 0 \\
& & & 600 & 0.1418 & 0.0625 & 0 & 0.0002 & 0 & 0 \\
\addlinespace
GN11 & 10 & 1 &  100 & 0.2266 & 0.0513 & 0.0353 & 0.0225 & 0.0042 & 0.0031 \\
& & & 200 & 0.139 & 0.0431 & 0.0121 & 0.0162 & 0.0066 & 0.0041 \\
& & & 600 & 0.0613 & 0.0197 & 0.0047 & 0.0043 & 0.0046 & 0.0039 \\
\addlinespace
GN8 & 8 & 2 &  100  & 0.2959 & 0.1517 & 0.0667 & 0.0666 & 0.0247 & 0.0526 \\
& & & 200 & 0.197 & 0.144 & 0.0692 & 0.0962 & 0.0186 & 0.0486 \\
& & & 600 & 0.1147 & 0.1073 & 0.0556 & 0.0858 & 0.0109 & 0.0467 \\
\bottomrule
\end{tabu}
\end{table}


\begin{table}[H]
\centering
\captionsetup{width = 0.9 \linewidth}
\caption{The mean and standard deviation of the edge-wise MSE for each edge in topology M1. We used the data sets previously simulated for M1 and included one false edge with the true edges in the input to baycn. We ran baycn with three different priors on edge states for each data set. The rows in red represent false edges.}
\begin{tabu}spread 0pt{@{} l l l l l l l l @{}}
\toprule
& & \multicolumn{6}{c}{eMSE: Topology M1} \\
 \cmidrule(lr){3-8}
& & \multicolumn{2}{c}{Prior 1} & \multicolumn{2}{c}{Prior 2}  & \multicolumn{2}{c}{Prior 3} \\
\cmidrule(lr){3-4} \cmidrule(lr){5-6} \cmidrule(lr){7-8}
$N$ & Edge & mean & sd & mean & sd & mean & sd \\
\midrule
100 & 1 & 0.0109 & 0.006 & 0.0078 & 0.0047 & 0.002 & 0.0026\\
 & \color{garnet} 2 & \color{garnet} 0.2853 & \color{garnet} 0.0568 & \color{garnet} 0.1859 & \color{garnet} 0.0618 & \color{garnet} 0.0131 & \color{garnet} 0.0139\\
 & 3 & 0.0107 & 0.0061 & 0.0091 & 0.0059 & 0.0014 & 0.0015\\
 \addlinespace
 200 & 1 & 0.0116 & 0.0067 & 0.0075 & 0.005 & 0.002 & 0.0027\\
 & \color{garnet} 2 & \color{garnet} 0.2861 & \color{garnet} 0.0816 & \color{garnet} 0.1953 & \color{garnet} 0.0992 & \color{garnet} 0.03 & \color{garnet} 0.0614\\
 & 3 & 0.0109 & 0.0063 & 0.0094 & 0.0046 & 0.0019 & 0.0025\\
 \addlinespace
 600 & 1 & 0.0127 & 0.0062 & 0.008 & 0.0056 & 0.0008 & 0.001\\
 & \color{garnet} 2 & \color{garnet} 0.2665 & \color{garnet} 0.038 & \color{garnet} 0.1593 & \color{garnet} 0.052 & \color{garnet} 0.0095 & \color{garnet} 0.0068\\
 & 3 & 0.0139 & 0.0085 & 0.0062 & 0.004 & 0.0016 & 0.0021\\
\bottomrule
\end{tabu}
\label{stab:m1fe}
\end{table}


\begin{table}[H]
\centering
\captionsetup{width = 0.9 \linewidth}
\caption{The mean and standard deviation of the edge-wise MSE for each edge in topology M2. We used the data sets previously simulated for M2 and included one false edge with the true edges in the input to baycn. We ran baycn with three different priors on edge states for each data set. The rows in red represent false edges.}
\label{stab:m2fe}
\begin{tabu}spread 0pt{@{} l l l l l l l l @{}}
\toprule
& & \multicolumn{6}{c}{eMSE: Topology M2} \\
 \cmidrule(lr){3-8}
& & \multicolumn{2}{c}{Prior 1} & \multicolumn{2}{c}{Prior 2}  & \multicolumn{2}{c}{Prior 3} \\
\cmidrule(lr){3-4} \cmidrule(lr){5-6} \cmidrule(lr){7-8}
$N$ & Edge & mean & sd & mean & sd & mean & sd \\
\midrule
100 & 1 & 0.1366 & 0.0269 & 0.1197 & 0.0199 & 0.0224 & 0.026\\
 & \color{garnet} 2 & \color{garnet} 0.4035 & \color{garnet} 0.0368 & \color{garnet} 0.3285 & \color{garnet} 0.0548 & \color{garnet} 0.0656 & \color{garnet} 0.0666\\
 & 3 & 0.1461 & 0.025 & 0.1241 & 0.0226 & 0.0232 & 0.028\\
 \addlinespace
 200 & 1 & 0.1329 & 0.0262 & 0.1215 & 0.0222 & 0.0297 & 0.0354\\
 & \color{garnet} 2 & \color{garnet} 0.4111 & \color{garnet} 0.0373 & \color{garnet} 0.3474 & \color{garnet} 0.0612 & \color{garnet} 0.0863 & \color{garnet} 0.0975\\
 & 3 & 0.1391 & 0.0227 & 0.1209 & 0.0245 & 0.0317 & 0.0396\\
 \addlinespace
 600 & 1 & 0.1369 & 0.0207 & 0.1183 & 0.0221 & 0.0236 & 0.0164\\
 & \color{garnet} 2 & \color{garnet} 0.4062 & \color{garnet} 0.0347 & \color{garnet} 0.3362 & \color{garnet} 0.0459 & \color{garnet} 0.0647 & \color{garnet} 0.0377\\
 & 3 & 0.1378 & 0.028 & 0.1186 & 0.0223 & 0.0224 & 0.0149\\
\bottomrule
\end{tabu}
\end{table}


\begin{table}[H]
\centering
\captionsetup{width = 0.9 \linewidth}
\caption{The mean and standard deviation of the edge-wise MSE for each edge in topology GN4. We used the data sets previously simulated for GN4 and included one false edge with the true edges in the input to baycn. We ran baycn with three different priors on edge states for each data set. The rows in red represent false edges.}
\label{stab:gn4fe}
\begin{tabu}spread 0pt{@{} l l l l l l l l @{}}
\toprule
& & \multicolumn{6}{c}{eMSE: Topology GN4} \\
 \cmidrule(lr){3-8}
& & \multicolumn{2}{c}{Prior 1} & \multicolumn{2}{c}{Prior 2}  & \multicolumn{2}{c}{Prior 3} \\
\cmidrule(lr){3-4} \cmidrule(lr){5-6} \cmidrule(lr){7-8}
$N$ & Edge & mean & sd & mean & sd & mean & sd \\
\midrule
100  & 1 & 0.0101 & 0.0182 & 0.011 & 0.0165 & 0.0418 & 0.0797 \\
 & 2 & 0.011 & 0.0204 & 0.0098 & 0.0192 & 0.0431 & 0.0821 \\
 & \color{garnet} 3 & \color{garnet} 0.2235 & \color{garnet} 0.1185 & \color{garnet} 0.1368 & \color{garnet} 0.1257 & \color{garnet} 0.0301 & \color{garnet} 0.1031 \\
 & 4 & 0.0085 & 0.0148 & 0.0078 & 0.0132 & 0.0462 & 0.0817 \\
 & 5 & 0.0087 & 0.0162 & 0.0107 & 0.0203 & 0.0459 & 0.0888 \\
 \addlinespace
200  & 1 & 0.0012 & 0.0023 & 0.0019 & 0.0036 & 0.0152 & 0.0414 \\
 & 2 & 0.0004 & 0.0013 & 0.0012 & 0.004 & 0.0156 & 0.0462 \\
 & \color{garnet} 3 & \color{garnet} 0.2154 & \color{garnet} 0.079 & \color{garnet} 0.1139 & \color{garnet} 0.0669 & \color{garnet} 0.005 & \color{garnet} 0.0083 \\
 & 4 & 0.0013 & 0.002 & 0.0023 & 0.0045 & 0.0181 & 0.0432 \\
 & 5 & 0.0003 & 0.001 & 0.0013 & 0.0043 & 0.012 & 0.0317 \\
\addlinespace
600  & 1 & 0.0008 & 0.0015 & 0.0009 & 0.001 & 0.0103 & 0.0387\\
 & 2 & 0 & 0 & 0 & 0 & 0.0102 & 0.0402\\
 & \color{garnet} 3 & \color{garnet} 0.2534 & \color{garnet} 0.1019 & \color{garnet} 0.1544 & \color{garnet} 0.1028 & \color{garnet} 0.0092 & \color{garnet} 0.0201\\
 & 4 & 0.0007 & 0.0012 & 0.0012 & 0.0015 & 0.011 & 0.038\\
 & 5 & 0 & 0 & 0 & 0 & 0.0112 & 0.0503 \\
\bottomrule
\end{tabu}
\end{table}


\begin{table}[H]
\centering
\captionsetup{width = 0.9 \linewidth}
\caption{The mean and standard deviation of the edge-wise MSE for each edge in topology GN11. We used the data sets previously simulated for GN11 and included two false edges with the true edges in the input to baycn. We ran baycn with three different priors on edge states for each data set. The rows in red represent false edges.}
\label{stab:gn11fe}
\begin{tabu}spread 0pt{@{} l l l l l l l l @{}}
\toprule
& & \multicolumn{6}{c}{eMSE: Topology GN11} \\
 \cmidrule(lr){3-8}
& & \multicolumn{2}{c}{Prior 1} & \multicolumn{2}{c}{Prior 2}  & \multicolumn{2}{c}{Prior 3} \\
\cmidrule(lr){3-4} \cmidrule(lr){5-6} \cmidrule(lr){7-8}
$N$ & Edge & mean & sd & mean & sd & mean & sd \\
\midrule
100 & 1 & 0.0023 & 0.0035 & 0.003 & 0.0038 & 0.0046 & 0.006\\
 & \color{garnet} 2 & \color{garnet} 0.2976 & \color{garnet} 0.1311 & \color{garnet} 0.2084 & \color{garnet} 0.1464 & \color{garnet} 0.0386 & \color{garnet} 0.1027\\
 & \color{garnet} 3 & \color{garnet} 0.2931 & \color{garnet} 0.0658 & \color{garnet} 0.2039 & \color{garnet} 0.0851 & \color{garnet} 0.0168 & \color{garnet} 0.0239\\
 & 4 & 0.0018 & 0.0028 & 0.0035 & 0.0033 & 0.009 & 0.0104\\
 & 5 & 0.0029 & 0.0046 & 0.0037 & 0.0048 & 0.0087 & 0.0116\\
 & 6 & 0.0045 & 0.007 & 0.0043 & 0.0064 & 0.0038 & 0.0044\\
 & 7 & 0 & 0 & 0 & 0 & 0 & 0\\
 & 8 & 0 & 0 & 0 & 0 & 0 & 0\\
 & 9 & 0.0009 & 0.0011 & 0.0014 & 0.0023 & 0.0037 & 0.005\\
 & 10 & 0.0018 & 0.002 & 0.002 & 0.0025 & 0.0107 & 0.0153\\
 & 11 & 0.002 & 0.0019 & 0.0024 & 0.0029 & 0.0097 & 0.0132\\
 & 12 & 0.0016 & 0.0018 & 0.0016 & 0.0026 & 0.0044 & 0.0057\\
 \addlinespace
 200 & 1 & 0.0048 & 0.0046 & 0.0033 & 0.0042 & 0.0029 & 0.0036\\
 & \color{garnet} 2 & \color{garnet} 0.286 & \color{garnet} 0.1233 & \color{garnet} 0.1938 & \color{garnet} 0.1334 & \color{garnet} 0.0338 & \color{garnet} 0.0753\\
 & \color{garnet} 3 & \color{garnet} 0.3161 & \color{garnet} 0.0676 & \color{garnet} 0.2138 & \color{garnet} 0.0676 & \color{garnet} 0.0188 & \color{garnet} 0.0152\\
 & 4 & 0.0032 & 0.0052 & 0.0028 & 0.0038 & 0.0076 & 0.0092\\
 & 5 & 0.0026 & 0.0043 & 0.0029 & 0.006 & 0.0069 & 0.0085\\
 & 6 & 0.0041 & 0.0056 & 0.0032 & 0.0056 & 0.0039 & 0.0052\\
 & 7 & 0 & 0 & 0 & 0 & 0 & 0\\
 & 8 & 0 & 0 & 0 & 0 & 0 & 0\\
 & 9 & 0.0006 & 0.0009 & 0.0004 & 0.0005 & 0.008 & 0.0137\\
 & 10 & 0.0013 & 0.0016 & 0.0007 & 0.0013 & 0.0115 & 0.0148\\
 & 11 & 0.0018 & 0.0021 & 0.0014 & 0.002 & 0.0093 & 0.0112\\
 & 12 & 0.0021 & 0.0025 & 0.0012 & 0.0018 & 0.0038 & 0.0041\\
 \addlinespace
 600 & 1 & 0.0069 & 0.0069 & 0.0052 & 0.0058 & 0.0048 & 0.0056\\
 & \color{garnet} 2 & \color{garnet} 0.2735 & \color{garnet} 0.0938 & \color{garnet} 0.1783 & \color{garnet} 0.0949 & \color{garnet} 0.0147 & \color{garnet} 0.0231\\
 & \color{garnet} 3 & \color{garnet} 0.3436 & \color{garnet} 0.1118 & \color{garnet} 0.2446 & \color{garnet} 0.129 & \color{garnet} 0.0557 & \color{garnet} 0.1172\\
 & 4 & 0.0049 & 0.0062 & 0.0028 & 0.003 & 0.0108 & 0.0147\\
 & 5 & 0.0024 & 0.0029 & 0.0017 & 0.0021 & 0.0077 & 0.0112\\
 & 6 & 0.0027 & 0.003 & 0.0015 & 0.0015 & 0.0024 & 0.0023\\
 & 7 & 0 & 0 & 0 & 0 & 0 & 0\\
 & 8 & 0 & 0 & 0 & 0 & 0 & 0\\
 & 9 & 0.0006 & 0.0008 & 0.0009 & 0.0013 & 0.0031 & 0.0036\\
 & 10 & 0.0015 & 0.0018 & 0.0023 & 0.0033 & 0.0058 & 0.0069\\
 & 11 & 0.0017 & 0.0021 & 0.0029 & 0.0047 & 0.0066 & 0.0087\\
 & 12 & 0.0026 & 0.005 & 0.0022 & 0.0041 & 0.0046 & 0.0047\\
\bottomrule
\end{tabu}
\end{table}


\begin{table}[H]
\centering
\caption{The mean and standard deviation of MSE$_2$ for GN4, GN8 and GN11 when the true skeleton was used as input. We simulated 25 data sets for each combination of topology, $N$, and $\beta$ and we ran each algorithm once per data set. MSE$_2$ is calculated only from the true edges.}
\label{stab:mse}
\begin{tabu}spread 0pt{@{} l l l l l l l l l @{}}
\toprule
& & & \multicolumn{6}{c}{MSE$_2$} \\
\cmidrule(lr){4-9}
& & & \multicolumn{2}{c}{GN4} & \multicolumn{2}{c}{GN8} & \multicolumn{2}{c}{GN11} \\
\cmidrule(lr){4-5} \cmidrule(lr){6-7} \cmidrule(lr){8-9}
Method & $N$ & $\beta$ & mean & sd & mean & sd & mean & sd \\
\midrule
baycn & 100 & 0.2 & 0.2305 & 0.0351 & 0.2707 & 0.0294 & 0.1743 & 0.0259 \\
 &  & 0.5 & 0.0987 & 0.0607 & 0.0951 & 0.0401 & 0.0514 & 0.0322 \\
 &  & 1 & 0.015 & 0.0414 & 0.0286 & 0.0549 & 0.0064 & 0.0047 \\

Gibbs & 100 & 0.2 & 0.2034 & 0.0292 & 0.2483 & 0.0276 & 0.1504 & 0.0215\\
 &  & 0.5 & 0.1009 & 0.0286 & 0.0873 & 0.0242 & 0.0439 & 0.03\\
 &  & 1 & 0.0221 & 0.0357 & 0.0313 & 0.0518 & 0.0005 & 0.0003\\

 MC$^3$ & 100 & 0.2 & 0.2039 & 0.0279 & 0.2488 & 0.0283 & 0.1508 & 0.0202 \\
 &  & 0.5 & 0.1022 & 0.0325 & 0.0888 & 0.027 & 0.0438 & 0.0299 \\
 &  & 1 & 0.1913 & 0.2189 & 0.1573 & 0.124 & 0.0003 & 0.0002 \\

order & 100 & 0.2 & 0.3144 & 0.0563 & 0.368 & 0.0375 & 0.271 & 0.0336 \\
 &  & 0.5 & 0.1187 & 0.0436 & 0.0997 & 0.0384 & 0.0657 & 0.0374 \\
 &  & 1 & 0.0146 & 0.0195 & 0.0309 & 0.0432 & 0.0069 & 0.0023 \\

partition & 100 & 0.2 & 0.2505 & 0.0423 & 0.316 & 0.025 & 0.2067 & 0.0247 \\
 &  & 0.5 & 0.1125 & 0.0388 & 0.1202 & 0.0478 & 0.0656 & 0.0329 \\
 &  & 1 & 0.0213 & 0.023 & 0.0393 & 0.05 & 0.0291 & 0.0223 \\
\addlinespace
baycn & 200 & 0.2 & 0.1796 & 0.0484 & 0.2129 & 0.0372 & 0.1308 & 0.0269 \\
 &  & 0.5 & 0.1132 & 0.0958 & 0.1302 & 0.0792 & 0.0182 & 0.0243 \\
 &  & 1 & 0.0069 & 0.0214 & 0.0649 & 0.1041 & 0.0098 & 0.0062 \\

Gibbs & 200 & 0.2 & 0.1716 & 0.0487 & 0.2071 & 0.0345 & 0.1239 & 0.0248 \\
 &  & 0.5 & 0.106 & 0.0683 & 0.0871 & 0.031 & 0.0134 & 0.0269 \\
 &  & 1 & 0.0015 & 0.0052 & 0.0005 & 0.0018 & 0.0004 & 0.0004 \\

MC$^3$ & 200 & 0.2 & 0.1725 & 0.0472 & 0.2071 & 0.035 & 0.1237 & 0.0252 \\
 &  & 0.5 & 0.1152 & 0.0966 & 0.1083 & 0.0626 & 0.0133 & 0.0268 \\
 &  & 1 & 0.2132 & 0.2264 & 0.2105 & 0.1217 & 0.0003 & 0.0002 \\

order & 200 & 0.2 & 0.2743 & 0.0793 & 0.2993 & 0.053 & 0.2052 & 0.0459 \\
 &  & 0.5 & 0.1089 & 0.072 & 0.0857 & 0.032 & 0.0322 & 0.0327 \\
 &  & 1 & 0.0067 & 0.0182 & 0.0051 & 0.0024 & 0.0087 & 0.004 \\

partition & 200 & 0.2 & 0.2035 & 0.045 & 0.2605 & 0.0325 & 0.1653 & 0.0327 \\
 &  & 0.5 & 0.1034 & 0.0628 & 0.1136 & 0.0561 & 0.0426 & 0.0259 \\
 &  & 1 & 0.0075 & 0.0133 & 0.013 & 0.0171 & 0.0327 & 0.0216 \\
\addlinespace
baycn & 600 & 0.2 & 0.1235 & 0.0372 & 0.1721 & 0.0813 & 0.0908 & 0.0288 \\
 &  & 0.5 & 0.0755 & 0.0903 & 0.1005 & 0.0992 & 0.0071 & 0.0064 \\
 &  & 1 & 0.0104 & 0.0273 & 0.0301 & 0.0686 & 0.0068 & 0.0059 \\

Gibbs & 600 & 0.2 & 0.125 & 0.0272 & 0.1674 & 0.0724 & 0.0864 & 0.0276\\
 &  & 0.5 & 0.0893 & 0.0835 & 0.0651 & 0.0431 & 0.0006 & 0.0012\\
 &  & 1 & 0 & 0 & 0 & 0 & 0.0005 & 0.0004\\

MC$^3$ & 600 & 0.2 & 0.1241 & 0.0282 & 0.1683 & 0.0741 & 0.0868 & 0.0274 \\
 &  & 0.5 & 0.229 & 0.2196 & 0.1316 & 0.1398 & 0.0005 & 0.0013 \\
 &  & 1 & 0.2311 & 0.2266 & 0.2232 & 0.1151 & 0.0003 & 0.0002 \\

order & 600 & 0.2 & 0.1351 & 0.0397 & 0.1648 & 0.069 & 0.1055 & 0.0382 \\
 &  & 0.5 & 0.0891 & 0.0898 & 0.0518 & 0.0433 & 0.0146 & 0.0158 \\
 &  & 1 & 0.0018 & 0.0006 & 0.0038 & 0.0005 & 0.0109 & 0.0052 \\

partition & 600 & 0.2 & 0.1257 & 0.028 & 0.1665 & 0.0615 & 0.1 & 0.0305 \\
 &  & 0.5 & 0.091 & 0.0803 & 0.0736 & 0.0792 & 0.0394 & 0.02 \\
 &  & 1 & 0.001 & 0.0015 & 0.0059 & 0.0146 & 0.0477 & 0.0389 \\
\bottomrule
\end{tabu}
\end{table}


\begin{longtable}{l l l l l l l l l}
\caption{\label{stab:pr} The mean and standard deviation of precision, power, and MSE$_2$ for topology GN4 when a fully connected graph was the input to each algorithm. When calculating precision and power we considered an edge present if the sum of the probability of the two directions was greater than 0.5.}\\
\toprule
& & & \multicolumn{2}{c}{Precision} & \multicolumn{2}{c}{Power} & \multicolumn{2}{c}{MSE$_2$} \\
\cmidrule(lr){4-5}  \cmidrule(lr){6-7} \cmidrule(lr){8-9}
Method & $N$ & $\beta$ & mean & sd & mean & sd & mean & sd \\
\midrule
\endfirsthead
\toprule
& & & \multicolumn{2}{c}{Precision} & \multicolumn{2}{c}{Power} & \multicolumn{2}{c}{MSE$_2$} \\
\cmidrule(lr){4-5}  \cmidrule(lr){6-7} \cmidrule(lr){8-9}
Method & $N$ & $\beta$ & mean & sd & mean & sd & mean & sd \\
\midrule
\endhead 
baycn & 100 & 0.2 & 0.94 & 0.2077 & 0.43 & 0.2654 & 0.2398 & 0.0326 \\
 &  & 0.5 & 0.982 & 0.0627 & 0.98 & 0.0692 & 0.1235 & 0.0478 \\
 &  & 1 & 0.9707 & 0.0841 & 1 & 0 & 0.0439 & 0.0387 \\
\addlinespace
Gibbs & 100 & 0.2 & 0.942 & 0.206 & 0.5 & 0.2602 & 0.2163 & 0.0303 \\
 &  & 0.5 & 0.984 & 0.0554 & 1 & 0 & 0.1212 & 0.0299 \\
 &  & 1 & 0.9093 & 0.1184 & 1 & 0 & 0.0938 & 0.0592 \\
\addlinespace
MC$^3$ & 100 & 0.2 & 0.942 & 0.206 & 0.49 & 0.265 & 0.2155 & 0.0312 \\
 &  & 0.5 & 0.976 & 0.0663 & 1 & 0 & 0.1215 & 0.0329 \\
 &  & 1 & 0.9093 & 0.1184 & 1 & 0 & 0.0945 & 0.0683 \\
\addlinespace
order & 100 & 0.2 & 0.80 & 0.4082 & 0.29 & 0.2126 & 0.315 & 0.0562 \\
 &  & 0.5 & 0.99 & 0.05 & 0.96 & 0.0935 & 0.1197 & 0.0502 \\
 &  & 1 & 0.992 & 0.04 & 1 & 0 & 0.0193 & 0.0287 \\
 \addlinespace
partition & 100 & 0.2 & 0.96 & 0.2 & 0.36 & 0.2051 & 0.2546 & 0.04\\
 &  & 0.5 & 0.982 & 0.0627 & 0.98 & 0.0692 & 0.1118 & 0.0357\\
 &  & 1 & 0.976 & 0.0663 & 1 & 0 & 0.0287 & 0.0311\\
 \addlinespace
scanBMA & 100 & 0.2 & 0.88 & 0.3317 & 0.35 & 0.25 & - & -\\
 &  & 0.5 & 0.948 & 0.0952 & 0.97 & 0.0829 & - & -\\
 &  & 1 & 0.7927 & 0.0281 & 0.99 & 0.05 & - & -\\
 
\addlinespace
 
baycn & 200 & 0.2 & 0.99 & 0.05 & 0.74 & 0.2222 & 0.1812 & 0.0427\\
 &  & 0.5 & 0.992 & 0.04 & 1 & 0 & 0.101 & 0.0769\\
 &  & 1 & 0.984 & 0.0554 & 1 & 0 & 0.0307 & 0.0427\\
\addlinespace
Gibbs & 200 & 0.2 & 0.98 & 0.0692 & 0.77 & 0.2155 & 0.1763 & 0.0383\\
 &  & 0.5 & 0.992 & 0.04 & 1 & 0 & 0.1134 & 0.0555\\
 &  & 1 & 0.952 & 0.0872 & 1 & 0 & 0.0521 & 0.0476\\
\addlinespace
MC$^3$ & 200 & 0.2 & 0.98 & 0.0692 & 0.77 & 0.2155 & 0.1765 & 0.0376\\
 &  & 0.5 & 0.992 & 0.04 & 1 & 0 & 0.1189 & 0.0619\\
 &  & 1 & 0.96 & 0.0816 & 1 & 0 & 0.0525 & 0.0452\\
\addlinespace
order & 200 & 0.2 & 0.9067 & 0.2809 & 0.48 & 0.2385 & 0.2768 & 0.0782\\
 &  & 0.5 & 1 & 0 & 1 & 0 & 0.1072 & 0.0721\\
 &  & 1 & 0.992 & 0.04 & 1 & 0 & 0.011 & 0.0349\\
\addlinespace
partition & 200 & 0.2 & 0.9867 & 0.0667 & 0.59 & 0.2026 & 0.2087 & 0.0448\\
 &  & 0.5 & 0.992 & 0.04 & 1 & 0 & 0.1043 & 0.0646\\
 &  & 1 & 0.992 & 0.04 & 1 & 0 & 0.012 & 0.0261\\
 \addlinespace
 scanBMA & 200 & 0.2 & 0.9787 & 0.0763 & 0.67 & 0.225 & - & -\\
 &  & 0.5 & 0.8747 & 0.1077 & 1 & 0 & - & -\\
 &  & 1 & 0.8 & 0 & 1 & 0 & - & -\\
 
 \addlinespace

baycn & 600 & 0.2 & 0.992 & 0.04 & 1 & 0 & 0.133 & 0.0358\\
 &  & 0.5 & 0.992 & 0.04 & 1 & 0 & 0.0954 & 0.096\\
 &  & 1 & 0.992 & 0.04 & 1 & 0 & 0.0125 & 0.0102\\
\addlinespace
Gibbs & 600 & 0.2 & 0.992 & 0.04 & 0.99 & 0.05 & 0.1274 & 0.0219\\
 &  & 0.5 & 0.992 & 0.04 & 1 & 0 & 0.0982 & 0.0741\\
 &  & 1 & 0.992 & 0.04 & 1 & 0 & 0.0163 & 0.0156\\
\addlinespace
MC$^3$ & 600 & 0.2 & 0.992 & 0.04 & 0.99 & 0.05 & 0.131 & 0.0225\\
 &  & 0.5 & 0.992 & 0.04 & 1 & 0 & 0.1098 & 0.093\\
 &  & 1 & 1 & 0 & 1 & 0 & 0.0153 & 0.0124\\
\addlinespace
order & 600 & 0.2 & 1 & 0 & 0.96 & 0.1181 & 0.1395 & 0.0391\\
 &  & 0.5 & 1 & 0 & 1 & 0 & 0.0909 & 0.0903\\
 &  & 1 & 1 & 0 & 1 & 0 & 0.002 & 0.0008\\
\addlinespace
partition & 600 & 0.2 & 1 & 0 & 0.97 & 0.1099 & 0.1288 & 0.0272\\
 &  & 0.5 & 1 & 0 & 1 & 0 & 0.0869 & 0.077\\
 &  & 1 & 1 & 0 & 1 & 0 & 0.0016 & 0.0018\\
 \addlinespace
 scanBMA & 600 & 0.2 & 0.976 & 0.0663 & 0.99 & 0.05 & - & -\\
 &  & 0.5 & 0.8 & 0 & 1 & 0 & - & -\\
 &  & 1 & 0.8 & 0 & 1 & 0 & - & -\\
\bottomrule
\end{longtable}


\begin{table}[H]
\centering
\caption{The posterior probability of the three edge states for baycn and partition MCMC from the first 7 simulated data sets of GN4  with $N = 100$ and $\beta = 0.2$. The edges in red are the false edges, edge numbers with -v are the edges that make a v-structure, and the edge numbers in bold are the edges where baycn inferred a true edge as present but partition MCMC did not.}
\label{stab:gn4.n100}
\begin{tabu}{@{} l l l l l l l l @{}}
\toprule
& & \multicolumn{6}{c}{Posterior Probability: GN4} \\
\cmidrule(lr){3-8}
& & \multicolumn{3}{c}{baycn} & \multicolumn{3}{c}{partition} \\
\cmidrule(lr){3-5} \cmidrule(lr){6-8}
data set & Edge & zero & one & two & zero & one & two \\
\midrule
1 & 1 & 0.08 & 0.045 & 0.875 & 0.032 & 0.04 & 0.928\\
 & 2-v & 0.12 & 0.085 & 0.795 & 0.045 & 0.047 & 0.908\\
 & \color{garnet} 3 & \color{garnet} 0.025 & \color{garnet} 0.055 & \color{garnet} 0.92 & \color{garnet} 0.03 & \color{garnet} 0.026 & \color{garnet} 0.944\\
 & \color{garnet} 4 & \color{garnet} 0.08 & \color{garnet} 0.045 & \color{garnet} 0.875 & \color{garnet} 0.044 & \color{garnet} 0.029 & \color{garnet} 0.928\\
 & 5 & 0.265 & 0.21 & 0.525 & 0.173 & 0.117 & 0.709\\
 & 6-v & 0.48 & 0.515 & 0.005 & 0.536 & 0.463 & 0.001\\
 \addlinespace
2 & 1 & 0.435 & 0.405 & 0.16 & 0.405 & 0.358 & 0.237\\
 & 2-v & 0.045 & 0.075 & 0.88 & 0.025 & 0.027 & 0.948\\
 & \color{garnet} 3 & \color{garnet} 0.035 & \color{garnet} 0.1 & \color{garnet} 0.865 & \color{garnet} 0.025 & \color{garnet} 0.04 & \color{garnet} 0.935\\
 & \color{garnet} 4 & \color{garnet} 0.245 & \color{garnet} 0.32 & \color{garnet} 0.435 & \color{garnet} 0.209 & \color{garnet} 0.289 & \color{garnet} 0.501\\
 & \large \textbf{5} & 0.2 & 0.36 & 0.44 & 0.145 & 0.244 & 0.611\\
 & 6-v & 0.505 & 0.44 & 0.055 & 0.516 & 0.434 & 0.05\\
\addlinespace
3 & 1 & 0.515 & 0.385 & 0.1 & 0.473 & 0.349 & 0.178\\
 & \large \textbf{2-v} & 0.285 & 0.285 & 0.43 & 0.177 & 0.226 & 0.597\\
 & \color{garnet} 3 & \color{garnet} 0.19 & \color{garnet} 0.34 & \color{garnet} 0.47 & \color{garnet} 0.161 & \color{garnet} 0.213 & \color{garnet} 0.626\\
 & \color{garnet} 4 & \color{garnet} 0.085 & \color{garnet} 0.045 & \color{garnet} 0.87 & \color{garnet} 0.039 & \color{garnet} 0.037 & \color{garnet} 0.924\\
 & 5 & 0.25 & 0.435 & 0.315 & 0.204 & 0.363 & 0.433\\
 & 6-v & 0.15 & 0.25 & 0.6 & 0.126 & 0.133 & 0.741\\
\addlinespace
4 & 1 & 0.14 & 0.105 & 0.755 & 0.101 & 0.097 & 0.802\\
 & 2-v & 0.185 & 0.19 & 0.625 & 0.126 & 0.132 & 0.742\\
 & \color{garnet} 3 & \color{garnet} 0.055 & \color{garnet} 0.095 & \color{garnet} 0.85 & \color{garnet} 0.051 & \color{garnet} 0.081 & \color{garnet} 0.868\\
 & \color{garnet} 4 & \color{garnet} 0.085 & \color{garnet} 0.06 & \color{garnet} 0.855 & \color{garnet} 0.056 & \color{garnet} 0.047 & \color{garnet} 0.897\\
 & \textbf{5} & 0.455 & 0.54 & 0.005 & 0.511 & 0.488 & 0.001\\
 & 6-v & 0.185 & 0.18 & 0.635 & 0.105 & 0.121 & 0.774\\
\addlinespace
5 & 1 & 0.16 & 0.175 & 0.665 & 0.142 & 0.152 & 0.706\\
 & 2-v & 0.175 & 0.115 & 0.71 & 0.076 & 0.06 & 0.864\\
 & \color{garnet} 3 & \color{garnet} 0.09 & \color{garnet} 0.09 & \color{garnet} 0.82 & \color{garnet} 0.047 & \color{garnet} 0.031 & \color{garnet} 0.921\\
 & \color{garnet} 4 & \color{garnet} 0.06 & \color{garnet} 0.095 & \color{garnet} 0.845 & \color{garnet} 0.025 & \color{garnet} 0.026 & \color{garnet} 0.949\\
 & 5 & 0.075 & 0.06 & 0.865 & 0.036 & 0.052 & 0.911\\
 & 6-v & 0.485 & 0.46 & 0.055 & 0.499 & 0.464 & 0.037\\
 \addlinespace
 6 & 1 & 0.165 & 0.25 & 0.585 & 0.153 & 0.132 & 0.714\\
 & 2-v & 0.52 & 0.475 & 0.005 & 0.469 & 0.491 & 0.04\\
 & \color{garnet} 3 & \color{garnet} 0.045 & \color{garnet} 0.08 & \color{garnet} 0.875 & \color{garnet} 0.032 & \color{garnet} 0.04 & \color{garnet} 0.928\\
 & \color{garnet} 4 & \color{garnet} 0.03 & \color{garnet} 0.025 & \color{garnet} 0.945 & \color{garnet} 0.03 & \color{garnet} 0.029 & \color{garnet} 0.941\\
 & 5 & 0.1 & 0.125 & 0.775 & 0.066 & 0.057 & 0.877\\
 & 6-v & 0.06 & 0.035 & 0.905 & 0.036 & 0.022 & 0.941\\
 \addlinespace
 7 & 1 & 0.48 & 0.43 & 0.09 & 0.416 & 0.383 & 0.201\\
 & 2-v & 0.065 & 0.085 & 0.85 & 0.05 & 0.045 & 0.905\\
 & \color{garnet} 3 & \color{garnet} 0.05 & \color{garnet} 0.05 & \color{garnet} 0.9 & \color{garnet} 0.039 & \color{garnet} 0.06 & \color{garnet} 0.901\\
 & \color{garnet} 4 & \color{garnet} 0.11 & \color{garnet} 0.04 & \color{garnet} 0.85 & \color{garnet} 0.059 & \color{garnet} 0.024 & \color{garnet} 0.918\\
 & 5 & 0.145 & 0.18 & 0.675 & 0.097 & 0.117 & 0.786\\
 & 6-v & 0.16 & 0.2 & 0.64 & 0.13 & 0.116 & 0.754\\
\bottomrule
\end{tabu}
\end{table}


\begin{table}[H]
\centering
\caption{The posterior probability of the three edge states for baycn and partition MCMC from the first 7 simulated data sets of GN4  with $N = 200$ and $\beta = 0.2$. The edges in red are the false edges, edge numbers with -v are the edges that make a v-structure, and the edge numbers in bold are the edges where baycn inferred a true edge as present but partition MCMC did not.}
\label{stab:gn4.n200}
\begin{tabu}{@{} l l l l l l l l @{}}
\toprule
& & \multicolumn{6}{c}{Posterior Probability: GN4} \\
\cmidrule(lr){3-8}
& & \multicolumn{3}{c}{baycn} & \multicolumn{3}{c}{partition} \\
\cmidrule(lr){3-5} \cmidrule(lr){6-8}
data set & Edge & zero & one & two & zero & one & two \\
\midrule
1 & 1 & 0.415 & 0.38 & 0.205 & 0.389 & 0.257 & 0.354\\
 & 2-v & 0.425 & 0.515 & 0.06 & 0.419 & 0.459 & 0.122\\
 & \color{garnet} 3 & \color{garnet} 0.095 & \color{garnet} 0.045 & \color{garnet} 0.86 & \color{garnet} 0.037 & \color{garnet} 0.032 & \color{garnet} 0.93\\
 & \color{garnet} 4 & \color{garnet} 0.075 & \color{garnet} 0.09 & \color{garnet} 0.835 & \color{garnet} 0.037 & \color{garnet} 0.054 & \color{garnet} 0.909\\
 & 5 & 0.14 & 0.115 & 0.745 & 0.056 & 0.091 & 0.853\\
 & 6-v & 0.605 & 0.395 & 0 & 0.599 & 0.392 & 0.01\\
 \addlinespace
2 & 1 & 0.33 & 0.67 & 0 & 0.272 & 0.728 & 0\\
 & 2-v & 0.335 & 0.645 & 0.02 & 0.231 & 0.692 & 0.077\\
 & \color{garnet} 3 & \color{garnet} 0.03 & \color{garnet} 0.075 & \color{garnet} 0.895 & \color{garnet} 0.015 & \color{garnet} 0.031 & \color{garnet} 0.954\\
 & \color{garnet} 4 & \color{garnet} 0.23 & \color{garnet} 0.16 & \color{garnet} 0.61 & \color{garnet} 0.121 & \color{garnet} 0.079 & \color{garnet} 0.8\\
 & \large \textbf{5} & 0.37 & 0.34 & 0.29 & 0.204 & 0.172 & 0.623\\
 & 6-v & 0.525 & 0.475 & 0 & 0.475 & 0.509 & 0.016\\
\addlinespace
3 & 1 & 0.37 & 0.385 & 0.245 & 0.298 & 0.294 & 0.408\\
 & \large \textbf{2-v} & 0.32 & 0.285 & 0.395 & 0.218 & 0.19 & 0.592\\
 & \color{garnet} 3 & \color{garnet} 0.025 & \color{garnet} 0.07 & \color{garnet} 0.905 & \color{garnet} 0.02 & \color{garnet} 0.016 & \color{garnet} 0.964\\
 & \color{garnet} 4 & \color{garnet} 0.13 & \color{garnet} 0.085 & \color{garnet} 0.785 & \color{garnet} 0.072 & \color{garnet} 0.065 & \color{garnet} 0.863\\
 & 5 & 0.335 & 0.535 & 0.13 & 0.349 & 0.43 & 0.221\\
 & 6-v & 0.32 & 0.44 & 0.24 & 0.253 & 0.286 & 0.461\\
\addlinespace
4 & 1 & 0.11 & 0.13 & 0.76 & 0.07 & 0.051 & 0.879\\
 & \large \textbf{2-v} & 0.415 & 0.29 & 0.295 & 0.3 & 0.181 & 0.519\\
 & \color{garnet} 3 & \color{garnet} 0.135 & \color{garnet} 0.07 & \color{garnet} 0.795 & \color{garnet} 0.052 & \color{garnet} 0.046 & \color{garnet} 0.901\\
 & \color{garnet} 4 & \color{garnet} 0.115 & \color{garnet} 0.19 & \color{garnet} 0.695 & \color{garnet} 0.076 & \color{garnet} 0.125 & \color{garnet} 0.799\\
 & 5 & 0.235 & 0.25 & 0.515 & 0.141 & 0.188 & 0.671\\
 & 6-v & 0.385 & 0.525 & 0.09 & 0.378 & 0.461 & 0.161\\
\addlinespace
5 & 1 & 0.51 & 0.48 & 0.01 & 0.498 & 0.47 & 0.032\\
 & 2-v & 0.41 & 0.4 & 0.19 & 0.383 & 0.312 & 0.305\\
 & \color{garnet} 3 & \color{garnet} 0.035 & \color{garnet} 0.045 & \color{garnet} 0.92 & \color{garnet} 0.019 & \color{garnet} 0.021 & \color{garnet} 0.96\\
 & \color{garnet} 4 & \color{garnet} 0.045 & \color{garnet} 0.035 & \color{garnet} 0.92 & \color{garnet} 0.016 & \color{garnet} 0.021 & \color{garnet} 0.963\\
 & 5 & 0.505 & 0.485 & 0.01 & 0.484 & 0.494 & 0.022\\
 & 6-v & 0.335 & 0.385 & 0.28 & 0.231 & 0.304 & 0.465\\
\addlinespace
6 & 1 & 0.225 & 0.17 & 0.605 & 0.146 & 0.103 & 0.751\\
 & 2-v & 0.56 & 0.44 & 0 & 0.591 & 0.403 & 0.006\\
 & \color{garnet} 3 & \color{garnet} 0.4 & \color{garnet} 0.48 & \color{garnet} 0.12 & \color{garnet} 0.47 & \color{garnet} 0.367 & \color{garnet} 0.163\\
 & \color{garnet} 4 & \color{garnet} 0.03 & \color{garnet} 0.05 & \color{garnet} 0.92 & \color{garnet} 0.021 & \color{garnet} 0.02 & \color{garnet} 0.959\\
 & 5 & 0.34 & 0.605 & 0.055 & 0.31 & 0.562 & 0.127\\
 & \large \textbf{6-v} & 0.285 & 0.295 & 0.42 & 0.17 & 0.19 & 0.641\\
\addlinespace
7 & \large \textbf{1} & 0.335 & 0.36 & 0.305 & 0.248 & 0.233 & 0.519\\
 & \large \textbf{2-v} & 0.27 & 0.235 & 0.495 & 0.166 & 0.157 & 0.677\\
 & \color{garnet} 3 & \color{garnet} 0.075 & \color{garnet} 0.035 & \color{garnet} 0.89 & \color{garnet} 0.02 & \color{garnet} 0.019 & \color{garnet} 0.961\\
 & \color{garnet} 4 & \color{garnet} 0.105 & \color{garnet} 0.12 & \color{garnet} 0.775 & \color{garnet} 0.059 & \color{garnet} 0.082 & \color{garnet} 0.859\\
 & 5 & 0.325 & 0.46 & 0.215 & 0.234 & 0.406 & 0.359\\
 & 6-v & 0.425 & 0.575 & 0 & 0.439 & 0.546 & 0.015\\
\bottomrule
\end{tabu}
\end{table}


\begin{table}[H]
\centering
\caption{The posterior probability of the three edge states for baycn and partition MCMC from the first 7 simulated data sets of GN4 with $N = 600$ and $\beta = 0.2$. The edges in red are the false edges, edge numbers with -v are the edges that make a v-structure, and the edge numbers in bold are the edges where baycn inferred a true edge as present but partition MCMC did not.}
\label{stab:gn4.n600}
\begin{tabu}{@{} l l l l l l l l @{}}
\toprule
& & \multicolumn{6}{c}{Posterior Probability: GN4} \\
\cmidrule(lr){3-8}
& & \multicolumn{3}{c}{baycn} & \multicolumn{3}{c}{partition} \\
\cmidrule(lr){3-5} \cmidrule(lr){6-8}
data set & Edge & zero & one & two & zero & one & two \\
\midrule
1 & 1 & 0.61 & 0.39 & 0 & 0.511 & 0.489 & 0\\
 & \large \textbf{2-v} & 0.345 & 0.325 & 0.33 & 0.224 & 0.148 & 0.627\\
 & \color{garnet} 3 & \color{garnet} 0.065 & \color{garnet} 0.03 & \color{garnet} 0.905 & \color{garnet} 0.02 & \color{garnet} 0.012 & \color{garnet} 0.968\\
 & \color{garnet} 4 & \color{garnet} 0.065 & \color{garnet} 0.045 & \color{garnet} 0.89 & \color{garnet} 0.016 & \color{garnet} 0.016 & \color{garnet} 0.968\\
 & 5 & 0.405 & 0.595 & 0 & 0.491 & 0.509 & 0\\
 & \large \textbf{6-v} & 0.29 & 0.26 & 0.45 & 0.091 & 0.142 & 0.767\\
 \addlinespace
2 & 1 & 0.61 & 0.39 & 0 & 0.632 & 0.368 & 0\\
 & 2-v & 0.585 & 0.415 & 0 & 0.641 & 0.359 & 0\\
 & \color{garnet} 3 & \color{garnet} 0.025 & \color{garnet} 0.055 & \color{garnet} 0.92 & \color{garnet} 0.009 & \color{garnet} 0.014 & \color{garnet} 0.978\\
 & \color{garnet} 4 & \color{garnet} 0.05 & \color{garnet} 0.05 & \color{garnet} 0.9 & \color{garnet} 0.017 & \color{garnet} 0.014 & \color{garnet} 0.969\\
 & 5 & 0.36 & 0.64 & 0 & 0.424 & 0.576 & 0\\
 & 6-v & 0.445 & 0.555 & 0 & 0.401 & 0.589 & 0.01\\
\addlinespace
3 & 1 & 0.63 & 0.37 & 0 & 0.615 & 0.385 & 0\\
 & 2-v & 0.465 & 0.465 & 0.07 & 0.379 & 0.384 & 0.237\\
 & \color{garnet} 3 & \color{garnet} 0.13 & \color{garnet} 0.095 & \color{garnet} 0.775 & \color{garnet} 0.037 & \color{garnet} 0.034 & \color{garnet} 0.929\\
 & \color{garnet} 4 & \color{garnet} 0.095 & \color{garnet} 0.15 & \color{garnet} 0.755 & \color{garnet} 0.019 & \color{garnet} 0.03 & \color{garnet} 0.951\\
 & 5 & 0.44 & 0.56 & 0 & 0.426 & 0.574 & 0\\
 & 6-v & 0.645 & 0.355 & 0 & 0.54 & 0.46 & 0\\
\addlinespace
4 & 1 & 0.57 & 0.43 & 0 & 0.469 & 0.53 & 0.001\\
 & 2-v & 0.555 & 0.39 & 0.055 & 0.368 & 0.431 & 0.201\\
 & \color{garnet} 3 & \color{garnet} 0.065 & \color{garnet} 0.065 & \color{garnet} 0.87 & \color{garnet} 0.007 & \color{garnet} 0.01 & \color{garnet} 0.983\\
 & \color{garnet} 4 & \color{garnet} 0.085 & \color{garnet} 0.055 & \color{garnet} 0.86 & \color{garnet} 0.007 & \color{garnet} 0.01 & \color{garnet} 0.983\\
 & 5 & 0.63 & 0.37 & 0 & 0.505 & 0.495 & 0\\
 & 6-v & 0.45 & 0.55 & 0 & 0.495 & 0.505 & 0\\
\addlinespace
5 & 1 & 0.545 & 0.455 & 0 & 0.586 & 0.414 & 0\\
 & 2-v & 0.515 & 0.485 & 0 & 0.602 & 0.392 & 0.006\\
 & \color{garnet} 3 & \color{garnet} 0.025 & \color{garnet} 0.045 & \color{garnet} 0.93 & \color{garnet} 0.007 & \color{garnet} 0.006 & \color{garnet} 0.986\\
 & \color{garnet} 4 & \color{garnet} 0.065 & \color{garnet} 0.04 & \color{garnet} 0.895 & \color{garnet} 0.012 & \color{garnet} 0.01 & \color{garnet} 0.978\\
 & 5 & 0.525 & 0.475 & 0 & 0.439 & 0.559 & 0.002\\
 & 6-v & 0.355 & 0.645 & 0 & 0.425 & 0.575 & 0\\
\addlinespace
6 & 1 & 0.765 & 0.23 & 0.005 & 0.758 & 0.233 & 0.009\\
 & 2-v & 0.645 & 0.355 & 0 & 0.682 & 0.314 & 0.004\\
 & \color{garnet} 3 & \color{garnet} 0.15 & \color{garnet} 0.07 & \color{garnet} 0.78 & \color{garnet} 0.02 & \color{garnet} 0.015 & \color{garnet} 0.965\\
 & \color{garnet} 4 & \color{garnet} 0.06 & \color{garnet} 0.05 & \color{garnet} 0.89 & \color{garnet} 0.007 & \color{garnet} 0.006 & \color{garnet} 0.986\\
 & 5 & 0.33 & 0.67 & 0 & 0.232 & 0.768 & 0\\
 & 6-v & 0.375 & 0.625 & 0 & 0.283 & 0.711 & 0.006\\
\addlinespace
7 & 1 & 0.425 & 0.575 & 0 & 0.46 & 0.54 & 0\\
 & 2-v & 0.4 & 0.6 & 0 & 0.5 & 0.5 & 0\\
 & \color{garnet} 3 & \color{garnet} 0.055 & \color{garnet} 0.06 & \color{garnet} 0.885 & \color{garnet} 0.006 & \color{garnet} 0.012 & \color{garnet} 0.981\\
 & \color{garnet} 4 & \color{garnet} 0.06 & \color{garnet} 0.06 & \color{garnet} 0.88 & \color{garnet} 0.012 & \color{garnet} 0.017 & \color{garnet} 0.97\\
 & 5 & 0.5 & 0.5 & 0 & 0.569 & 0.431 & 0\\
 & 6-v & 0.45 & 0.55 & 0 & 0.534 & 0.46 & 0.006\\
\bottomrule
\end{tabu}
\end{table}


\begin{table}[H]
\caption{\label{stab:q8.baycn} Posterior probabilities from baycn on the GEUVADIS eQTL-gene set Q8 with five associated PCs included in the network as confounding variables. A fully connected graph (excluding the edges between PC nodes) was used as the input to baycn. The rows highlighted in yellow indicate the edges between the nodes of interest.}
\centering
\begin{tabu}{@{} l l l l @{}}
\toprule
edge & zero & one & two \\
\midrule
\rowcolor{yellow!50}
rs11305802-TMEM55B & 0.170 & 0.000 & 0.830\\
\rowcolor{yellow!50}
rs11305802-RP11-203M5.8 & 0.805 & 0.000 & 0.195\\
\rowcolor{yellow!50}
rs11305802-PNP & 1.000 & 0.000 & 0.000\\
rs11305802-PC1 & 0.240 & 0.000 & 0.760\\
rs11305802-PC2 & 0.195 & 0.000 & 0.805\\
\addlinespace
rs11305802-PC6 & 0.255 & 0.000 & 0.745\\
rs11305802-PC7 & 0.125 & 0.000 & 0.875\\
rs11305802-PC9 & 0.370 & 0.000 & 0.630\\
\rowcolor{yellow!50}
TMEM55B-RP11-203M5.8 & 0.150 & 0.820 & 0.030\\
\rowcolor{yellow!50}
TMEM55B-PNP & 0.330 & 0.670 & 0.000\\
\addlinespace
TMEM55B-PC1 & 0.385 & 0.615 & 0.000\\
TMEM55B-PC2 & 0.085 & 0.550 & 0.365\\
TMEM55B-PC6 & 0.015 & 0.055 & 0.930\\
TMEM55B-PC7 & 0.050 & 0.030 & 0.920\\
TMEM55B-PC9 & 0.260 & 0.730 & 0.010\\
\addlinespace
\rowcolor{yellow!50}
RP11-203M5.8-PNP & 0.760 & 0.240 & 0.000\\
RP11-203M5.8-PC1 & 0.075 & 0.145 & 0.780\\
RP11-203M5.8-PC2 & 0.085 & 0.240 & 0.675\\
RP11-203M5.8-PC6 & 0.315 & 0.685 & 0.000\\
RP11-203M5.8-PC7 & 0.175 & 0.115 & 0.710\\
\addlinespace
RP11-203M5.8-PC9 & 0.035 & 0.060 & 0.905\\
PNP-PC1 & 0.190 & 0.715 & 0.095\\
PNP-PC2 & 0.135 & 0.845 & 0.020\\
PNP-PC6 & 0.040 & 0.065 & 0.895\\
PNP-PC7 & 0.565 & 0.180 & 0.255\\
\addlinespace
PNP-PC9 & 0.165 & 0.695 & 0.140\\
\bottomrule
\end{tabu}
\end{table}


\begin{table}[H]
\centering
\caption{\label{stab:q8.order} Posterior probabilities from order MCMC on the GEUVADIS eQTL-gene set Q8 with associated PCs.  
A fully connected graph (excluding the edges between PC nodes) was used as the input.}
\begin{tabular}{lrrr}
\toprule
edge & zero & one & two \\ 
  \hline
 rs11305802-TMEM55B & 0.00 & 0.00 & 1.00 \\ 
 rs11305802-RP11-203M5.8 & 0.25 & 0.00 & 0.75 \\ 
 rs11305802-PNP & 1.00 & 0.00 & 0.00 \\ 
 rs11305802-PC1 & 0.00 & 0.00 & 1.00 \\ 
 rs11305802-PC2 & 0.00 & 0.00 & 1.00 \\ 
 rs11305802-PC6 & 0.00 & 0.00 & 1.00 \\ 
 rs11305802-PC7 & 0.00 & 0.00 & 1.00 \\ 
 rs11305802-PC9 & 0.00 & 0.00 & 1.00 \\ 
 TMEM55B-RP11-203M5.8 & 0.37 & 0.63 & -0.00 \\ 
 TMEM55B-PNP & 0.08 & 0.92 & 0.00 \\ 
 TMEM55B-PC1 & 0.00 & 0.00 & 1.00 \\ 
 TMEM55B-PC2 & 0.00 & 0.00 & 1.00 \\ 
 TMEM55B-PC6 & 0.00 & 0.00 & 1.00 \\ 
 TMEM55B-PC7 & 0.00 & 0.00 & 1.00 \\ 
 TMEM55B-PC9 & 0.00 & 0.00 & 1.00 \\ 
 RP11-203M5.8-PNP & 0.25 & 0.75 & 0.00 \\ 
 RP11-203M5.8-PC1 & 0.00 & 0.00 & 1.00 \\ 
 RP11-203M5.8-PC2 & 0.00 & 0.00 & 1.00 \\ 
 RP11-203M5.8-PC6 & 0.00 & 0.22 & 0.78 \\ 
 RP11-203M5.8-PC7 & 0.00 & 0.00 & 1.00 \\ 
 RP11-203M5.8-PC9 & 0.00 & 0.00 & 1.00 \\ 
 PNP-PC1 & 0.00 & 0.00 & 1.00 \\ 
 PNP-PC2 & 0.00 & 0.00 & 1.00 \\ 
 PNP-PC6 & 0.00 & 0.00 & 1.00 \\ 
 PNP-PC7 & 0.00 & 0.00 & 1.00 \\ 
 PNP-PC9 & 0.00 & 0.00 & 1.00 \\ 
\bottomrule
\end{tabular}
\end{table}


\begin{table}[H]
\centering
\caption{\label{stab:q8.partition} Posterior probabilities from partition MCMC on the GEUVADIS eQTL-gene set Q8 with associated PCs. 
A fully connected graph (excluding the edges between PC nodes) was used as the input.}
\begin{tabular}{lrrr}
  \toprule
 edge & zero & one & two \\ 
  \hline
rs11305802-TMEM55B & 0.02 & 0.00 & 0.98 \\ 
  rs11305802-RP11-203M5.8 & 0.32 & 0.00 & 0.68 \\ 
  rs11305802-PNP & 1.00 & 0.00 & 0.00 \\ 
  rs11305802-PC1 & 0.00 & 0.00 & 1.00 \\ 
  rs11305802-PC2 & 0.00 & 0.00 & 1.00 \\ 
  rs11305802-PC6 & 0.00 & 0.00 & 1.00 \\ 
  rs11305802-PC7 & 0.00 & 0.00 & 1.00 \\ 
  rs11305802-PC9 & 0.00 & 0.00 & 1.00 \\ 
  TMEM55B-RP11-203M5.8 & 0.34 & 0.51 & 0.16 \\ 
  TMEM55B-PNP & 0.06 & 0.94 & 0.00 \\ 
  TMEM55B-PC1 & 0.15 & 0.00 & 0.85 \\ 
  TMEM55B-PC2 & 0.00 & 0.00 & 1.00 \\ 
  TMEM55B-PC6 & 0.00 & 0.00 & 1.00 \\ 
  TMEM55B-PC7 & 0.00 & 0.00 & 1.00 \\ 
  TMEM55B-PC9 & 0.00 & 0.04 & 0.96 \\ 
  RP11-203M5.8-PNP & 0.22 & 0.78 & 0.00 \\ 
  RP11-203M5.8-PC1 & 0.00 & 0.00 & 1.00 \\ 
  RP11-203M5.8-PC2 & 0.00 & 0.00 & 1.00 \\ 
  RP11-203M5.8-PC6 & 0.34 & 0.26 & 0.40 \\ 
  RP11-203M5.8-PC7 & 0.00 & 0.00 & 1.00 \\ 
  RP11-203M5.8-PC9 & 0.00 & 0.00 & 1.00 \\ 
  PNP-PC1 & 0.00 & 0.01 & 0.99 \\ 
  PNP-PC2 & 0.00 & 0.00 & 1.00 \\ 
  PNP-PC6 & 0.00 & 0.00 & 1.00 \\ 
  PNP-PC7 & 0.00 & 0.00 & 1.00 \\ 
  PNP-PC9 & 0.00 & 0.00 & 1.00 \\ 
   \bottomrule
\end{tabular}
\end{table}


\begin{table}[H]
\centering
\caption{\label{stab:q8.scan} Posterior probabilities from scanBMA on the GEUVADIS eQTL-gene set Q8 with associated PCs. 
A fully connected graph (excluding the edges between PC nodes) was used as the input.}
\begin{tabular}{lrrr}
  \toprule
 edge & zero & one & two \\ 
  \hline
rs11305802-TMEM55B & 0.00 & 0.00 & NA \\ 
  rs11305802-RP11-203M5.8 & 0.07 & 0.00 & NA \\ 
  rs11305802-PNP & 1.00 & 0.00 & NA \\ 
  rs11305802-PC1 & 0.07 & 0.00 & NA \\ 
  rs11305802-PC2 & 0.00 & 0.00 & NA \\ 
  rs11305802-PC6 & 0.20 & 0.00 & NA \\ 
  rs11305802-PC7 & 0.00 & 0.00 & NA \\ 
  rs11305802-PC9 & 0.27 & 0.00 & NA \\ 
  TMEM55B-RP11-203M5.8 & 0.84 & 0.80 & NA \\ 
  TMEM55B-PNP & 1.00 & 1.00 & NA \\ 
  TMEM55B-PC1 & 1.00 & 1.00 & NA \\ 
  TMEM55B-PC2 & 0.28 & 0.23 & NA \\ 
  TMEM55B-PC6 & 0.00 & 0.00 & NA \\ 
  TMEM55B-PC7 & 0.00 & 0.00 & NA \\ 
  TMEM55B-PC9 & 0.94 & 1.00 & NA \\ 
  RP11-203M5.8-PNP & 1.00 & 1.00 & NA \\ 
  RP11-203M5.8-PC1 & 0.00 & 0.00 & NA \\ 
  RP11-203M5.8-PC2 & 0.50 & 0.77 & NA \\ 
  RP11-203M5.8-PC6 & 1.00 & 1.00 & NA \\ 
  RP11-203M5.8-PC7 & 0.12 & 0.00 & NA \\ 
  RP11-203M5.8-PC9 & 0.02 & 0.03 & NA \\ 
  PNP-PC1 & 0.19 & 0.68 & NA \\ 
  PNP-PC2 & 0.76 & 0.93 & NA \\ 
  PNP-PC6 & 0.04 & 0.00 & NA \\ 
  PNP-PC7 & 0.57 & 0.17 & NA \\ 
  PNP-PC9 & 0.61 & 0.96 & NA \\ 
  \bottomrule
\end{tabular}
\end{table}

\begin{table}[H]
\centering
\caption{\label{stab:q8} Posterior probabilities from baycn on the GEUVADIS eQTL-gene set Q8 without the associated PCs. A fully connected graph was used as  the input.}
\begin{tabu}{@{} l l l l @{}}
\toprule
edge & zero & one & two \\
\midrule
rs11305802-TMEM55B & 0.210 & 0.000 & 0.79 \\
rs11305802-RP11-203M5.8 & 0.410 & 0.000 & 0.59 \\
rs11305802-PNP & 1.000 & 0.000 & 0.00 \\
TMEM55B-RP11-203M5.8 & 0.525 & 0.465 & 0.01 \\
TMEM55B-PNP & 0.155 & 0.845 & 0.00 \\
RP11-203M5.8-PNP & 0.185 & 0.815 & 0.00 \\
\bottomrule
\end{tabu}
\end{table}

\begin{table}[H]
\centering
\caption{\label{stab:q20} Posterior probabilities from baycn on the GEUVADIS eQTL-gene set Q20. A fully connected graph was used as  the input.}
\begin{tabu}{@{} l l l l @{}}
\toprule
edge & zero & one & two \\
\midrule
rs142060986-RP11292F9.1 & 0 & 0 & 1 \\
rs142060986-FAM27E1 & 1 & 0 & 0 \\
RP11292F9.1-FAM27E1 & 0 & 1 & 0 \\
\bottomrule
\end{tabu}
\end{table}


\begin{table}[H]
\caption{\label{stab:q21_pcs} Posterior probabilities from baycn for the GEUVADIS eQTL-gene set Q21 with two PCs are included in the network as confounding variables. A fully connected graph was used as the input. The rows highlighted in yellow indicate the edges between the nodes of interest.}
\centering
\begin{tabu}{@{} l l l l @{}}
\toprule
edge & zero & one & two \\
\midrule
\rowcolor{yellow!50}
rs147156488-FAM27C & 0.000 & 0.000 & 1.000\\
\rowcolor{yellow!50}
rs147156488-FAM27A & 0.000 & 0.000 & 1.000\\
\rowcolor{yellow!50}
rs147156488-FAM27D1 & 1.000 & 0.000 & 0.000\\
rs147156488-PC1 & 1.000 & 0.000 & 0.000\\
rs147156488-PC3 & 1.000 & 0.000 & 0.000\\
\addlinespace
\rowcolor{yellow!50}
FAM27C-FAM27A & 0.510 & 0.030 & 0.460\\
\rowcolor{yellow!50}
FAM27C-FAM27D1 & 0.000 & 1.000 & 0.000\\
FAM27C-PC1 & 0.250 & 0.030 & 0.720\\
FAM27C-PC3 & 0.180 & 0.140 & 0.680\\
\rowcolor{yellow!50}
FAM27A-FAM27D1 & 0.000 & 0.510 & 0.490\\
\addlinespace
FAM27A-PC1 & 0.145 & 0.050 & 0.805\\
FAM27A-PC3 & 0.555 & 0.115 & 0.330\\
FAM27D1-PC1 & 0.245 & 0.220 & 0.535\\
FAM27D1-PC3 & 0.160 & 0.125 & 0.715\\
\bottomrule
\end{tabu}
\end{table}


\begin{table}[H]
\caption{\label{stab:q23_pcs} Posterior probabilities from baycn on the GEUVADIS eQTL-gene set Q23 with two PCs included in the network as confounding variables. A fully connected graph was used as input. The rows highlighted in yellow indicate the edges between the nodes of interest.}
\centering
\begin{tabu}{@{} l l l l @{}}
\toprule
edge & zero & one & two \\
\midrule
\rowcolor{yellow!50}
rs150605045-AGAP9 & 1.000 & 0.000 & 0.000\\
\rowcolor{yellow!50}
rs150605045-AGAP10 & 1.000 & 0.000 & 0.000\\
rs150605045-PC1 & 1.000 & 0.000 & 0.000\\
rs150605045-PC8 & 1.000 & 0.000 & 0.000\\
\rowcolor{yellow!50}
AGAP9-AGAP10 & 0.720 & 0.280 & 0.000\\
\addlinespace
AGAP9-PC1 & 0.595 & 0.385 & 0.020\\
AGAP9-PC8 & 0.495 & 0.455 & 0.050\\
AGAP10-PC1 & 0.065 & 0.065 & 0.870\\
AGAP10-PC8 & 0.035 & 0.080 & 0.885\\
\bottomrule
\end{tabu}
\end{table}


\begin{table}[H]
\caption{\label{stab:q37_pcs} Posterior probabilities from baycn on the GEUVADIS eQTL-gene set Q37 with one PC included in the network as a confounding variable. A fully connected graph was used as input. The rows highlighted in yellow indicate the edges between the nodes of interest.}
\centering
\begin{tabu}{@{} l l l l @{}}
\toprule
edge & zero & one & two \\
\midrule
\rowcolor{yellow!50}
rs3858954-GOLGA6L19 & 0.660 & 0.000 & 0.340\\
\rowcolor{yellow!50}
rs3858954-GOLGA6L9 & 0.335 & 0.000 & 0.665\\
\rowcolor{yellow!50}
rs3858954-GOLGA6L20 & 0.965 & 0.000 & 0.035\\
rs3858954-PC1 & 0.100 & 0.000 & 0.900\\
\rowcolor{yellow!50}
GOLGA6L19-GOLGA6L9 & 0.035 & 0.055 & 0.910\\
\addlinespace
\rowcolor{yellow!50}
GOLGA6L19-GOLGA6L20 & 0.335 & 0.665 & 0.000\\
GOLGA6L19-PC1 & 0.665 & 0.210 & 0.125\\
\rowcolor{yellow!50}
GOLGA6L9-GOLGA6L20 & 0.175 & 0.825 & 0.000\\
GOLGA6L9-PC1 & 0.110 & 0.030 & 0.860\\
GOLGA6L20-PC1 & 0.135 & 0.050 & 0.815\\
\bottomrule
\end{tabu}
\end{table}


\begin{table}[H]
\caption{\label{stab:q50_pcs} Posterior probabilities from baycn on the GEUVADIS eQTL-gene set Q50 with two PCs included in the network as confounding variables. A fully connected graph was used as input. The rows highlighted in yellow indicate the edges between the nodes of interest.}
\centering
\begin{tabu}{@{} l l l l @{}}
\toprule
edge & zero & one & two \\
\midrule
\rowcolor{yellow!50}
rs7124238-SBF2-AS1 & 0.195 & 0.000 & 0.805\\
\rowcolor{yellow!50}
rs7124238-SWAP70 & 1.000 & 0.000 & 0.000\\
rs7124238-PC1 & 0.135 & 0.000 & 0.865\\
rs7124238-PC5 & 0.465 & 0.000 & 0.535\\
\rowcolor{yellow!50}
SBF2-AS1-SWAP70 & 0.035 & 0.965 & 0.000\\
\addlinespace
SBF2-AS1-PC1 & 0.035 & 0.020 & 0.945\\
SBF2-AS1-PC5 & 0.015 & 0.040 & 0.945\\
SWAP70-PC1 & 0.550 & 0.415 & 0.035\\
SWAP70-PC5 & 0.280 & 0.720 & 0.000\\
\bottomrule
\end{tabu}
\end{table}


\begin{table}[H]
\caption{\label{stab:q62} Posterior probabilities from baycn on the GEUVADIS eQTL-gene set Q62. A fully connected graph was used as input.}
\centering
\begin{tabu}{@{} l l l l @{}}
\toprule
edge & zero & one & two \\
\midrule
rs9426902-S100A6 & 0.845 & 0.00 & 0.155 \\
rs9426902-S100A4 & 0.420 & 0.00 & 0.580 \\
S100A6-S100A4 & 0.720 & 0.28 & 0.000 \\
\bottomrule
\end{tabu}
\end{table}


\begin{sidewaystable}[ph!]
\caption{{Posterior probability for each edge state for baycn, order MCMC, partition MCMC, and scanBMA on the Drosophila data.}}
\label{stab:drosophila.all}
\centering
\begin{tabular}{lrrrrrrrrrrrr}
\toprule
& \multicolumn{3}{c}{baycn} & \multicolumn{3}{c}{order MCMC} & \multicolumn{3}{c}{partition MCMC} & \multicolumn{3}{c}{scanBMA} \\
\cmidrule(lr){2-4} \cmidrule(lr){5-7} \cmidrule(lr){8-10} \cmidrule(lr){11-13}
edge & zero & one & two & zero & one & two & zero & one & two & zero & one & two\\
\midrule
Meso-Twi\_2\_4h & 0.282 & 0.644 & 0.074 & 0.026 & 0.747 & 0.227 & 0.084 & 0.385 & 0.531 & 0.248 & 0.800 & NA\\
VM-Bin\_6\_8h & 0.126 & 0.462 & 0.412 & 0.006 & 0.161 & 0.833 & 0.004 & 0.388 & 0.608 & 0.032 & 0.208 & NA\\
VM-Bin\_8\_10h & 0.172 & 0.474 & 0.354 & 0.087 & 0.677 & 0.236 & 0.012 & 0.387 & 0.601 & 0.392 & 0.913 & NA\\
VM-Bin\_10\_12h & 0.076 & 0.054 & 0.870 & 0.000 & 0.000 & 1.000 & 0.000 & 0.030 & 0.970 & 0.000 & 0.000 & NA\\
SM-Mef2\_10\_12h & 0.440 & 0.208 & 0.352 & 0.000 & 0.000 & 1.000 & 0.046 & 0.062 & 0.892 & 0.321 & 0.235 & NA\\
\addlinespace
CM-Bin\_10\_12h & 0.420 & 0.196 & 0.384 & 0.000 & 0.000 & 1.000 & 0.075 & 0.081 & 0.844 & 0.767 & 0.327 & NA\\
Meso\_SM-Tin\_2\_4h & 0.102 & 0.534 & 0.364 & 0.212 & 0.163 & 0.625 & 0.027 & 0.395 & 0.577 & 0.296 & 0.757 & NA\\
Meso\_SM-Tin\_4\_6h & 0.024 & 0.164 & 0.812 & 0.004 & 0.027 & 0.969 & 0.011 & 0.087 & 0.901 & 0.000 & 0.000 & NA\\
Meso\_SM-Tin\_6\_8h & 0.180 & 0.040 & 0.780 & 0.000 & 0.000 & 1.000 & 0.005 & 0.004 & 0.991 & 0.000 & 0.000 & NA\\
Meso\_SM-Mef2\_2\_4h & 0.226 & 0.556 & 0.218 & 0.002 & 0.389 & 0.608 & 0.059 & 0.337 & 0.605 & 1.000 & 1.000 & NA\\
\addlinespace
Meso\_SM-Mef2\_4\_6h & 0.174 & 0.282 & 0.544 & 0.004 & 0.392 & 0.605 & 0.016 & 0.313 & 0.671 & 0.094 & 0.062 & NA\\
Meso\_SM-Mef2\_6\_8h & 0.024 & 0.050 & 0.926 & 0.000 & 0.000 & 1.000 & 0.001 & 0.000 & 0.999 & 0.000 & 0.000 & NA\\
Meso\_SM-Mef2\_8\_10h & 0.052 & 0.056 & 0.892 & 0.000 & 0.000 & 1.000 & 0.002 & 0.007 & 0.990 & 0.000 & 0.000 & NA\\
Meso\_SM-Mef2\_10\_12h & 0.036 & 0.048 & 0.916 & 0.000 & 0.000 & 1.000 & 0.007 & 0.010 & 0.983 & 0.000 & 0.000 & NA\\
VM\_SM-Bin\_10\_12h & 0.504 & 0.282 & 0.214 & 0.000 & 0.000 & 1.000 & 0.000 & 0.000 & 1.000 & 0.640 & 0.787 & NA\\
\addlinespace
VM\_SM-Twi\_2\_4h & 0.378 & 0.622 & 0.000 & 0.180 & 0.820 & 0.000 & 0.707 & 0.288 & 0.005 & 0.734 & 0.526 & NA\\
VM\_SM-Mef2\_10\_12h & 0.738 & 0.262 & 0.000 & 0.712 & 0.288 & 0.000 & 0.414 & 0.579 & 0.007 & 1.000 & 1.000 & NA\\
Tin\_2\_4h-Tin\_4\_6h & 0.332 & 0.668 & 0.000 & 0.461 & 0.539 & 0.000 & 0.943 & 0.057 & 0.000 & 1.000 & 1.000 & NA\\
Tin\_2\_4h-Twi\_4\_6h & 0.526 & 0.464 & 0.010 & 0.197 & 0.349 & 0.454 & 0.001 & 0.945 & 0.054 & 0.970 & 1.000 & NA\\
Tin\_4\_6h-Twi\_4\_6h & 0.548 & 0.382 & 0.070 & 0.362 & 0.094 & 0.545 & 0.001 & 0.055 & 0.944 & 0.219 & 0.342 & NA\\
\addlinespace
Tin\_4\_6h-Twi\_6\_8h & 0.522 & 0.476 & 0.002 & 0.445 & 0.448 & 0.107 & 0.000 & 0.944 & 0.056 & 1.000 & 1.000 & NA\\
Tin\_4\_6h-Mef2\_2\_4h & 0.932 & 0.066 & 0.002 & 0.939 & 0.016 & 0.045 & 0.796 & 0.000 & 0.204 & 0.780 & 0.167 & NA\\
Tin\_4\_6h-Mef2\_4\_6h & 0.972 & 0.028 & 0.000 & 0.675 & 0.090 & 0.236 & 0.002 & 0.057 & 0.940 & 1.000 & 0.952 & NA\\
Tin\_6\_8h-Bin\_6\_8h & 0.300 & 0.576 & 0.124 & 0.559 & 0.056 & 0.385 & 0.247 & 0.000 & 0.753 & 0.921 & 0.682 & NA\\
Tin\_6\_8h-Bap\_6\_8h & 0.494 & 0.506 & 0.000 & 0.894 & 0.106 & 0.000 & 0.980 & 0.020 & 0.000 & 1.000 & 1.000 & NA\\
\addlinespace
Bin\_6\_8h-Bin\_8\_10h & 0.604 & 0.396 & 0.000 & 0.546 & 0.454 & 0.000 & 0.961 & 0.039 & 0.000 & 1.000 & 1.000 & NA\\
Bin\_6\_8h-Bin\_10\_12h & 0.494 & 0.408 & 0.098 & 0.000 & 0.000 & 1.000 & 0.001 & 0.000 & 0.999 & 0.973 & 1.000 & NA\\
Bin\_6\_8h-Bap\_6\_8h & 0.746 & 0.216 & 0.038 & 0.030 & 0.355 & 0.615 & 0.009 & 0.761 & 0.231 & 0.974 & 0.571 & NA\\
Bin\_8\_10h-Bin\_10\_12h & 0.500 & 0.500 & 0.000 & 0.894 & 0.106 & 0.000 & 1.000 & 0.000 & 0.000 & 1.000 & 1.000 & NA\\
Twi\_2\_4h-Twi\_4\_6h & 0.224 & 0.776 & 0.000 & 0.319 & 0.680 & 0.001 & 0.556 & 0.443 & 0.001 & 1.000 & 1.000 & NA\\
\addlinespace
Twi\_4\_6h-Twi\_6\_8h & 0.400 & 0.596 & 0.004 & 0.509 & 0.145 & 0.347 & 0.678 & 0.318 & 0.004 & 1.000 & 1.000 & NA\\
Twi\_4\_6h-Mef2\_4\_6h & 0.992 & 0.008 & 0.000 & 0.640 & 0.360 & 0.000 & 0.772 & 0.228 & 0.000 & 1.000 & 1.000 & NA\\
Twi\_6\_8h-Mef2\_6\_8h & 0.890 & 0.110 & 0.000 & 0.824 & 0.176 & 0.000 & 0.642 & 0.358 & 0.000 & 1.000 & 1.000 & NA\\
Mef2\_2\_4h-Mef2\_6\_8h & 0.604 & 0.284 & 0.112 & 0.002 & 0.005 & 0.993 & 0.015 & 0.126 & 0.859 & 0.170 & 0.686 & NA\\
Mef2\_2\_4h-Mef2\_10\_12h & 0.310 & 0.610 & 0.080 & 0.175 & 0.782 & 0.044 & 0.041 & 0.607 & 0.352 & 0.705 & 0.891 & NA\\
\addlinespace
Mef2\_4\_6h-Mef2\_6\_8h & 0.008 & 0.992 & 0.000 & 0.107 & 0.125 & 0.768 & 0.054 & 0.946 & 0.000 & 1.000 & 1.000 & NA\\
Mef2\_6\_8h-Mef2\_8\_10h & 0.624 & 0.376 & 0.000 & 0.138 & 0.862 & 0.000 & 0.264 & 0.736 & 0.000 & 1.000 & 1.000 & NA\\
Mef2\_8\_10h-Mef2\_10\_12h & 0.172 & 0.828 & 0.000 & 0.137 & 0.863 & 0.000 & 0.455 & 0.545 & 0.000 & 1.000 & 1.000 & NA\\
\bottomrule
\end{tabular}
\end{sidewaystable}

\clearpage



\begin{sidewaystable}[ph!]
\caption{{Posterior probability for each edge state for baycn, order MCMC, partition MCMC, and scanBMA on the Q8 network from GEUVADIS.} All possible edges were input to each method.}
\centering
\begin{tabular}{lrrrrrrrrrrrr}
\toprule
& \multicolumn{3}{c}{baycn} & \multicolumn{3}{c}{order MCMC}  & \multicolumn{3}{c}{partition MCMC}  & \multicolumn{3}{c}{scanBMA}\\
\cmidrule(lr){2-4} \cmidrule(lr){5-7} \cmidrule(lr){8-10} \cmidrule(lr){11-13}
edges & zero & one & two & zero & one & two & zero & one & two & zero & one & two\\
\midrule
rs11305802-TMEM55B & 0.210 & 0.00 & 0.79 & 0.002 & 0.000 & 0.998 & 0.027 & 0.000 & 0.973 & 0.000 & 0.000 & NA\\
rs11305802-RP11-203M5.8 & 0.410 & 0.00 & 0.59 & 0.089 & 0.000 & 0.911 & 0.177 & 0.000 & 0.823 & 0.063 & 0.072 & NA\\
rs11305802-PNP & 1.000 & 0.00 & 0.000 & 1.000 & 0.000 & 0.000 & 1.000 & 0.000 & 0.000 & 1.000 & 1.000 & NA\\
TMEM55B-RP11-203M5.8 & 0.525 & 0.465 & 0.01 & 0.445 & 0.555 & 0.000 & 0.423 & 0.424 & 0.153 & 0.844 & 0.850 & NA\\
TMEM55B-PNP & 0.155 & 0.845 & 0.000 & 0.040 & 0.960 & 0.000 & 0.025 & 0.975 & 0.000 & 1.000 & 1.000 & NA\\
RP11-203M5.8-PNP & 0.185 & 0.815 & 0.000 & 0.089 & 0.911 & 0.000 & 0.064 & 0.936 & 0.000 & 1.000 & 1.000 & NA\\
\bottomrule
\end{tabular}
\end{sidewaystable}


\begin{table}[H]
\caption{\label{tab:drosophila} Posterior probabilities inferred by baycn for the Drosophila data set. The graph output from MRPC was used as the input to baycn with the addition of the edges: VM-bin\_6.8, VM-bin\_10.12, Meso\_SM-tin\_4.6, Meso\_SM-tin\_6.8, Meso\_SM-mef2\_4.6, Meso\_SM-mef2\_6.8, Meso\_SM-mef2\_8.10, and Meso\_SM-mef2\_10.12.}
\centering
\begin{tabu}{@{} l l l l @{}}
\toprule
& \multicolumn{3}{c}{Posterior Probability} \\
\cmidrule(lr){2-4}
edge & zero & one & two \\
\midrule
Meso-twi\_2.4 & 0.282 & 0.644 & 0.074\\
VM-bin\_6.8 & 0.126 & 0.462 & 0.412\\
VM-bin\_8.10 & 0.172 & 0.474 & 0.354\\
VM-bin\_10.12 & 0.076 & 0.054 & 0.870\\
SM-mef2\_10.12 & 0.440 & 0.208 & 0.352\\
\addlinespace
CM-bin\_10.12 & 0.420 & 0.196 & 0.384\\
Meso\_SM-tin\_2.4 & 0.102 & 0.534 & 0.364\\
Meso\_SM-tin\_4.6 & 0.024 & 0.164 & 0.812\\
Meso\_SM-tin\_6.8 & 0.180 & 0.040 & 0.780\\
Meso\_SM-mef2\_2.4 & 0.226 & 0.556 & 0.218\\
\addlinespace
Meso\_SM-mef2\_4.6 & 0.174 & 0.282 & 0.544\\
Meso\_SM-mef2\_6.8 & 0.024 & 0.050 & 0.926\\
Meso\_SM-mef2\_8.10 & 0.052 & 0.056 & 0.892\\
Meso\_SM-mef2\_10.12 & 0.036 & 0.048 & 0.916\\
VM\_SM-bin\_10.12 & 0.504 & 0.282 & 0.214\\
\addlinespace
VM\_SM-twi\_2.4 & 0.378 & 0.622 & 0.000\\
VM\_SM-mef2\_10.12 & 0.738 & 0.262 & 0.000\\
tin\_2.4-tin\_4.6 & 0.332 & 0.668 & 0.000\\
tin\_2.4-twi\_4.6 & 0.526 & 0.464 & 0.010\\
tin\_4.6-twi\_4.6 & 0.548 & 0.382 & 0.070\\
\addlinespace
tin\_4.6-twi\_6.8 & 0.522 & 0.476 & 0.002\\
tin\_4.6-mef2\_2.4 & 0.932 & 0.066 & 0.002\\
tin\_4.6-mef2\_4.6 & 0.972 & 0.028 & 0.000\\
tin\_6.8-bin\_6.8 & 0.300 & 0.576 & 0.124\\
tin\_6.8-bap\_6.8 & 0.494 & 0.506 & 0.000\\
\addlinespace
bin\_6.8-bin\_8.10 & 0.604 & 0.396 & 0.000\\
bin\_6.8-bin\_10.12 & 0.494 & 0.408 & 0.098\\
bin\_6.8-bap\_6.8 & 0.746 & 0.216 & 0.038\\
bin\_8.10-bin\_10.12 & 0.500 & 0.500 & 0.000\\
twi\_2.4-twi\_4.6 & 0.224 & 0.776 & 0.000\\
\addlinespace
twi\_4.6-twi\_6.8 & 0.400 & 0.596 & 0.004\\
twi\_4.6-mef2\_4.6 & 0.992 & 0.008 & 0.000\\
twi\_6.8-mef2\_6.8 & 0.890 & 0.110 & 0.000\\
mef2\_2.4-mef2\_6.8 & 0.604 & 0.284 & 0.112\\
mef2\_2.4-mef2\_10.12 & 0.310 & 0.610 & 0.080\\
\addlinespace
mef2\_4.6-mef2\_6.8 & 0.008 & 0.992 & 0.000\\
mef2\_6.8-mef2\_8.10 & 0.624 & 0.376 & 0.000\\
mef2\_8.10-mef2\_10.12 & 0.172 & 0.828 & 0.000\\
\bottomrule
\end{tabu}
\end{table}



\FloatBarrier

\bibliographystyle{imsart-nameyear} 
\bibliography{arxiv_2023_11}       

\end{document}